\documentclass[twocolumn,showpacs,aps,floatfix,prd,nofootinbib]{revtex4}
\usepackage{graphicx}
\usepackage{dcolumn}
\usepackage{array, longtable}
\usepackage{epsfig}
\usepackage{amsmath}
\usepackage{colordvi}
\usepackage{hhline}
\usepackage{color}

\newcommand{\BaBarYear}{18}
\newcommand{\BaBarNumber}{008}
\newcommand{\SLACPubNumber}{17344}

 \newcommand{\BaBarType}      {PUB}  

\input babarsym

\def\Ecm          {\ensuremath {E_{\rm c.m.}}\xspace}

\def\mgg  {\ensuremath {m(\gamma\gamma)}\xspace}

\long\def\inst#1{\par\nobreak\kern 4pt\nobreak
    {\it #1}\par\vskip 10pt plus 3pt minus 3pt}

\begin{document}

\begin{flushleft}
\babar-\BaBarType-\BaBarYear/\BaBarNumber \\
SLAC-PUB-\SLACPubNumber \\
\end{flushleft}


\title{\large \bf
\boldmath
Study of the reactions  $\epem\to\pipi\pi^0\pi^0\pi^0$ and
$\pipi\pi^0\pi^0\eta$  at
 center-of-mass energies from threshold to 4.35 GeV using initial-state
radiation
} 

%
\author{J.~P.~Lees}
\author{V.~Poireau}
\author{V.~Tisserand}
\affiliation{Laboratoire d'Annecy-le-Vieux de Physique des Particules (LAPP), Universit\'e de Savoie, CNRS/IN2P3,  F-74941 Annecy-Le-Vieux, France}
\author{E.~Grauges}
\affiliation{Universitat de Barcelona, Facultat de Fisica, Departament ECM, E-08028 Barcelona, Spain }
\author{A.~Palano}
\affiliation{INFN Sezione di Bari and Dipartimento di Fisica, Universit\`a di Bari, I-70126 Bari, Italy }
\author{G.~Eigen}
\affiliation{University of Bergen, Institute of Physics, N-5007 Bergen, Norway }
\author{D.~N.~Brown}
\author{Yu.~G.~Kolomensky}
\affiliation{Lawrence Berkeley National Laboratory and University of California, Berkeley, California 94720, USA }
\author{M.~Fritsch}
\author{H.~Koch}
\author{T.~Schroeder}
\affiliation{Ruhr Universit\"at Bochum, Institut f\"ur Experimentalphysik 1, D-44780 Bochum, Germany }
% ...
\collaboration{The \babar\ Collaboration}
\noaffiliation


\begin{abstract}
We study the processes $\epem\to
\pipi\ppz\piz\gamma$ and $\pipi\ppz\eta\gamma$ in which an energetic
photon is radiated from the initial state. 
The data were collected with the \babar~ detector at SLAC.
 About 14\,000 and 4700 events, respectively, are
selected from a data sample corresponding to an integrated
luminosity of 469~\invfb.
The invariant mass of the hadronic final state defines the effective \epem
center-of-mass energy.  From the mass spectra, 
the first precise measurement of
the $\epem\to\pipi\ppz\piz$ cross section and the first measurement
ever of the $\epem\to\pipi\ppz\eta$  cross section are  performed.
The center-of-mass energies range from threshold to 4.35~\gev. The
systematic uncertainty  is typically between 10 and 13\%.
The contributions from  $\omega\ppz$, $\eta\pipi$, and other
intermediate  states are presented. 
We observe the $J/\psi$ and $\psi(2S)$ in most of these final states and
measure the
corresponding branching fractions, many of them for the first time.
\end{abstract}


\vfill
\maketitle

\setcounter{footnote}{0}


\section{Introduction}
\label{sec:Introduction}

Electron-positron annihilation events with initial-state radiation
(ISR) can be used to study processes over a wide range of energies
below the nominal \epem center-of-mass (c.m.) energy (\Ecm),
as proposed in Ref.~\cite{baier}.
The possibility of exploiting ISR to make precise measurements of
low-energy cross sections at high-luminosity $\phi$ and $B$ factories
is discussed in Refs.~\cite{arbus, kuehn, ivanch},
and motivates the studies described in this paper.  
Such measurements are of particular  interest because of a $\sim$3.5
standard-deviation discrepancy between the measured value of the
muon anomalous magnetic moment ($g_\mu-2$) and 
the Standard Model value~\cite{dehz},
 where the Standard Model calculation requires input
from experimental \epem hadronic cross section
data in order to account for hadronic vacuum polarization (HVP) terms.
The calculation is most sensitive to the low-energy region, 
where the inclusive hadronic cross section cannot be measured
reliably and a sum of exclusive states must be used.
Not all accessible states have yet been measured, 
and new measurements will improve the reliability of the calculation.
In addition, studies of ISR events at $B$ factories 
are interesting in their own right, because they provide
information on resonance spectroscopy for masses up to the
charmonium region. 
                                
Studies of the ISR processes $e^+e^-\to\mumu\gamma$~\cite{Druzhinin1,isr2pi}
and $\epem \to X_h\gamma$,  
using data from the \babar\ experiment at SLAC,
have been previously reported.
Here $X_h$ represents any of several exclusive hadronic final states.
The $X_h$ studied to date include:
charged hadron pairs $\pip\pim$~\cite{isr2pi}, $\Kp\Km$~\cite{isr2k}, and
$p\overline{p}$~\cite{isr2p};
four or six charged mesons~\cite{isr4pi,isr2k2pi,isr6pi};
charged mesons plus one or two \piz mesons~\cite{isr2k2pi,isr6pi,isr3pi,isr5pi,isr2pi2pi0};
a \KS meson plus charged and neutral mesons~\cite{isrkkpi}; 
and channels with  \KL mesons~\cite{isrkskl}. 
The ISR events are characterized by good
reconstruction efficiency and by well understood kinematics (see for
example Ref.~\cite{isr3pi}), tracking,
 particle identification, and \piz, \KS, and
 \KL reconstruction, demonstrated in above references.
 
This paper reports analyses of the $\pipi3\piz$ and
$\pipi2\piz\eta$ final states produced in
conjunction with a hard photon, assumed to result from ISR.
While \babar\ data are available at effective c.m.\@ energies up to
10.58  \gev, 
the present analysis is restricted to 
energies below 4.35 \gev because of backgrounds from $\Upsilon(4S)$
decays. As part of the analysis, we search for and observe
intermediate states, including the $\eta$,
$\omega$, $\rho$, $a_0(980)$, and $a_1(1260)$ resonances.
A clear $J/\psi$ signal is observed for both the $\pipi3\piz$ and
$\pipi2\piz\eta$ channels, and the
corresponding $J/\psi$ branching fractions are measured.
The decay $\psi(2S)\to\pipi\ppz\piz$  is observed and its branching fraction is measured.

Previous measurements of the $\epem\to\pipi\ppz\piz$
 cross section were reported by the M3N~\cite{M3N} and MEA~\cite{MEA}
 experiments, but with very limited precision, leading
to a large uncertainty in the corresponding HVP contribution.
The \babar~ experiment previously measured the $\epem\to\eta\pipi$ reaction
in the  $\eta\to\pipi\piz$~\cite{isr5pi} and
$\eta\to\gamma\gamma$~\cite{isretapipi}  decay channels.
Below, we present the 
measurement of $\epem\to\eta\pipi$ with $\eta\to\ppz\piz$:
this process contributes to $\epem\to\pipi\ppz\piz$.
There are no previous results for $\epem\to\pipi\ppz\eta$.

\section{\boldmath The \babar\ detector and dataset}
\label{sec:babar}

The data used in this analysis were collected with the \babar\ detector at
the \pep2\ asymmetric-energy \epem\ storage ring. 
The total integrated luminosity used is 468.6~\invfb~\cite{lumi}, 
which includes data collected at the $\Upsilon(4S)$
resonance (424.7~\invfb) and at a c.m.\ energy 40~\mev below this
resonance (43.9~\invfb).

The \babar\ detector is described in detail elsewhere~\cite{babar}. 
Charged particles are reconstructed using the \babar\ tracking system,
which is comprised of the silicon vertex tracker (SVT) and the drift
chamber (DCH), both located 
inside the 1.5 T solenoid.
Separation of pions and kaons is accomplished by means of the detector of
internally reflected Cherenkov light (DIRC) and energy-loss measurements in
the SVT and DCH. 
Photons and \KL mesons are detected in the electromagnetic calorimeter (EMC).  
Muon identification is provided by the instrumented flux return.

To evaluate the detector acceptance and efficiency, 
we have developed a special package of Monte Carlo (MC) simulation programs for
radiative processes based on 
the approach of K\"uhn  and Czy\.z~\cite{kuehn2}.  
Multiple collinear soft-photon emission from the initial \epem state 
is implemented with the structure function technique~\cite{kuraev,strfun}, 
while additional photon radiation from final-state particles is
simulated using the PHOTOS package~\cite{PHOTOS}.  
The precision of the radiative simulation is such that it contributes less than 1\% to
the uncertainty in the measured hadronic cross sections.

We simulate $\epem\to\pipi\ppz\piz\gamma$ events assuming production
through the $\omega(782)\ppz$ and $\eta\rho(770)$ intermediate channels,
with decay of  the $\omega$ to three pions and
decay of the $\eta$ to all  its measured decay modes~\cite{PDG}.
The two neutral pions in the  $\omega\ppz$ system are in an
S-wave state and are described by a combination of phase space and $f_0(980)\to\ppz$,
based on our study of the $\omega\pipi$ state~\cite{isr5pi}.
The simulation of $\epem\to\pipi\ppz\eta\gamma$ events
is similarly based on two production channels: a phase space model,  and
 a model with an $\omega\piz\eta$ intermediate state with 
a $\piz\eta$ S-wave system.

A sample of 100-200k simulated events is generated  for each
signal reaction  and
processed through the detector response simulation, based on the GEANT4 package~\cite{GEANT4}. These
events are reconstructed using
the same software chain as the data. Variations in detector
and background conditions are taken into account.

For the purpose of background estimation,  large samples of events from the
main relevant ISR processes ($2\pi\gamma$,
$3\pi\gamma$, $4\pi\gamma$, $5\pi\gamma$, 
$2K\pi\gamma$, and  $\pipi\ppz\gamma$)
are simulated.  
To evaluate the background from the relevant
 non-ISR processes, namely $\epem\to\qqbar$ $(q=u, d, s)$ and
 $\epem\to\tau^+\tau^-$, 
 simulated samples with integrated luminosities about
twice that of the data are generated 
using the \textsc{jetset}~\cite{jetset}
and \textsc{koralb}~\cite{koralb} programs,
respectively.
The cross sections for the above processes
are  known with an accuracy slightly better than
10\%, which is sufficient for the present purposes.

\begin{figure}[tbh]
\begin{center}
\includegraphics[width=0.79\linewidth]{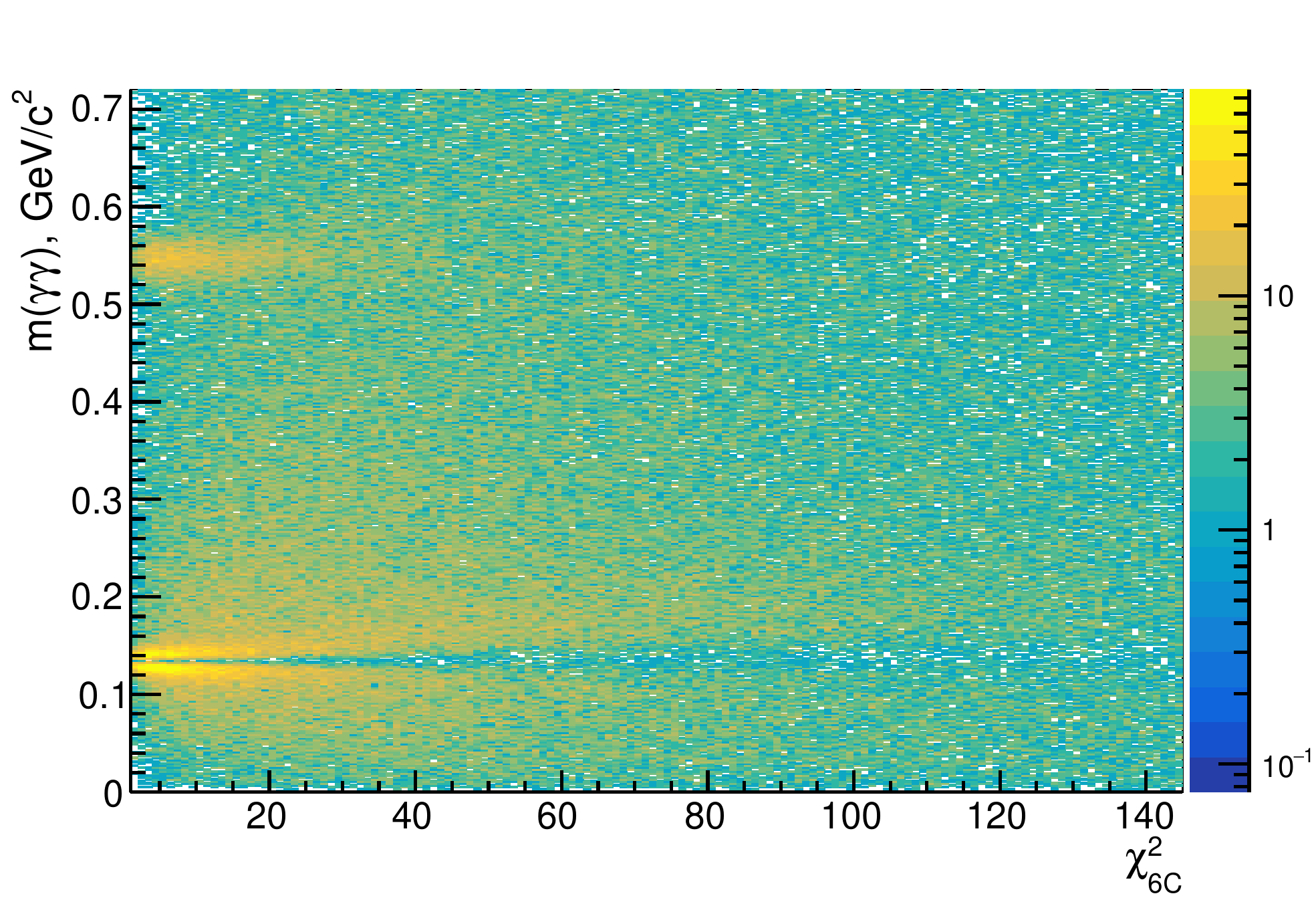}
\put(-50,100){\makebox(0,0)[lb]{\textcolor{white}{\bf (a)}}}\\
\includegraphics[width=0.79\linewidth]{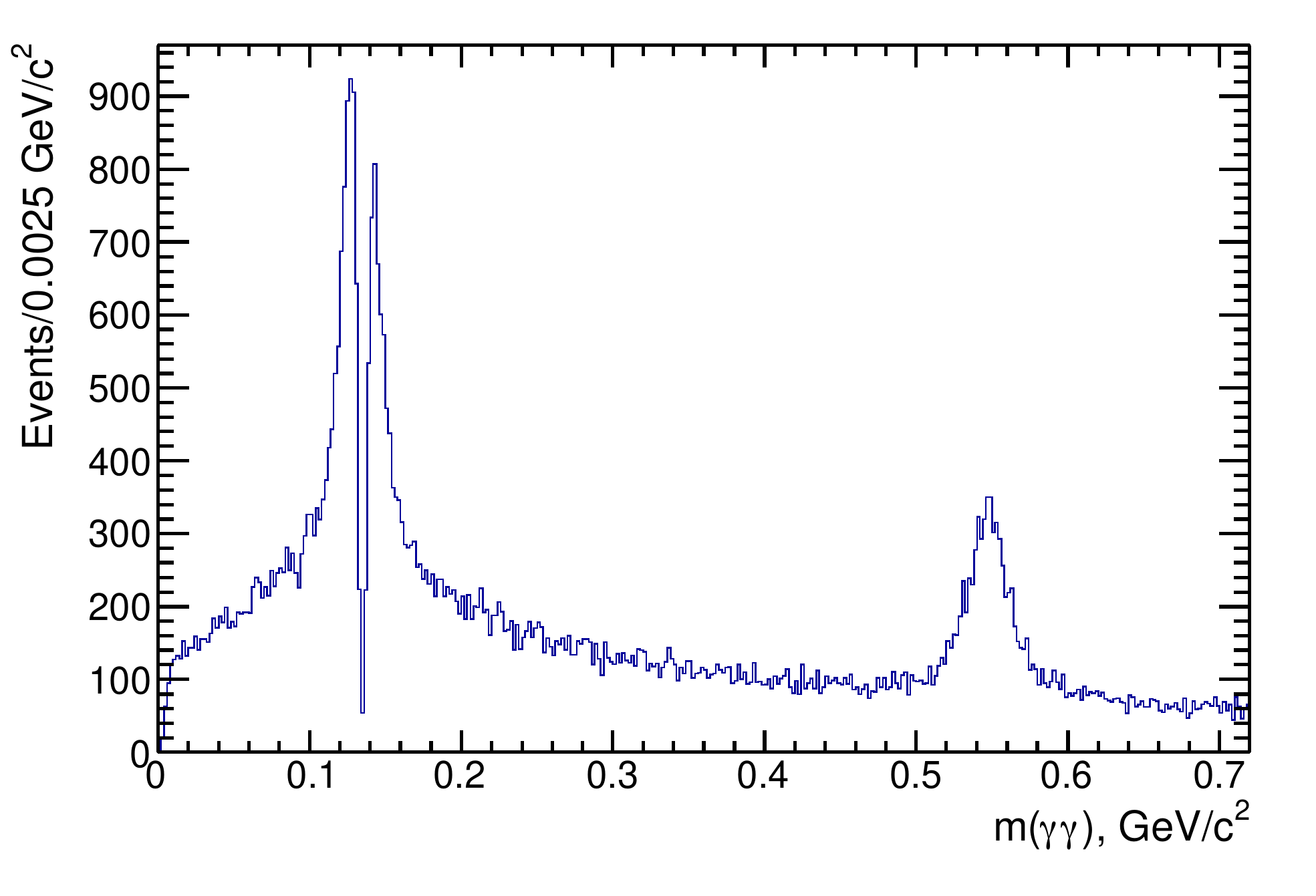}
\put(-50,100){\makebox(0,0)[lb]{\bf(b)}}
\caption{(a) The invariant mass $\mgg$ of the third photon pair  vs
  $\chisq_{2\pi2\piz\gamma\gamma}$.
(b) The $\mgg$ distribution for 
$\chisq_{2\pi2\piz\gamma\gamma} <60$ and with additional selection
criteria applied as described in the text. 
}
\label{2pi3pi0_chi2_all}
\end{center}
\end{figure}
\begin{figure}[t]
\begin{center}
\includegraphics[width=0.79\linewidth]{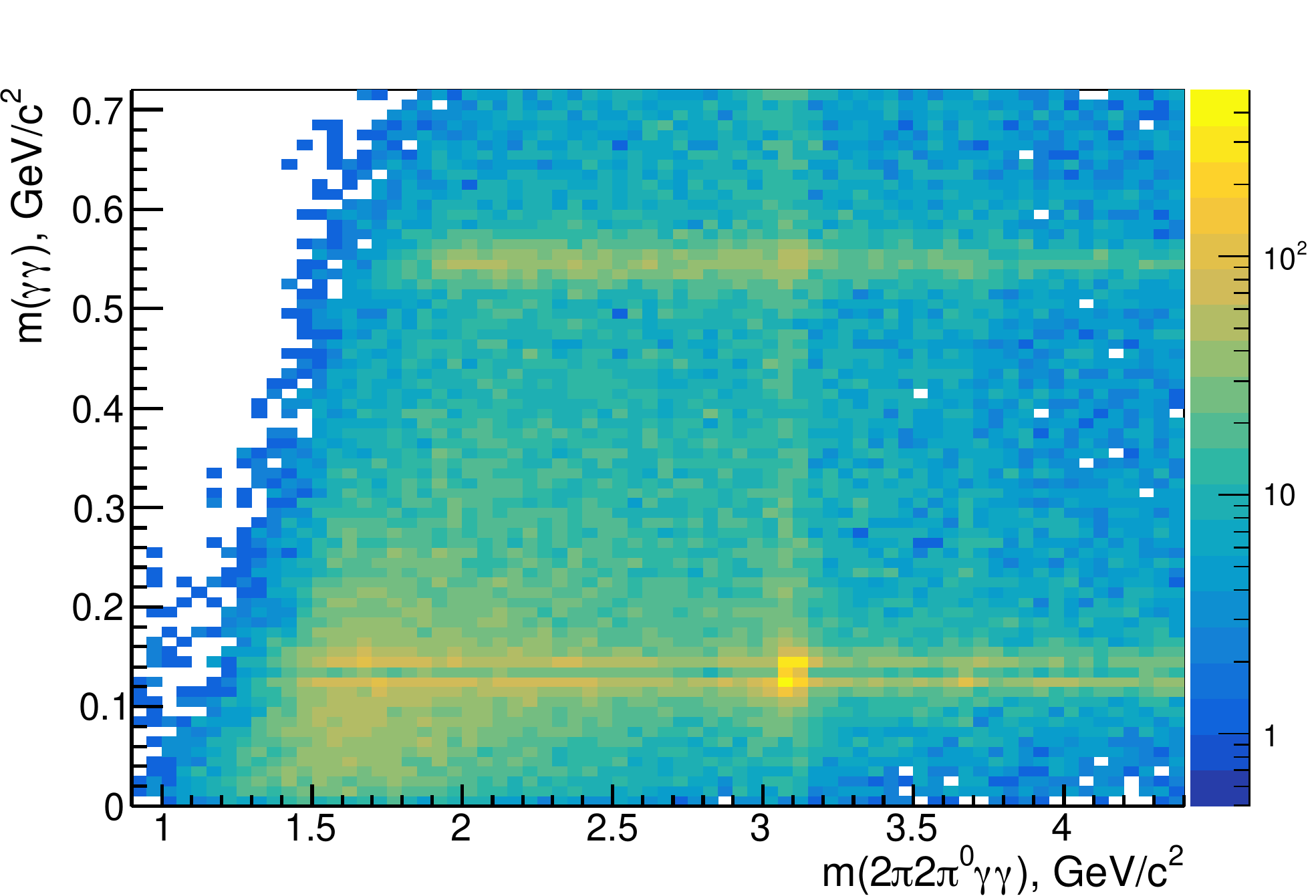}
\put(-165,100){\makebox(0,0)[lb]{\bf(a)}}\\
\includegraphics[width=0.79\linewidth]{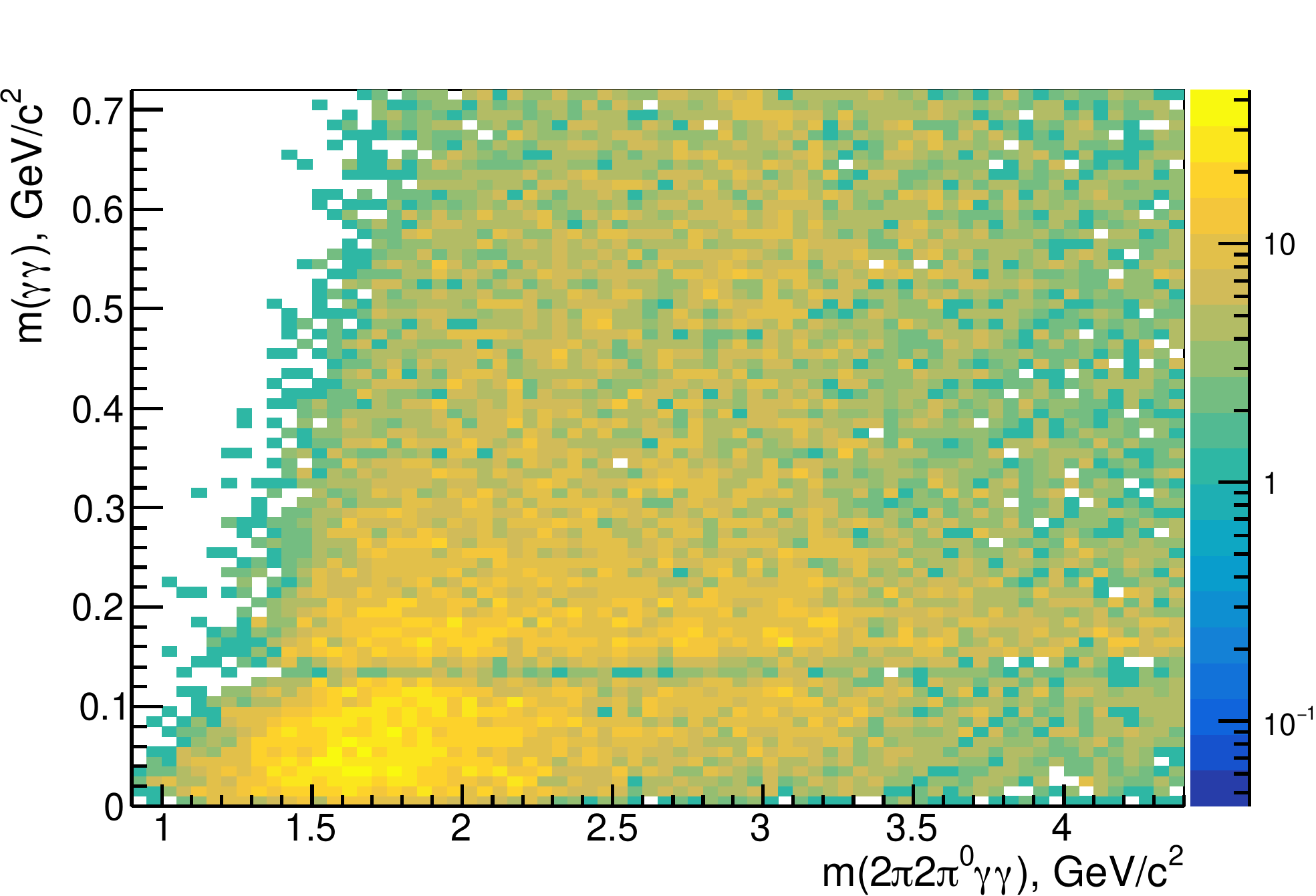}
\put(-165,100){\makebox(0,0)[lb]{\bf(b)}}
\caption{(a) The third-photon-pair invariant mass $\mgg$ vs
  $m(2\pi2\piz\gamma\gamma)$ for (a) $\chisq_{2\pi2\piz\gamma\gamma}<
  60$ and
(b) $60< \chisq_{2\pi2\piz\gamma\gamma}< 120$.
}
\label{2pi3pi0_mass_all}
\end{center}
\end{figure}

\section{\boldmath Event Selection and Kinematic Fit}
\label{sec:Fits}

A relatively clean sample of 
$\pipi3\piz\gamma$ and $\pipi2\piz\eta\gamma$  events is selected
by requiring that there be two tracks reconstructed in the DCH,
SVT, or both, and seven or more photons, with an energy above
0.02~\gev,  in the EMC. We assume the photon with the highest energy to be  the ISR
photon, and we require its c.m. energy to be larger than 3~\gev.
 
We allow either exactly two or exactly three tracks
in an event, but only two that extrapolate to within
0.25 cm of the beam axis and 3.0 cm 
of the nominal collision point along that axis.
The reason a third track is allowed is to capture a
relatively small fraction of signal events that contain a
background track.
The two tracks that satisfy the extrapolation criteria are
fit to a vertex, which is used as the point of origin
in the calculation of the photon directions.

We subject each candidate event to a set of constrained 
kinematic fits and use the fit results,
along with charged-particle identification,
to select the final states of interest and evaluate backgrounds
from other processes.
 The kinematic fits make use of the
four-momenta and covariance matrices of the initial $e^+$, $e^-$, and the
set of selected tracks and photons.
The fitted three-momenta of each track and photon are then used 
in further kinematical calculations.

Excluding the photon with the highest c.m.\  energy,
which is assumed to arise from ISR, six other
photons are combined into three pairs.
For each set of six photons,
there are 15 independent combinations of photon pairs.
 We retain those combinations in which the diphoton mass
 of at least two pairs lies within 35~\mevcc of the
\piz mass $m_{\piz}$.
The selected combinations are subjected to a fit
 in which the diphoton masses of the two pairs with
$|m(\gamma\gamma)-m_{\pi^0}|<35~\mevcc$
 are constrained to $m_{\piz}$.
In combination with the constraints
due to four-momentum conservation, there are thus six
 constraints (6C) in the fit.  The photons in the
 remaining (``third'') pair are treated as being independent.
If all three photon pairs in the combination
satisfy $|m(\gamma\gamma)-m_{\pi^0}|<35~\mevcc$,
 we test all possible combinations,
  allowing each of the three diphoton pairs in turn
 to be the third pair, i.e., the pair without the
 $m_{\piz}$ constraint.

 The above procedure allows us not only to search 
for events with  $\piz\to\gamma\gamma$  in the third photon pair, but
also for events with $\eta\to\gamma\gamma$.

The 6C fit is performed under the signal hypothesis
$\epem\to\pipi\ppz\gamma\gamma\gamma_{ISR}$.
The combination with the smallest \chisq is retained, along with the obtained
$\chisq_{2\pi2\piz\gamma\gamma}$ value and the fitted three-momenta of each
track and photon. 
Each selected event is also subjected to a 6C fit under the
$\epem\to\pipi\ppz\gamma_{ISR}$ background hypothesis, and the
 $\chisq_{2\pi2\piz}$ value is retained.  
The $\pipi\ppz$ process has a larger
cross section than the $\pipi3\piz$ signal process and can
contribute to the background when two background photons are present.
Most events contain additional soft photons due to machine background
or interactions in the detector material.

\section{The {\boldmath $\pipi3\piz$} final state}
\subsection{Additional selection criteria}

The results of the 6C fit     
to events with two tracks and at least seven photon candidates
are used to perform the final selection of the five-pion sample. 
We require the tracks to lie within the fiducial
region of the DCH (0.45-2.40 radians) and to be
inconsistent with being a kaon or muon.
The photon candidates are required to lie within the
fiducial region of the EMC (0.35-2.40 radians) and to have
an energy larger than 0.035 GeV.
A requirement that there be no charged
tracks within 1 radian of the ISR photon reduces the $\tau^+\tau^-$ background
to a negligible level. 
A requirement that any extra photons in an event each
have an energy below 0.7 GeV slightly reduces the multi-photon
background. 

Figure~\ref{2pi3pi0_chi2_all} (a) shows the invariant mass $\mgg$ of the third
photon pair vs $\chisq_{2\pi2\piz\gamma\gamma}$. Clear $\piz$
and $\eta$ peaks are visible at small  \chisq values.
We require $\chisq_{2\pi2\piz\gamma\gamma}< 60$ for the signal  hypothesis
and $\chi_{2\pi 2\piz}^2 >30$ for the 
$2\pi 2\pi^0$ background hypothesis. 
This requirement  reduces the contamination due to    
 $2\pi 2\piz$ events from ~30\% to about 1-2\% while reducing
the signal efficiency by only 5\%.

Figure~\ref{2pi3pi0_chi2_all} (b)
shows the $\mgg$ distribution  after the
above requirements have been applied. 
The dip in this distribution at the $\piz$ mass value is a
consequence of the kinematic fit constraint of the best
two photon pairs to the $\piz$ mass. Also, because of this constraint, the third photon pair is
sometimes formed from photon candidates that are less
well measured.

Figure~\ref{2pi3pi0_mass_all} shows the $\mgg$ 
distribution vs the invariant mass $m(2\pi 2\piz\gamma\gamma)$ 
 for events (a) in the signal region
$\chisq_{2\pi2\piz\gamma\gamma}< 60$ and (b) in a control
region defined by $60 < \chisq_{2\pi2\piz\gamma\gamma}< 120$. Events from the
$\epem\to\pipi\ppz\piz$ and $\pipi2\piz\eta$ processes are
clearly seen in the signal region, as well as $J/\psi$ decays to these final states. In the
control region no significant structures are seen and we use these
events to evaluate background.

Our strategy to extract the signals for the $\epem\to\pipi\ppz\piz$
and $\pipi\ppz\eta$ processes 
is to perform a fit for the $\piz$ and $\eta$ yields
in intervals of 0.05~\gevcc in the distribution of
the $\pipi2\piz\gamma\gamma$ invariant mass $m(\pipi2\piz\gamma\gamma)$.

\begin{figure}[tbh]
\begin{center}
\includegraphics[width=0.79\linewidth]{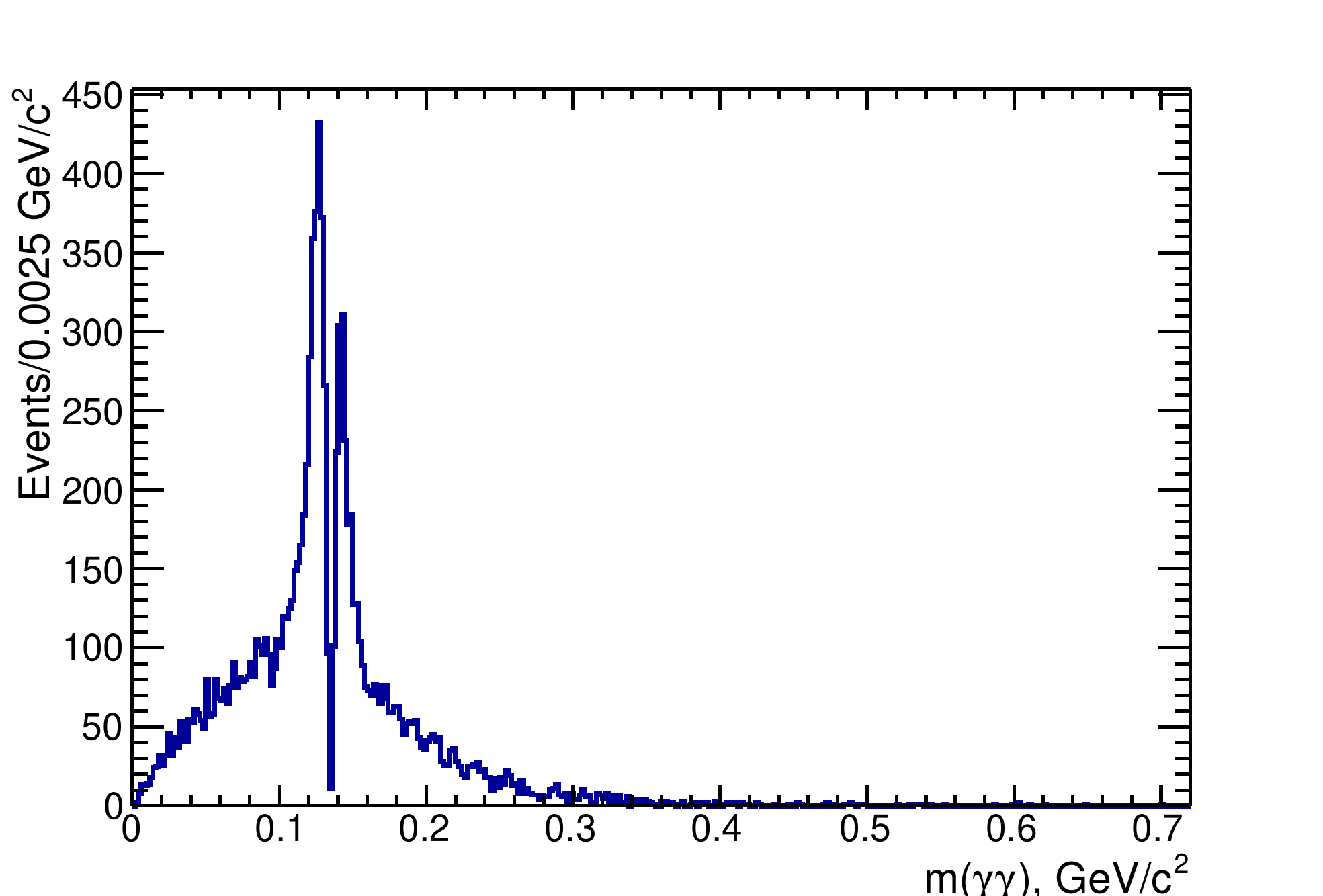}
\put(-50,100){\makebox(0,0)[lb]{\bf(a)}}\\
\includegraphics[width=0.79\linewidth]{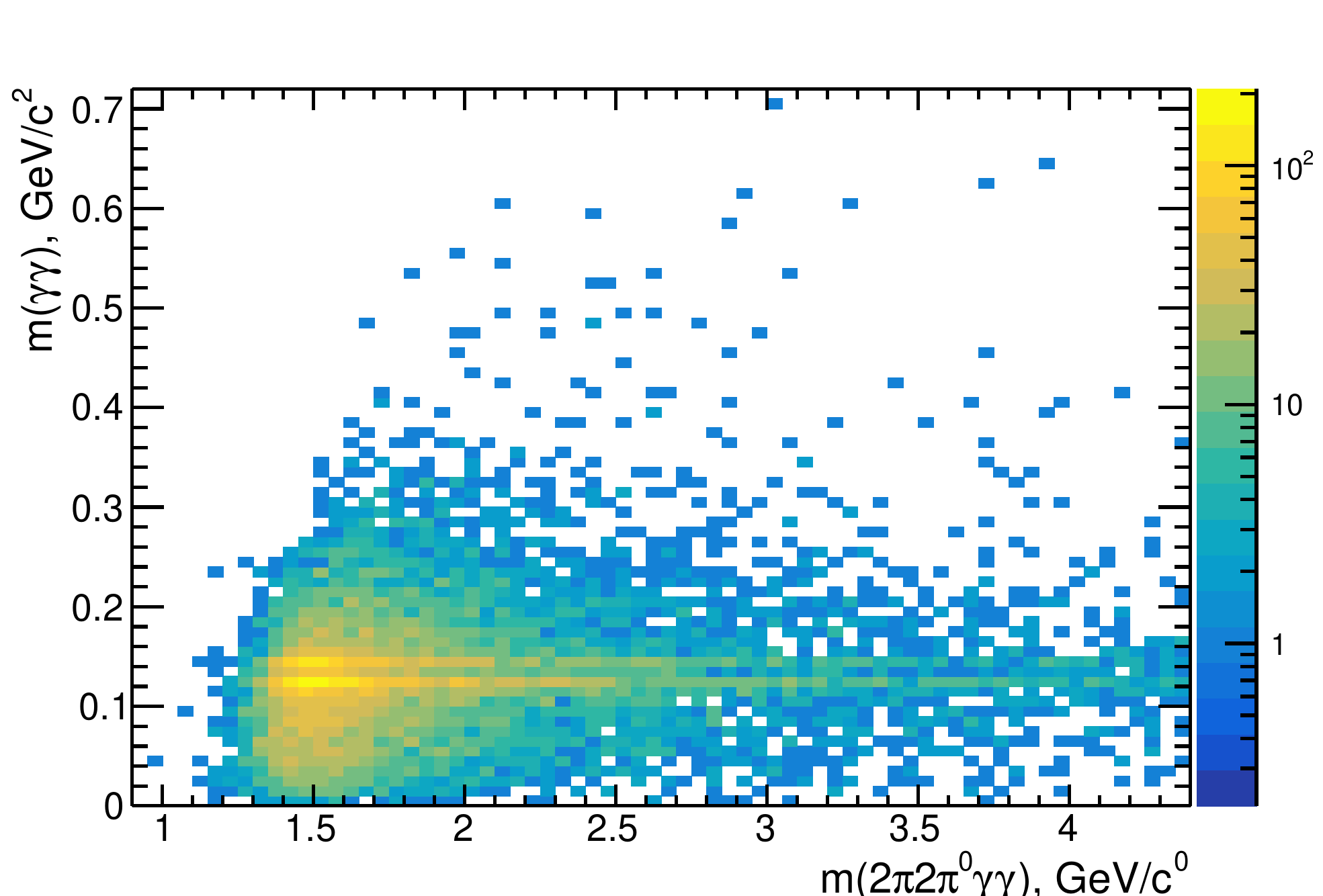}
\put(-50,100){\makebox(0,0)[lb]{\bf(b)}}
\caption{
The  MC-simulated distribution for $\epem\to\eta\pipi$ events of (a)
the third-photon-pair invariant mass $\mgg$, and (b) $\mgg$ vs $m(\pipi2\piz\gamma\gamma)$.
}
\label{mgg_eta2pi}
\end{center}
\end{figure}
\begin{figure}[tbh]
\begin{center}
\includegraphics[width=0.79\linewidth]{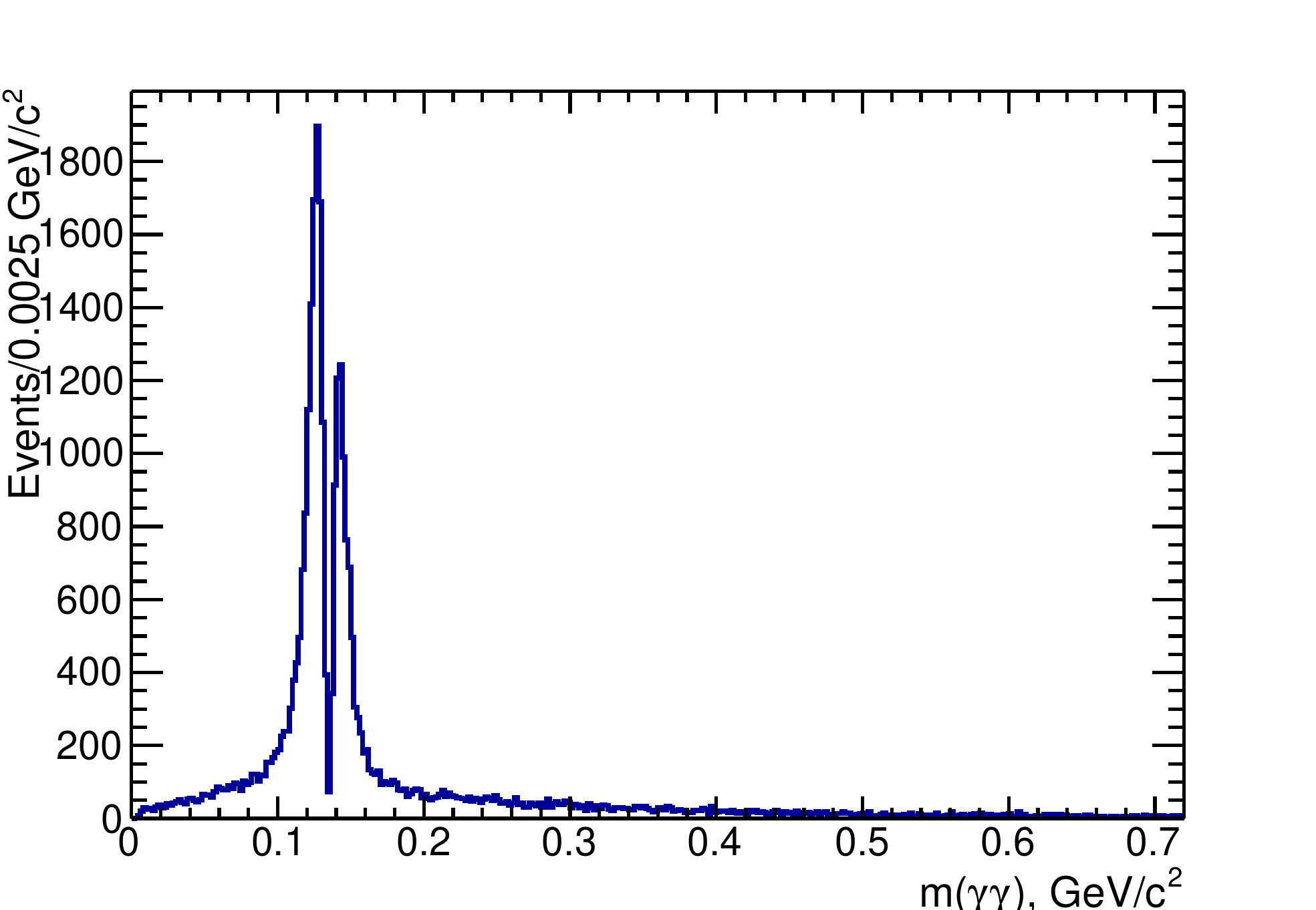}
\put(-50,100){\makebox(0,0)[lb]{\bf(a)}}\\
\includegraphics[width=0.79\linewidth]{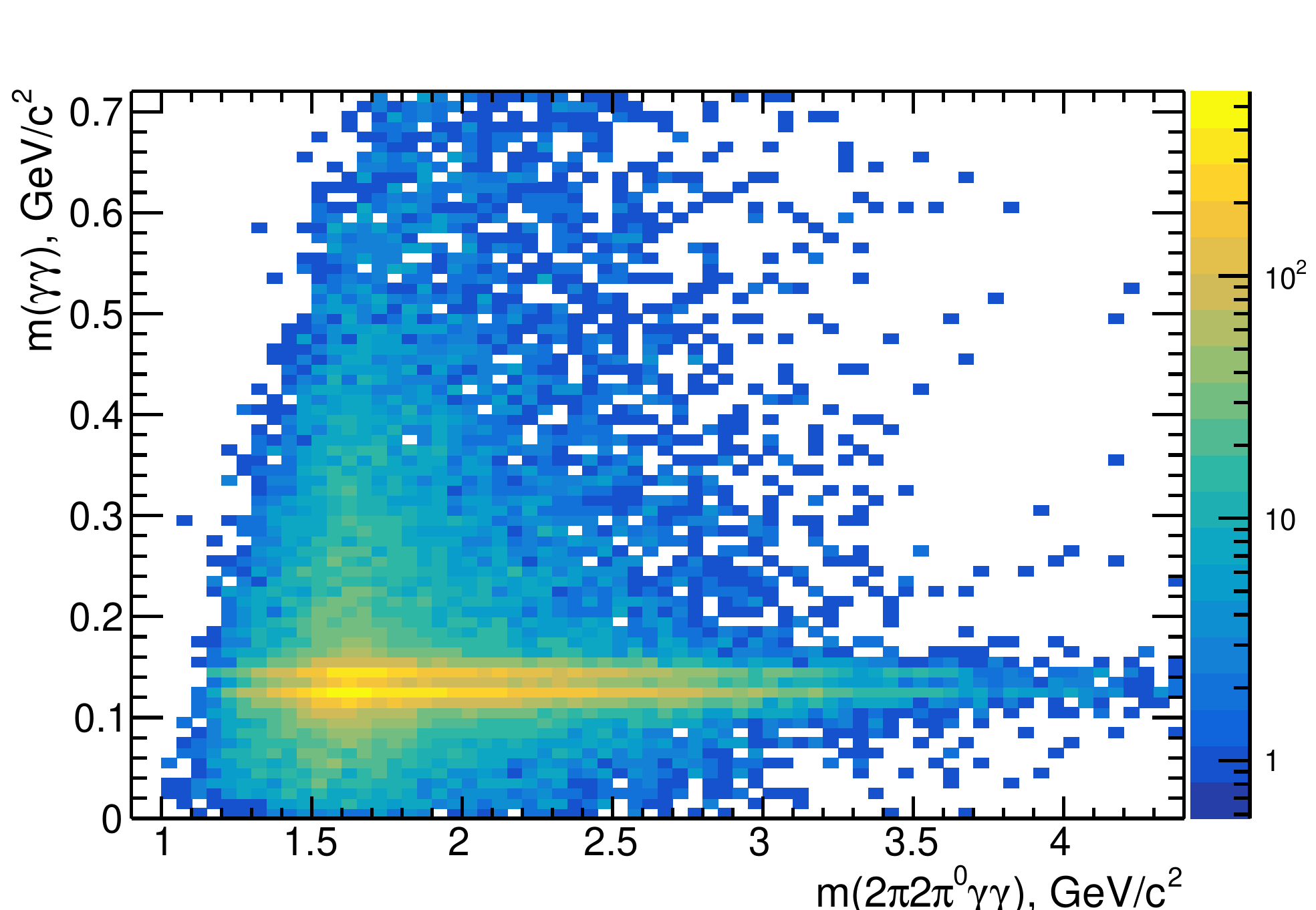}
\put(-50,100){\makebox(0,0)[lb]{\bf(b)}}
\caption{
The  MC-simulated distribution for $\epem\to\omega\ppz$ events of (a)
the third-photon-pair invariant mass $\mgg$, and (b) $\mgg$ vs $m(\pipi2\piz\gamma\gamma)$.
}
\label{mgg_omega2pi}
\end{center}
\end{figure}

\subsection{Detection efficiency}\label{sec:efficiency}
As mentioned in Sec.~\ref{sec:babar}, the model used in the MC simulation 
assumes that the five-pion final state results predominantly from 
$\omega\ppz$ and $\eta\pipi$ production,  with $\omega$
decays to three pions and $\eta$ decays to all modes.  As shown below, these  two final
states dominate the observed cross section. 

The selection procedure applied to the data is also applied to the         
MC-simulated events. Figures~\ref{mgg_eta2pi}
and \ref{mgg_omega2pi} show
(a) the  $\mgg$ distribution and (b) the distribution of
$\mgg$ vs $m(2\pi2\piz\gamma\gamma)$ for the simulated  $\eta\pipi$
and $\omega\ppz$ events, respectively. 
The $\piz$ peak is not Gaussian in either reaction and
is broader for $\eta\pipi$ events than for $\omega\ppz$ events
because the photon energies are lower.
Background photons are included in the simulation.
Thus these distributions include
simulation of the combinatoric background that arises when background
photons are combined with photons from
the signal reactions.

\begin{figure}[tbh]
\begin{center}
\includegraphics[width=0.79\linewidth]{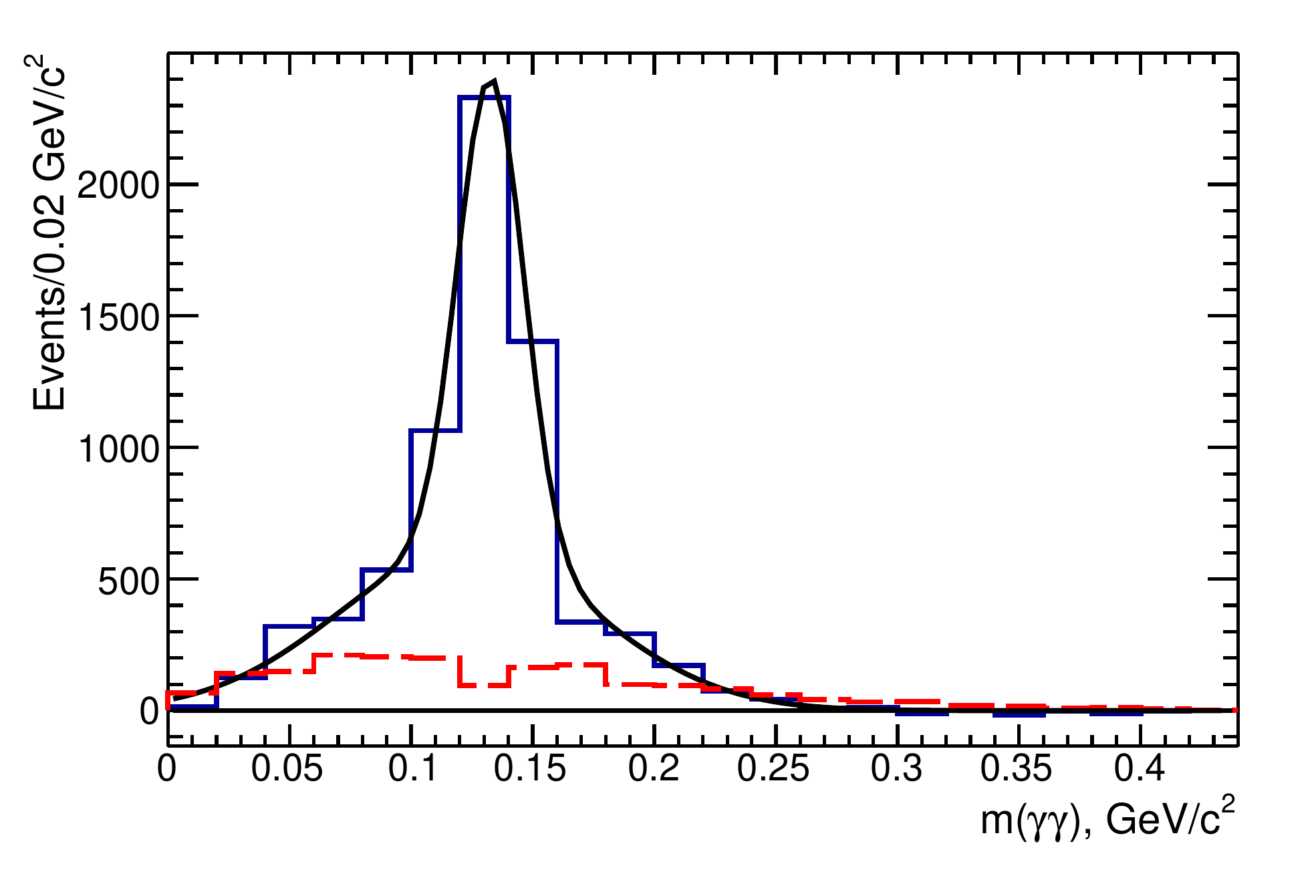}
\put(-50,100){\makebox(0,0)[lb]{\bf(a)}}\\
\includegraphics[width=0.79\linewidth]{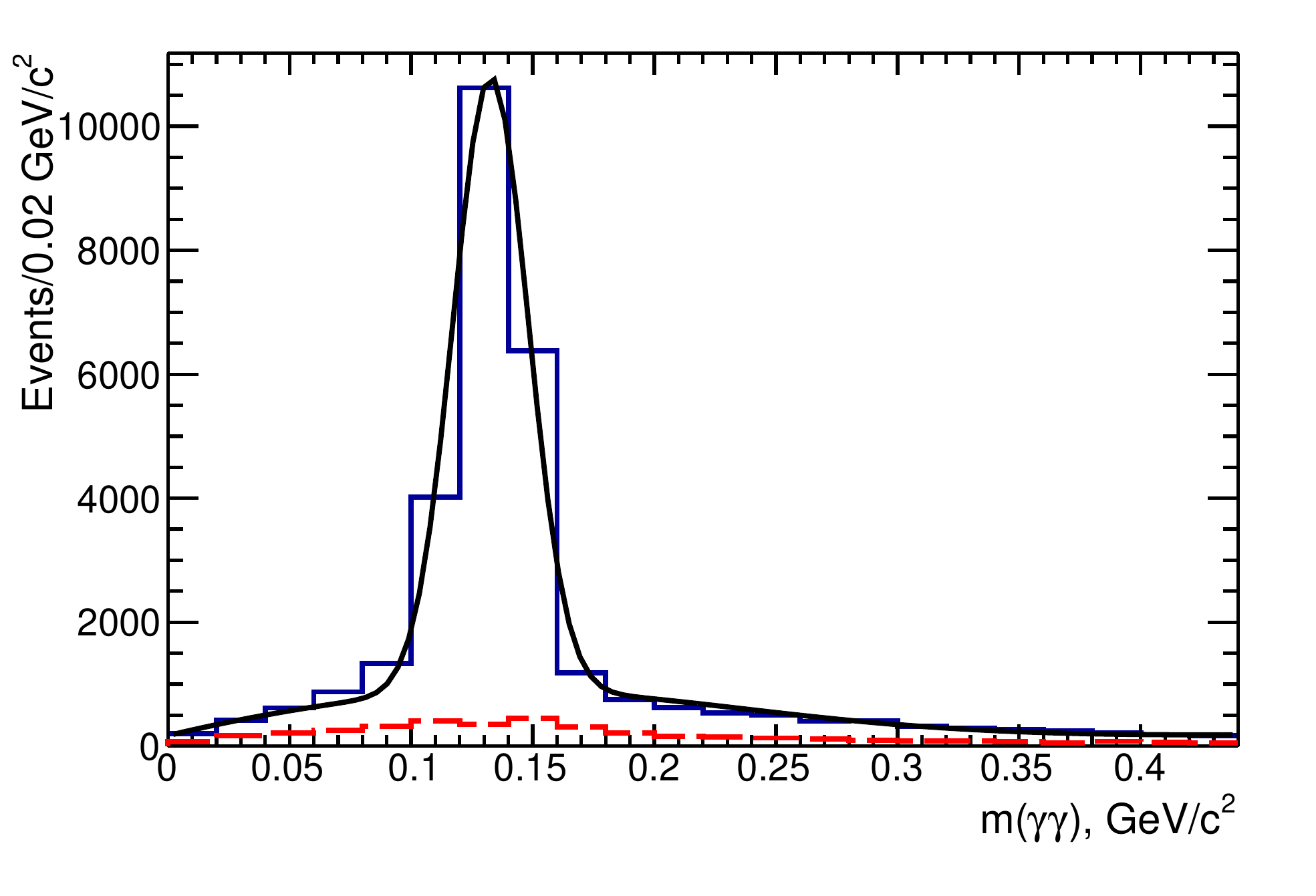}
\put(-50,100){\makebox(0,0)[lb]{\bf(b)}}
\caption{
The background subtracted MC-simulated $\mgg$ distribution for
(a) $\epem\to\eta\pipi$ and (b) $\epem\to\omega\ppz$ events.
The dashed histogram shows the simulated distribution
from the \chisq control region,  used for subtraction. The fit function
is described in the text.
}
\label{mgg_eta2pi_fit}
\end{center}
\end{figure}

The combinatoric background  is subtracted using the data
 from the \chisq control region.
 The method is illustrated using simulation in Fig.~\ref{mgg_eta2pi_fit},
which shows the $\mgg$ distribution with a bin width of 0.02~\gevcc.
The dashed histograms show the simulated combinatoric background.
The solid histograms show the simulated results from the
signal region after subtraction of the simulated
combinatoric background.
The sum of three Gaussian functions with a common mean
is used to describe the $\piz$ signal shape. 
  The fitted fit function is shown by the smooth curve in Fig.~\ref{mgg_eta2pi_fit}.
 We perform a fit of the $\piz$ signal in every
0.05~\gevcc interval in the $m(2\pi2\piz\gamma\gamma)$ invariant mass for
the two different simulated channels.

\begin{figure}[tbh]
\begin{center}
\includegraphics[width=0.79\linewidth]{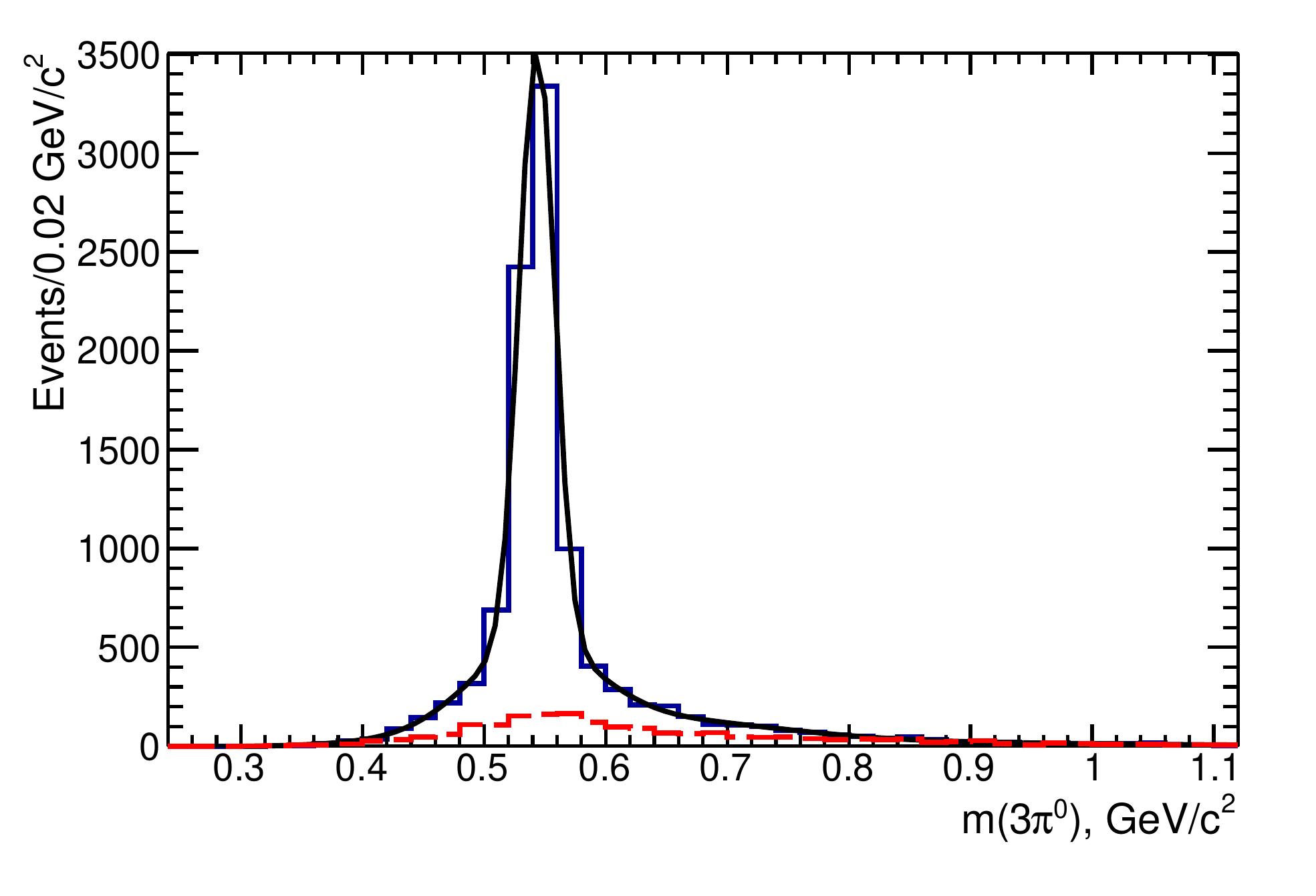}
\put(-50,100){\makebox(0,0)[lb]{\bf(a)}}\\
\includegraphics[width=0.79\linewidth]{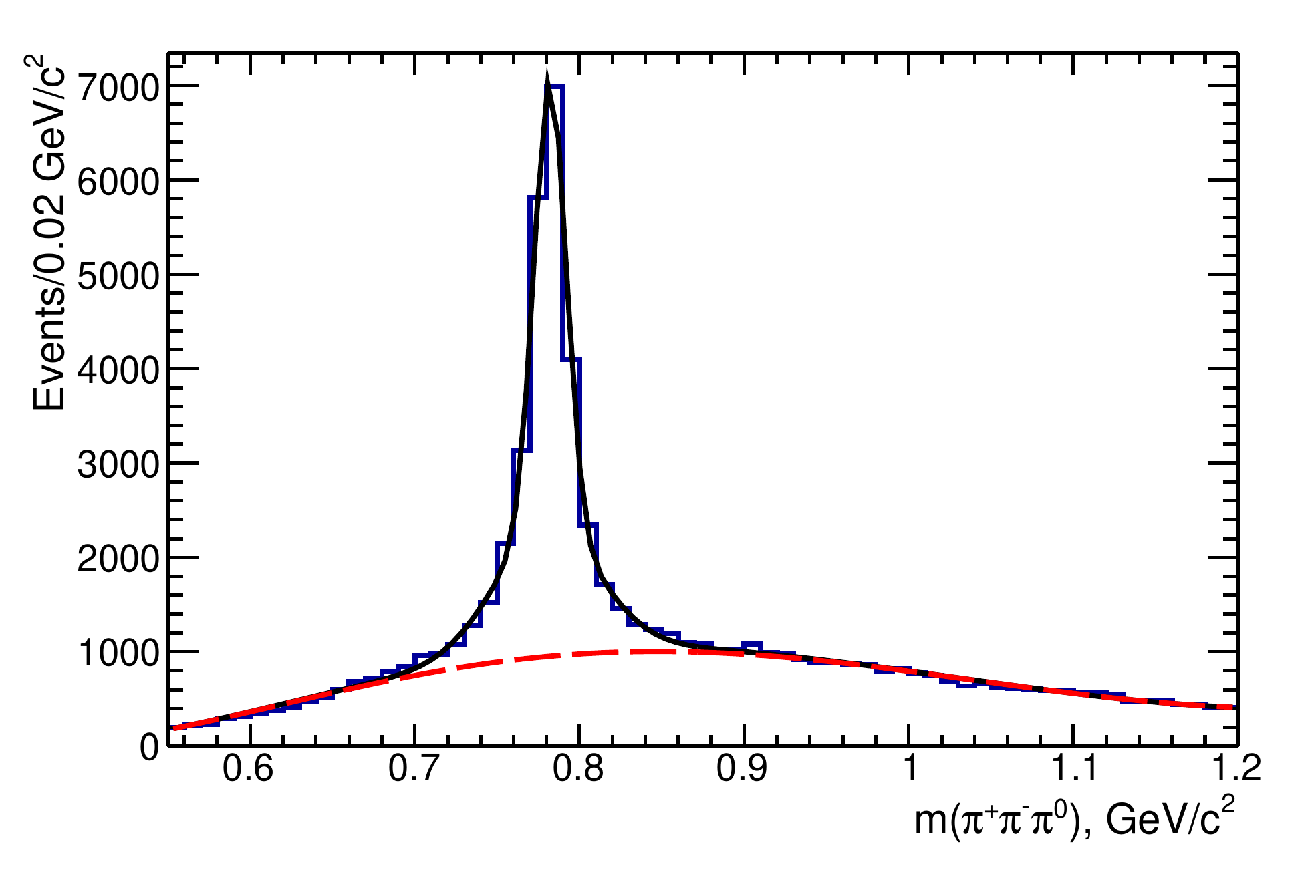}
\put(-50,100){\makebox(0,0)[lb]{\bf(b)}}
\vspace{-0.5cm}
\caption{(a) The background subtracted MC-simulated
3$\piz$ invariant mass for the $\epem\to\eta\pipi$ events. 
The dashed distribution is from  the simulated \chisq control region, used
for background subtraction. 
(b) The $\pipi\piz$  invariant mass for the
  MC-simulated $\epem\to\omega\ppz$ events (three entries per
  event). The solid curve shows the fit function used to obtain number of signal
  events. The dashed curve shows the fit function for the combinatorial background.
}
\label{m3pi_omega_eta_mc}
\end{center}
\end{figure}

Alternatively, for the $\eta\pipi$ events, we determine the number of
events vs the $m(2\pi2\piz\gamma\gamma)$ invariant mass by fitting the
$\eta$ signal from the $\eta\to\ppz\piz$ decay: the simulated
background-subtracted distribution is shown  in
Fig.~\ref{m3pi_omega_eta_mc}(a).
The fit function is again the sum of three Gaussian functions with a common mean.

Similarly, as an alternative for the $\omega\ppz$ events,
the $\omega$ mass peak can be used.  The $\omega$ mass peak in
simulation is shown in Fig.~\ref{m3pi_omega_eta_mc}(b), with three entries per event.
We obtain the number of events by fitting
$m(\pipi\piz)$ in 0.05~\gevcc intervals of the 
$m(\pipi2\piz\gamma\gamma)$ invariant mass.
A Breit-Wigner (BW) function,
convoluted with a Gaussian distribution to account
for the detector resolution, is used to describe the $\omega$ signal.
A second-order polynomial is used to describe the background.

The mass-dependent  detection
efficiency is obtained by dividing the number of fitted MC
events in each 0.05~\gevcc mass interval by the number generated in          
the same interval. 
Although the signal simulation accounts for all
$\eta$ decay modes, the efficiency calculation
considers the signal $\eta\to\ppz\piz$ decay mode only.
This efficiency estimate takes into account the geometrical acceptance of the detector 
for the final-state photons and the charged pions, the inefficiency of 
the  detector subsystems, and  the event loss due to additional
soft-photon  emission from the initial and final states.
Corrections that account for data-MC differences are discussed below.

\begin{figure}[tbh]
\begin{center}
\includegraphics[width=0.9\linewidth]{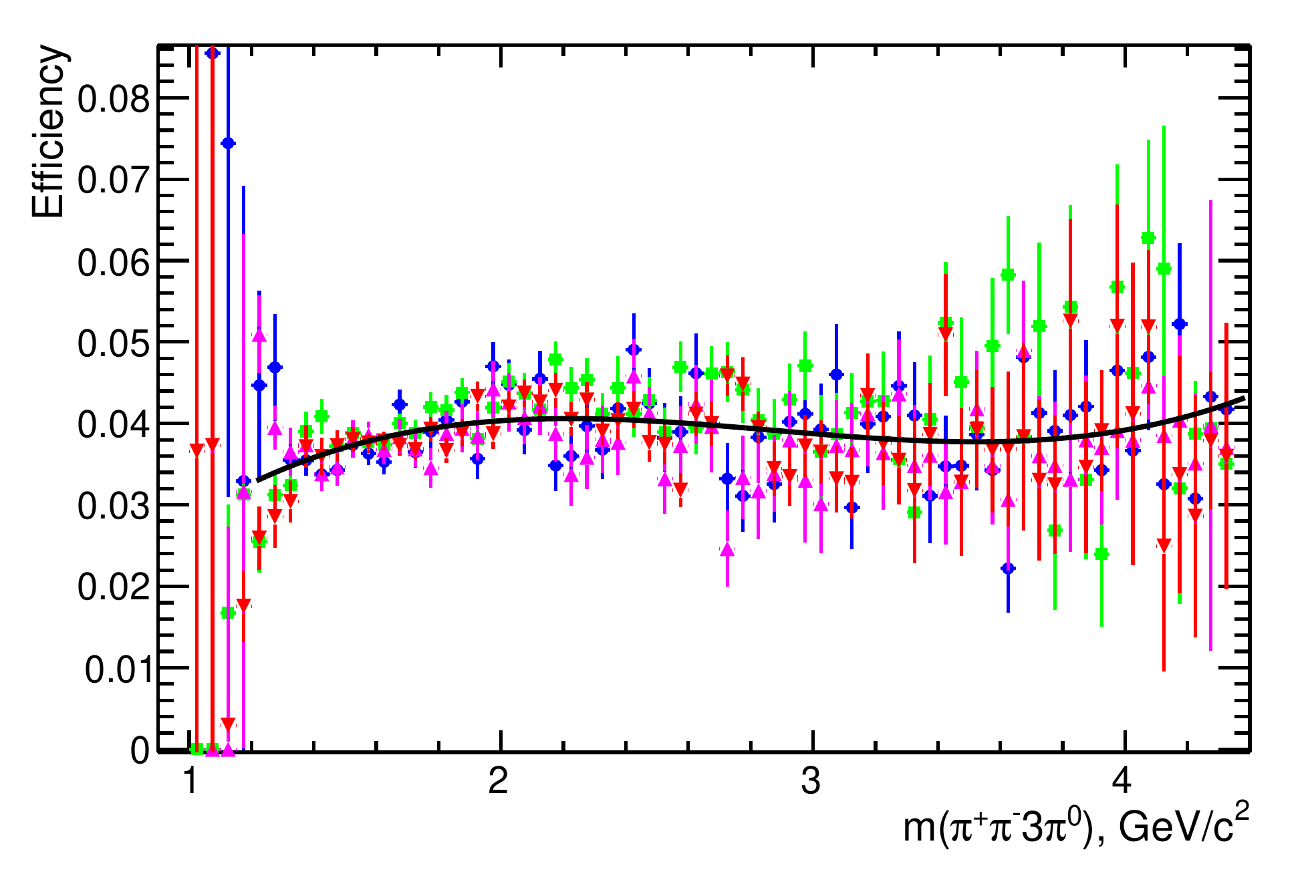}
\vspace{-0.5cm}
\caption{ The energy-dependent reconstruction efficiency for 
  $\epem\to\pipi\ppz\piz$ events, determined using
  four different methods: see text. 
The curve shows the results of a fit to the
average values, which is used in the cross
section calculation.
}
\label{mc_acc}
\end{center}
\end{figure}

The mass-dependent efficiencies from 
the $\piz$ fit are shown in Fig.~\ref{mc_acc} by 
points for the $\eta\pipi$ and by squares for the $\omega\ppz$
intermediate states, respectively. The efficiencies
determined from the $\eta$ and $\omega$ fits
are shown in Fig.~\ref{mc_acc} by the triangles and upside-down
triangles, respectively. 
These results are very similar to those obtained from the $\piz$ fits.

\begin{figure}[tbh]
\begin{center}
\includegraphics[width=0.85\linewidth]{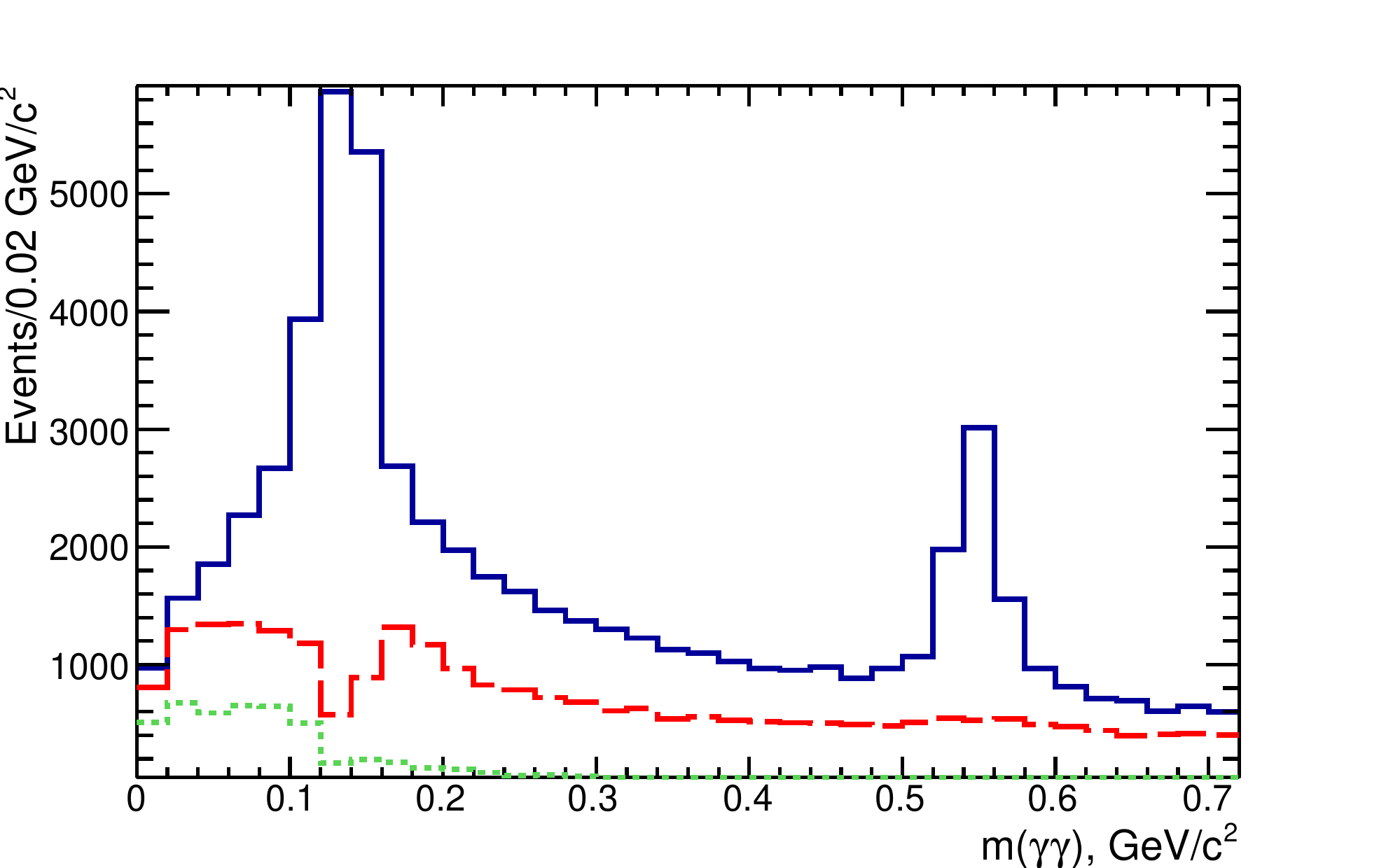}
\put(-50,90){\makebox(0,0)[lb]{\bf(a)}}\\
\includegraphics[width=0.85\linewidth]{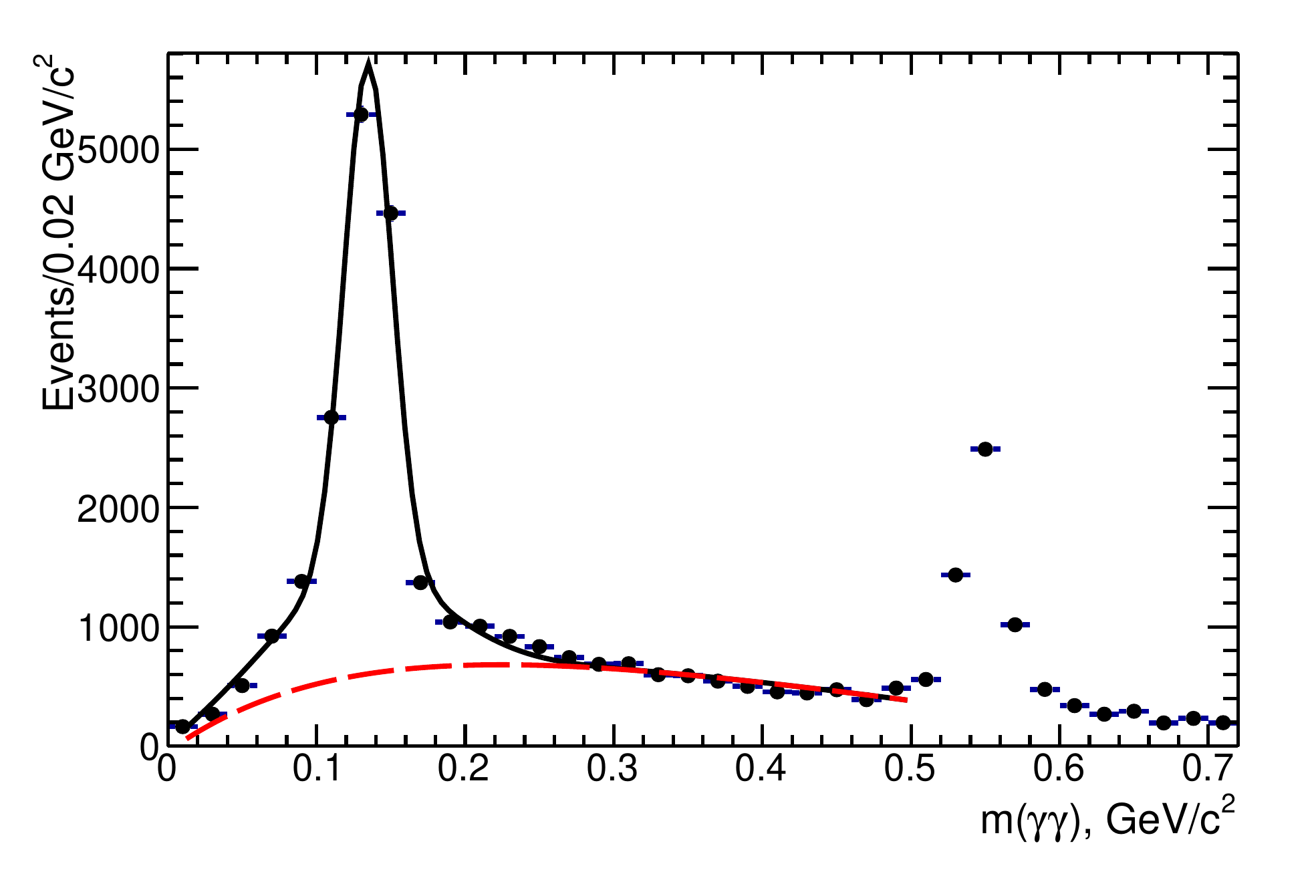}
\put(-50,90){\makebox(0,0)[lb]{\bf(b)}}
\vspace{-0.5cm}
\caption{
(a) The third-photon-pair invariant mass $\mgg$ for data in
the signal (solid) and \chisq control (dashed) regions. 
The dotted histogram shows the estimated background from
$\epem\to\pipi\ppz$.
(b) The $\mgg$ invariant mass for data after background
subtraction. The curves are the fit results as described in the text.
}
\label{2pi2pi0_bkg}
\end{center}
\end{figure}
From Fig.~\ref{mc_acc} it is seen that the reconstruction
efficiency is about 4\%, roughly independent of mass.
By comparing the results of the four different
methods used to evaluate the efficiency, 
we conclude that the overall acceptance does not change by more than 
5\% because of variations of the functions used to extract the number
of events or the use of different models. This
value is taken as an estimate of the systematic uncertainty in the acceptance
associated with the simulation model used and with the fit procedure. We
average the four efficiencies in each 0.05~\gevcc mass interval and
fit the result with a third order polynomial function, shown in
Fig.~\ref{mc_acc}. The result of this fit is used for the cross section calculation.

\subsection{\boldmath Number of $\pipi3\piz$ events}\label{sec:signal}

The solid histogram in Fig.~\ref{2pi2pi0_bkg} (a) shows the $\mgg$
data of Fig.~\ref{2pi3pi0_chi2_all} (b) binned in mass interval of 0.02~\gevcc.
The dashed histogram shows the distribution of data from the 
\chisq control region. The dotted
histogram is the estimated
remaining background from the $\epem\to\pipi\ppz$ process. 
No evidence for a peaking background is seen in either
of the two background distributions.
We subtract the background evaluated using the \chisq control region.
The resulting $\mgg$ distribution is shown in Fig.~\ref{2pi2pi0_bkg} (b).

\begin{figure}[tbh]
\begin{center}
\includegraphics[width=0.9\linewidth]{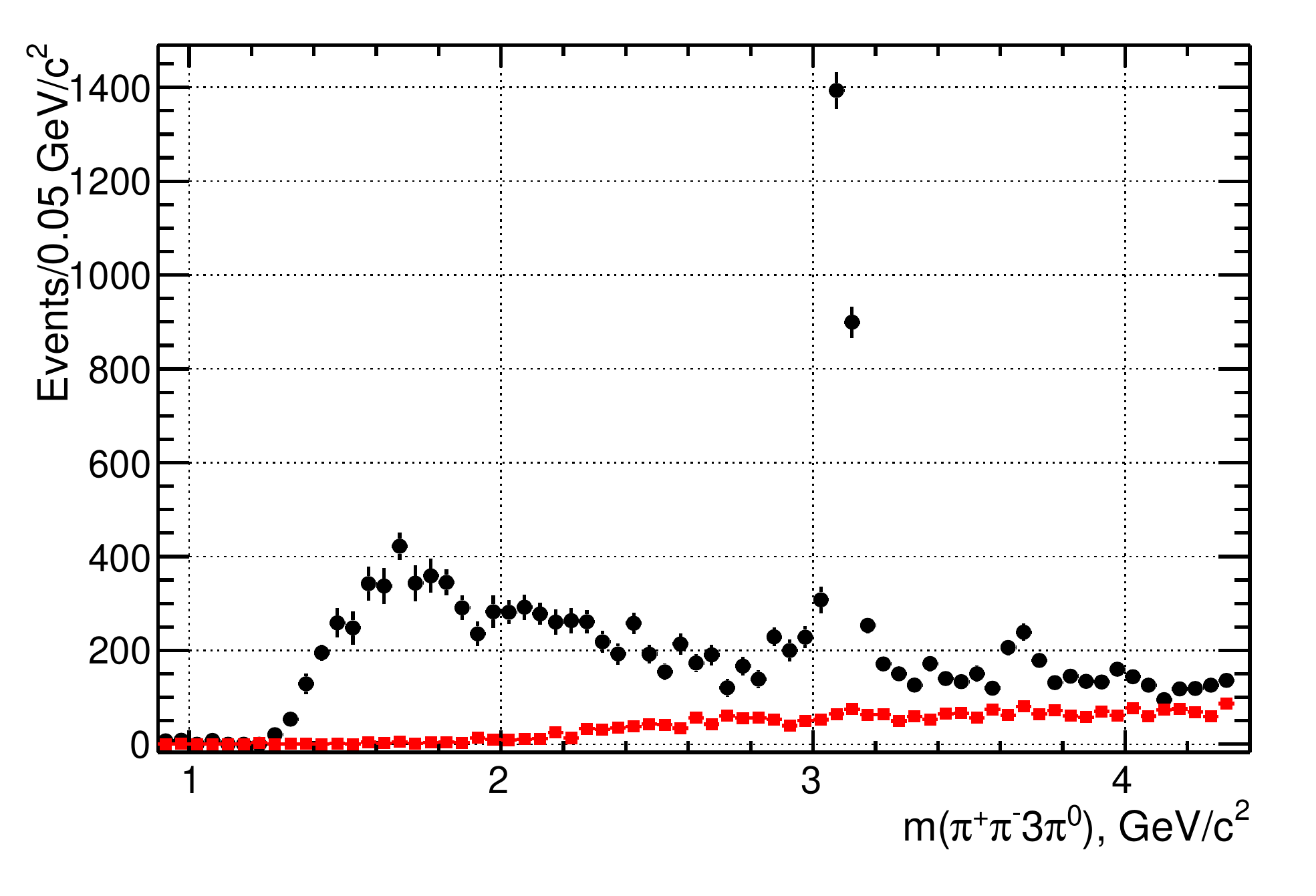}
\vspace{-0.5cm}
\caption{The invariant mass distribution of 
$\pipi3\piz$ events, obtained from the fit to the $\piz$ mass peak.
The contribution from non-ISR $uds$ background is shown by squares.
}
\label{nev_2pi3pi0_data}
\end{center}
\end{figure}

We fit the data of Fig.~\ref{2pi2pi0_bkg} (b) with a combination of a
signal function, taken from simulation, and a
background function, taken to be a third-order polynomial.
The fit is performed in the $\mgg$ mass range from
0.0 to 0.5~\gevcc.
The result of the fit is shown by the solid and
dashed curves in Fig.~\ref{2pi2pi0_bkg} (b). In total $14\,390\pm182$
events are obtained.
Note that this number includes a relatively small
peaking background component, due to $\qqbar$ events,
which is discussed in Sect.~\ref{sec:udsbkg}.
The same fit is applied to the corresponding \mgg
distribution in each 0.05~\gevcc interval in the
$\pipi2\piz\gamma\gamma$  invariant mass.
The resulting number of $\pipi3\piz$ event candidates
as a function of $m(\pipi3\piz)$, including the
peaking $\qqbar$ background, is shown by the data
points in Fig.~\ref{nev_2pi3pi0_data}.

\begin{figure}[tbh]
\begin{center}
\includegraphics[width=0.95\linewidth]{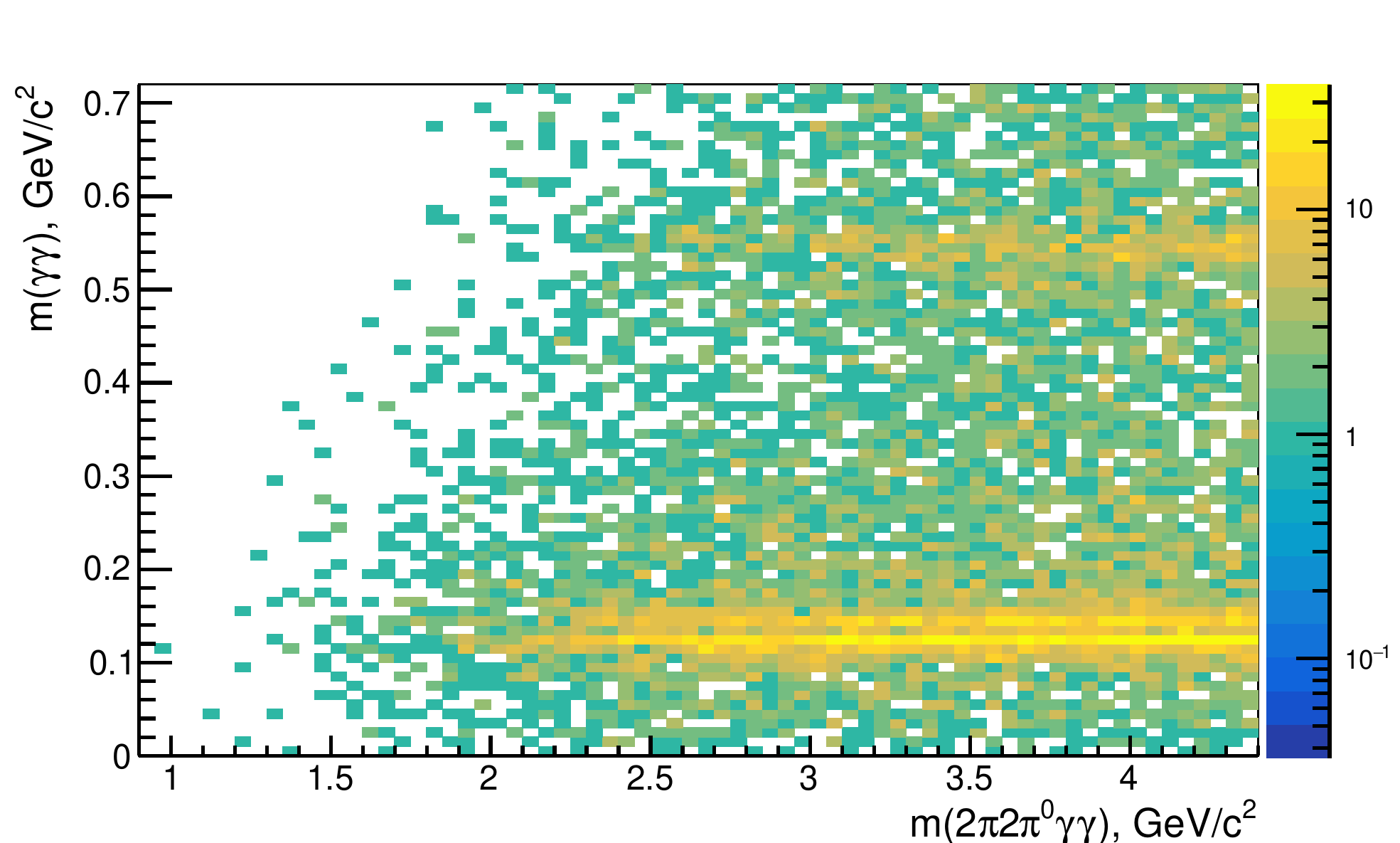}
\put(-190,100){\makebox(0,0)[lb]{\bf(a)}}\\
\includegraphics[width=0.95\linewidth]{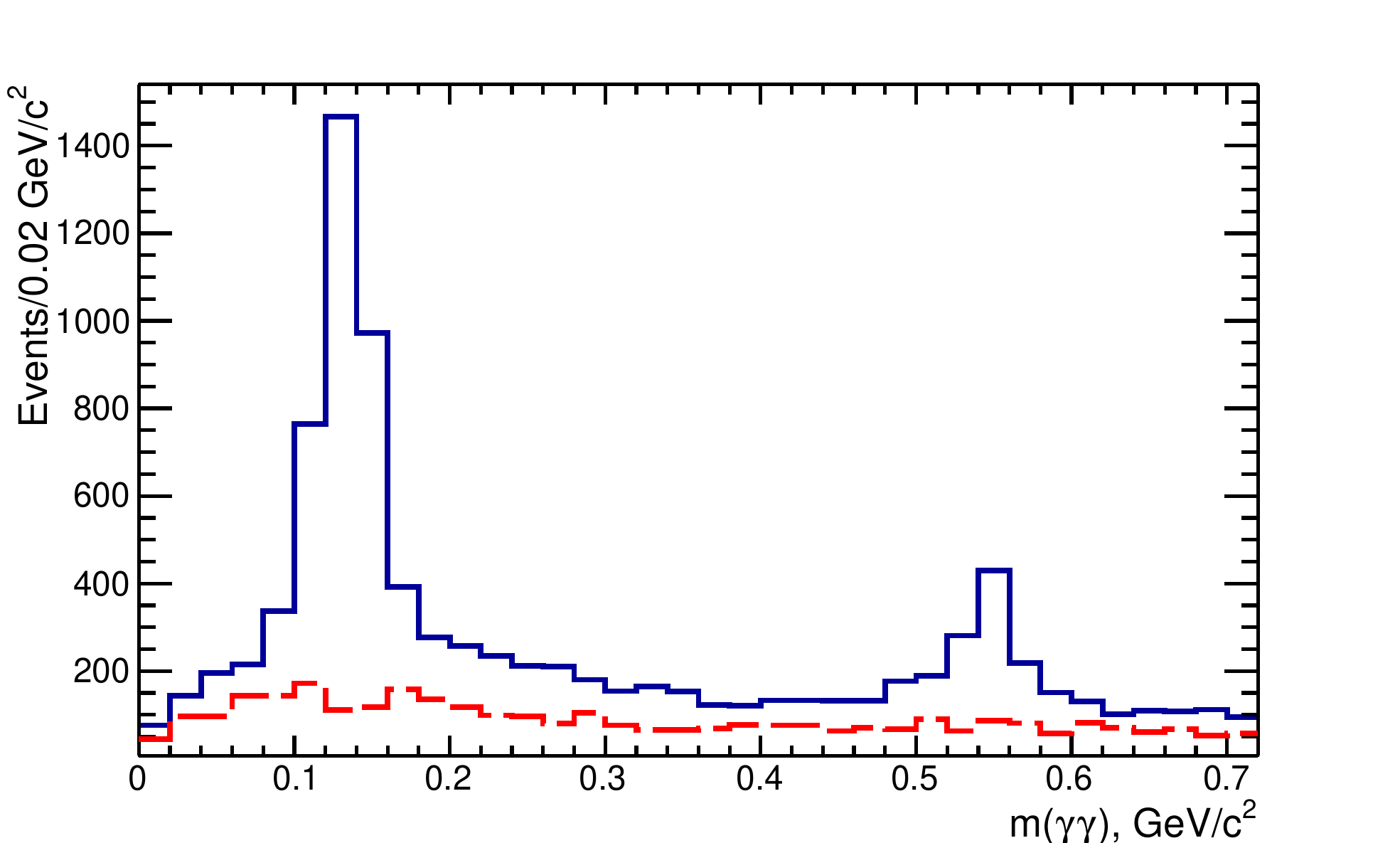}
\put(-100,100){\makebox(0,0)[lb]{\bf(b)}}
\vspace{-0.2cm}
\caption{(a) The third-photon-pair invariant mass vs
  $m(\pipi\ppz\gamma\gamma)$ for the $uds$ simulation.
(b) The projection plot for (a) the signal region
$\chisq_{2\pi2\piz\gamma\gamma} <60$ (solid histogram), and
the control region $60<\chisq_{2\pi2\piz\gamma\gamma} <120$
(dashed histogram). 
}
\label{udsbkg}
\end{center}
\end{figure}
\subsection{Peaking background}\label{sec:udsbkg}
The major background producing a $\piz$ peak
following application of the
selection criteria of Sect. IV.A is
from non-ISR \qqbar events, the most important
channel being $\epem\to\pipi\ppz\ppz$
in which one of the
neutral pions decays asymmetrically, 
yielding a high energy photon
that mimics an ISR photon.
Figure~\ref{udsbkg} (a) shows the third-photon-pair invariant
mass vs $m(\pipi\ppz\gamma\gamma)$ for the non-ISR light quark
$\qqbar$ ($uds$) simulation:
clear signals from $\piz$ and 
$\eta$ are seen. Figure~\ref{udsbkg}(b) shows the projection plots
for  $\chisq_{2\pi2\piz\gamma\gamma} <60$ and 
$60<\chisq_{2\pi2\piz\gamma\gamma} <120$. 

To normalize the $uds$ simulation, we calculate the diphoton invariant mass
distribution of the ISR candidate with all the remaining photons in
the event.  A $\piz$ peak is observed, with approximately the same
number of events in data and simulation, leading
to a normalization factor of $1.0\pm0.1$.
The resulting $uds$ background is shown by the squares
in Fig.~\ref{nev_2pi3pi0_data}: the $uds$ background is negligible
below 2~\gevcc, but accounts for more than half 
 the total background for around 4~\gevcc and above.

\begin{figure}[tbh]
\begin{center}
\includegraphics[width=0.9\linewidth]{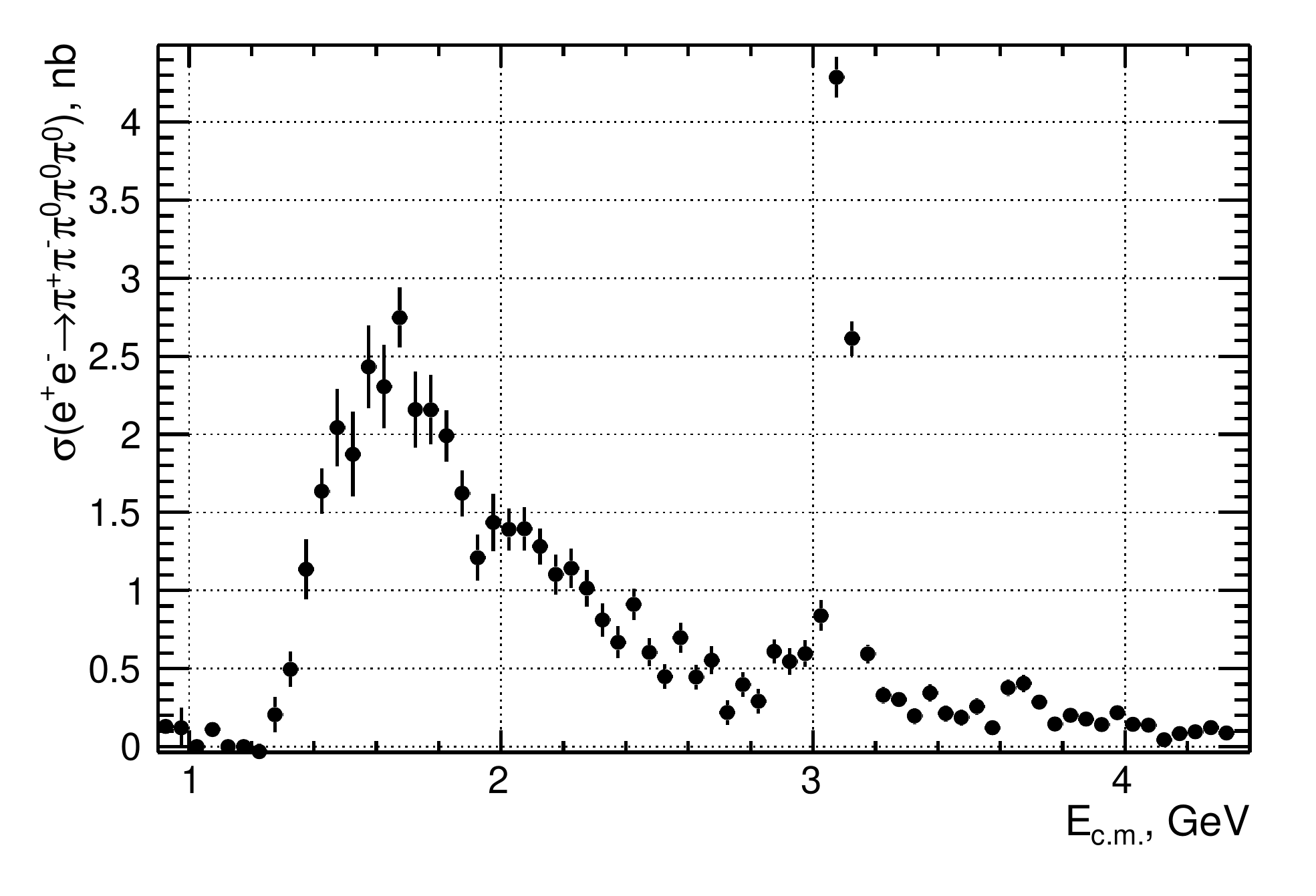}
\vspace{-0.5cm}
\caption{
The  measured $\epem\to\pipi\ppz\piz$ cross section.
The uncertainties are statistical only.
}
\label{2pi3pi0_ee_babar}
\end{center}
\end{figure} 
\begin{table*}
\caption{Summary of the $\epem\to\pipi\ppz\piz$ 
cross section measurement. The uncertainties are statistical only.}
\label{2pi3pi0_tab}
\begin{tabular}{c c c c c c c c c c}
$E_{\rm c.m.}$, GeV & $\sigma$, nb  
& $E_{\rm c.m.}$, GeV & $\sigma$, nb 
& $E_{\rm c.m.}$, GeV & $\sigma$, nb 
& $E_{\rm c.m.}$, GeV & $\sigma$, nb  
& $E_{\rm c.m.}$, GeV & $\sigma$, nb  
\\
\hline
1.125 & 0.00 $\pm$ 0.02 &1.775 & 2.20 $\pm$ 0.23 &2.425 & 0.92 $\pm$ 0.10 &3.075 & 4.36 $\pm$ 0.13 &3.725 & 0.29 $\pm$ 0.05 \\ 
1.175 & 0.00 $\pm$ 0.03 &1.825 & 2.03 $\pm$ 0.17 &2.475 & 0.61 $\pm$ 0.09 &3.125 & 2.66 $\pm$ 0.11 &3.775 & 0.15 $\pm$ 0.04 \\ 
1.225 & --0.03 $\pm$ 0.05 &1.875 & 1.65 $\pm$ 0.15 &2.525 & 0.45 $\pm$ 0.08 &3.175 & 0.60 $\pm$ 0.06 &3.825 & 0.20 $\pm$ 0.04 \\ 
1.275 & 0.21 $\pm$ 0.12 &1.925 & 1.23 $\pm$ 0.15 &2.575 & 0.71 $\pm$ 0.10 &3.225 & 0.33 $\pm$ 0.05 &3.875 & 0.18 $\pm$ 0.04 \\ 
1.325 & 0.51 $\pm$ 0.12 &1.975 & 1.46 $\pm$ 0.19 &2.625 & 0.45 $\pm$ 0.08 &3.275 & 0.31 $\pm$ 0.05 &3.925 & 0.14 $\pm$ 0.04 \\ 
1.375 & 1.17 $\pm$ 0.20 &2.025 & 1.41 $\pm$ 0.14 &2.675 & 0.56 $\pm$ 0.09 &3.325 & 0.20 $\pm$ 0.05 &3.975 & 0.22 $\pm$ 0.04 \\ 
1.425 & 1.68 $\pm$ 0.15 &2.075 & 1.42 $\pm$ 0.14 &2.725 & 0.22 $\pm$ 0.08 &3.375 & 0.35 $\pm$ 0.05 &4.025 & 0.14 $\pm$ 0.04 \\ 
1.475 & 2.10 $\pm$ 0.26 &2.125 & 1.30 $\pm$ 0.12 &2.775 & 0.40 $\pm$ 0.08 &3.425 & 0.22 $\pm$ 0.05 &4.075 & 0.14 $\pm$ 0.03 \\ 
1.525 & 1.92 $\pm$ 0.28 &2.175 & 1.12 $\pm$ 0.13 &2.825 & 0.29 $\pm$ 0.08 &3.475 & 0.19 $\pm$ 0.05 &4.125 & 0.04 $\pm$ 0.03 \\ 
1.575 & 2.49 $\pm$ 0.27 &2.225 & 1.16 $\pm$ 0.13 &2.875 & 0.62 $\pm$ 0.08 &3.525 & 0.26 $\pm$ 0.05 &4.175 & 0.08 $\pm$ 0.03 \\ 
1.625 & 2.36 $\pm$ 0.27 &2.275 & 1.03 $\pm$ 0.12 &2.925 & 0.55 $\pm$ 0.08 &3.575 & 0.12 $\pm$ 0.05 &4.225 & 0.09 $\pm$ 0.03 \\ 
1.675 & 2.81 $\pm$ 0.20 &2.325 & 0.82 $\pm$ 0.11 &2.975 & 0.60 $\pm$ 0.09 &3.625 & 0.38 $\pm$ 0.05 &4.275 & 0.12 $\pm$ 0.03 \\ 
1.725 & 2.20 $\pm$ 0.25 &2.375 & 0.68 $\pm$ 0.10 &3.025 & 0.85 $\pm$ 0.10 &3.675 & 0.41 $\pm$ 0.06 &4.325 & 0.09 $\pm$ 0.03 \\ 
\hline
\end{tabular}
\end{table*}

\begin{figure*}[p]
\begin{center}
\includegraphics[width=0.315\linewidth]{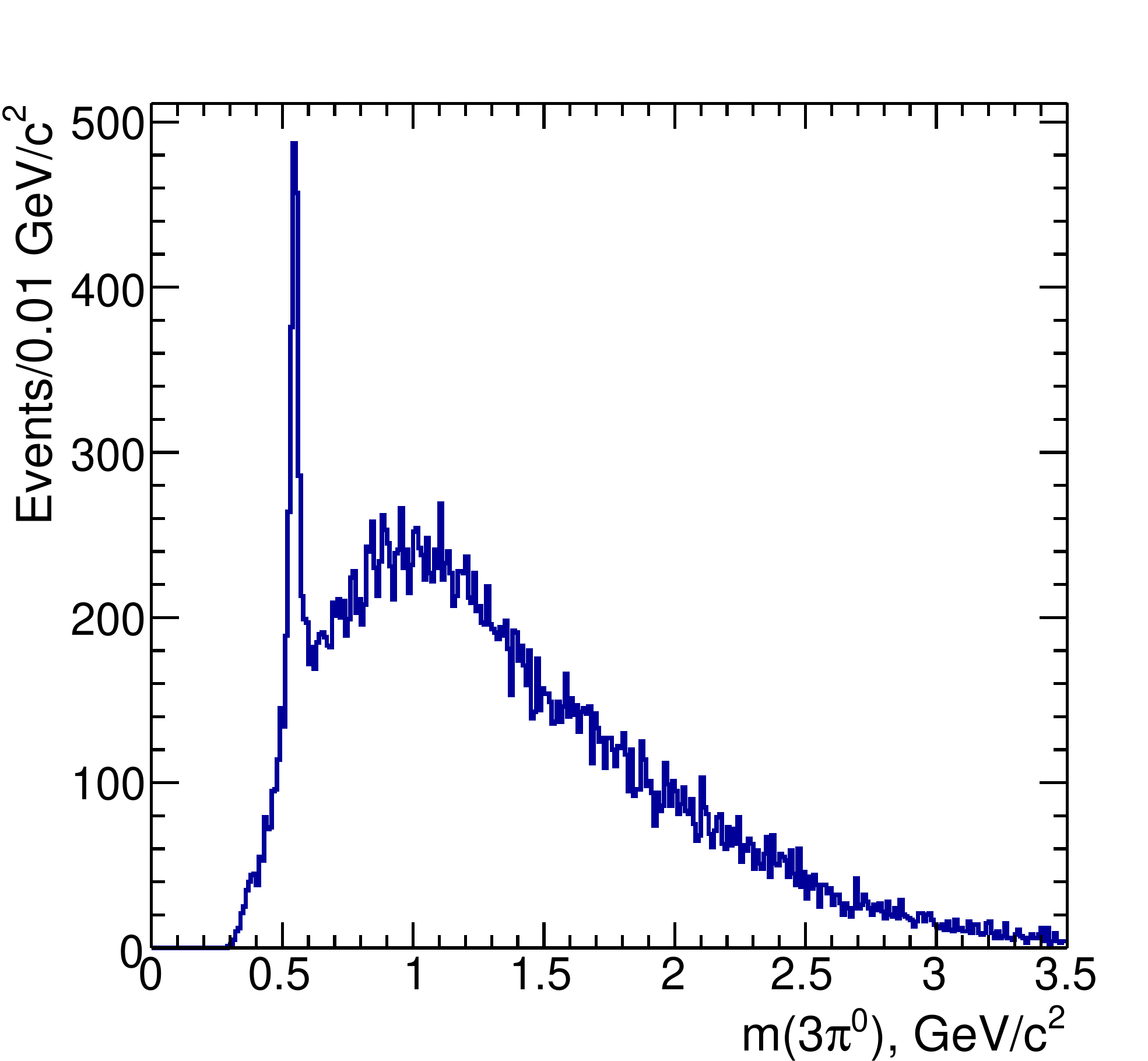}
\put(-50,120){\makebox(0,0)[lb]{\bf(a)}}
\includegraphics[width=0.315\linewidth]{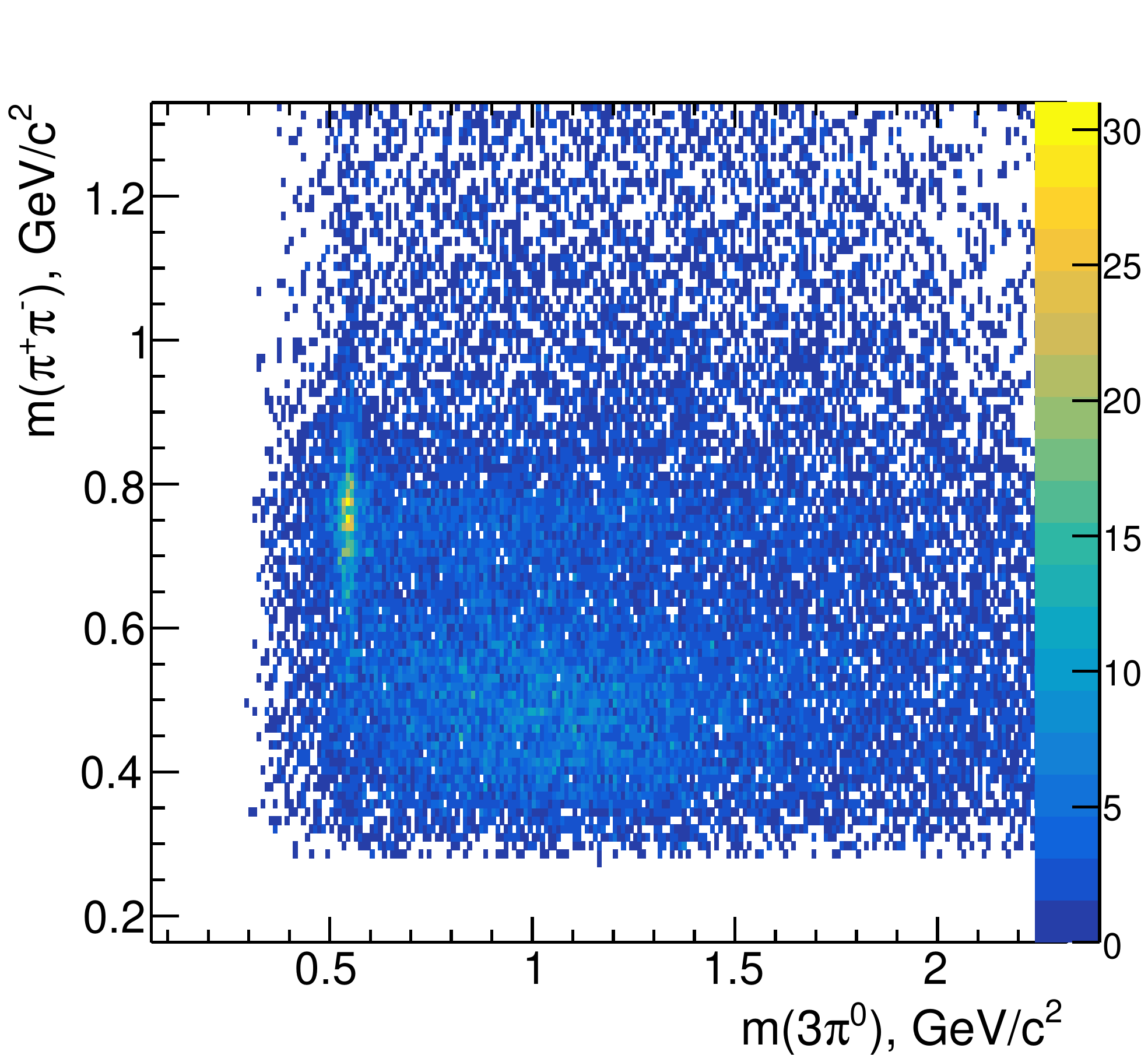}
\put(-138,120){\makebox(0,0)[lb]{\bf(b)}}
\includegraphics[width=0.315\linewidth]{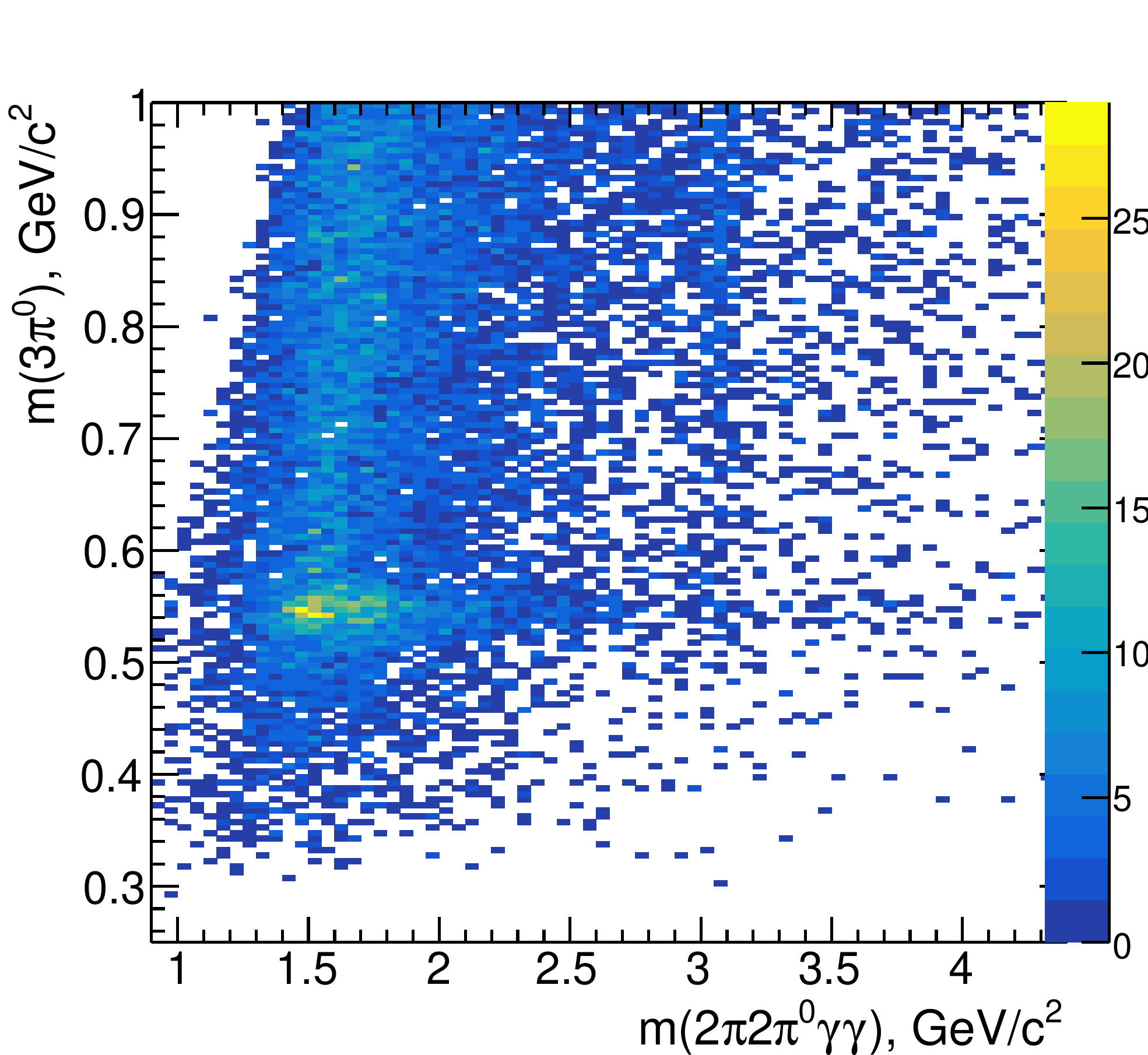}
\put(-138,120){\makebox(0,0)[lb]{\bf(c)}}
\caption{
(a) The $\ppz\pi^0$  invariant mass.
(b) The $\pipi$ vs the $\ppz\pi^0$ invariant mass.
(c)  The $\ppz\pi^0$  invariant mass vs the five-pion invariant mass.
}
\label{3pi0vs5pi}  
\end{center}
%
\begin{center}
\includegraphics[width=0.315\linewidth]{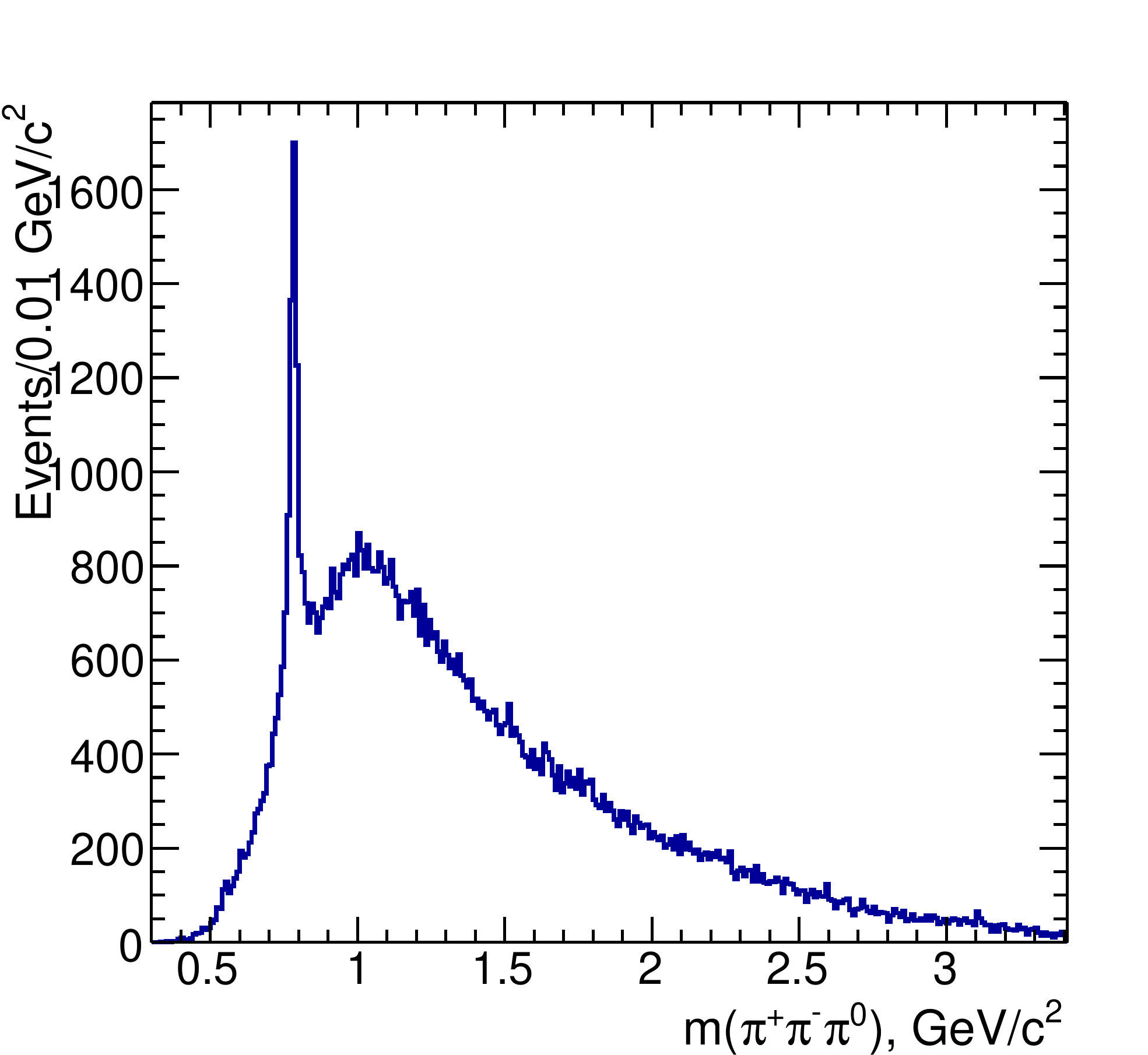}
\put(-50,120){\makebox(0,0)[lb]{\bf(a)}}
\includegraphics[width=0.315\linewidth]{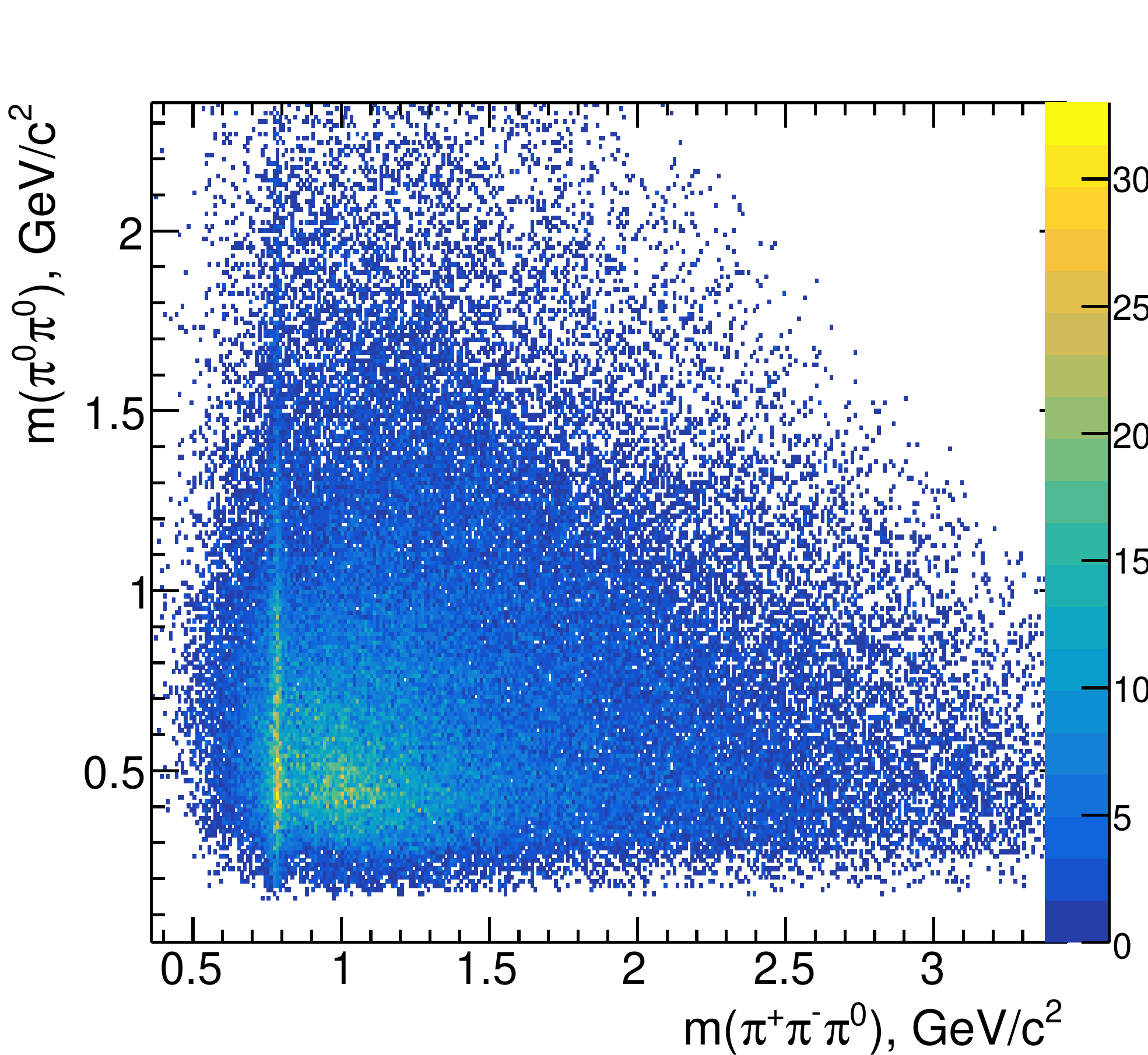}
\put(-50,120){\makebox(0,0)[lb]{\bf(b)}}
\includegraphics[width=0.315\linewidth]{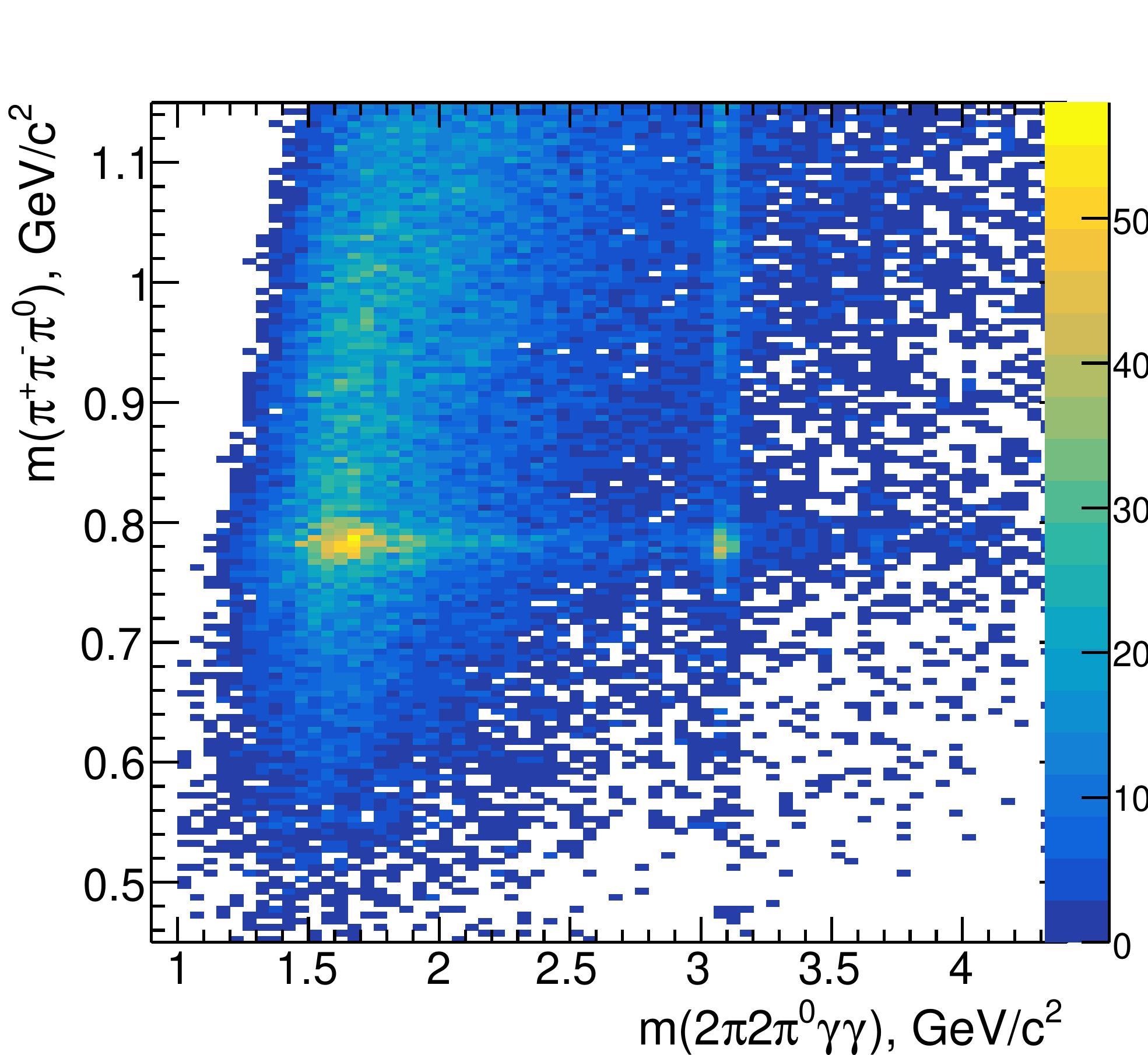}
\put(-138,120){\makebox(0,0)[lb]{\bf(c)}}
\caption{
(a) The $\pipi\pi^0$  invariant mass (three combinations per event).
(b) The $\ppz$ vs the $\pipi\pi^0$ invariant mass.
(c)  The $\pipi\pi^0$  invariant mass vs the five-pion invariant mass.
}
\label{3pivs5pi}  
\end{center}
%
\begin{center}
\includegraphics[width=0.315\linewidth]{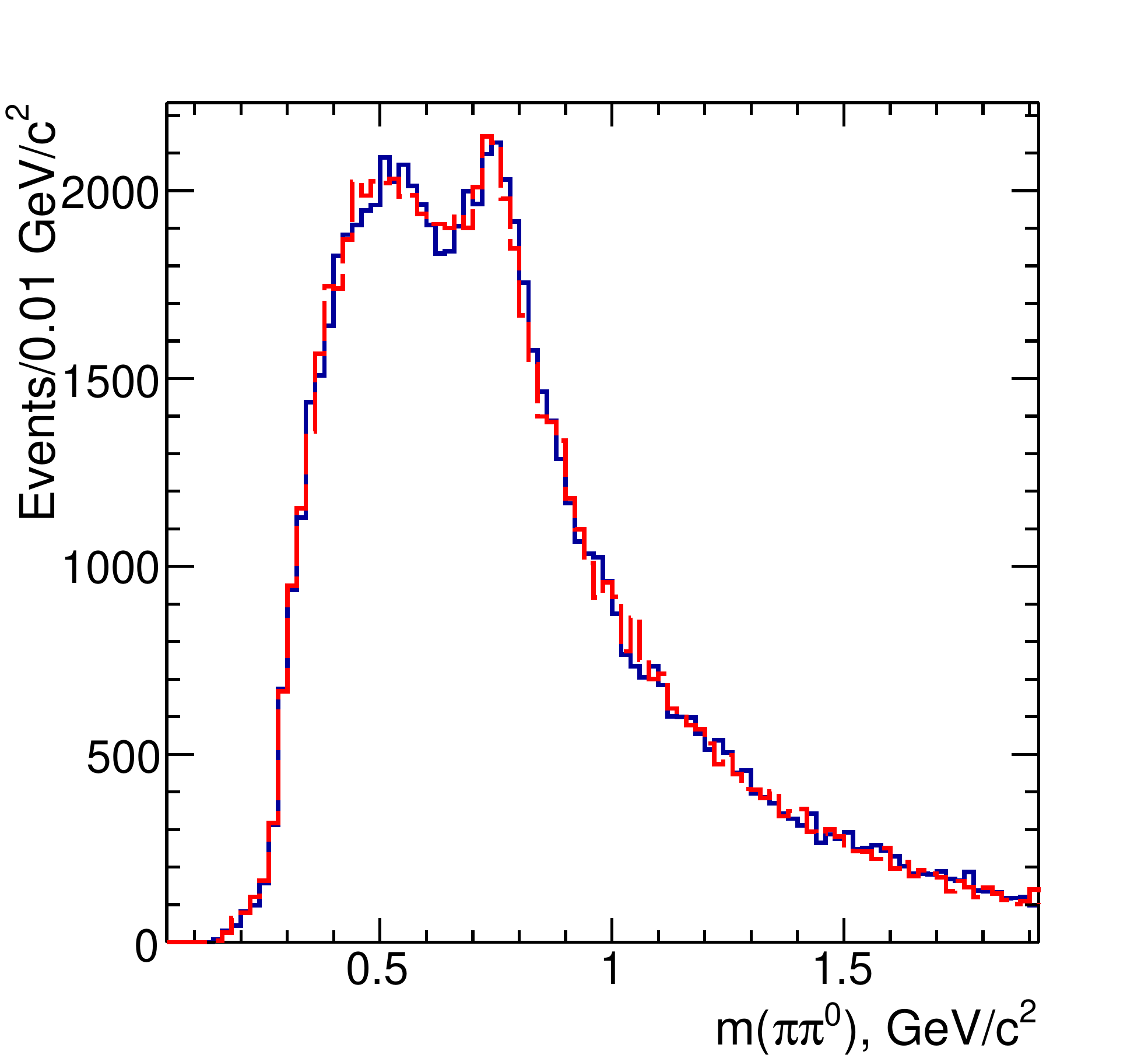}
\put(-50,120){\makebox(0,0)[lb]{\bf(a)}}
\includegraphics[width=0.315\linewidth]{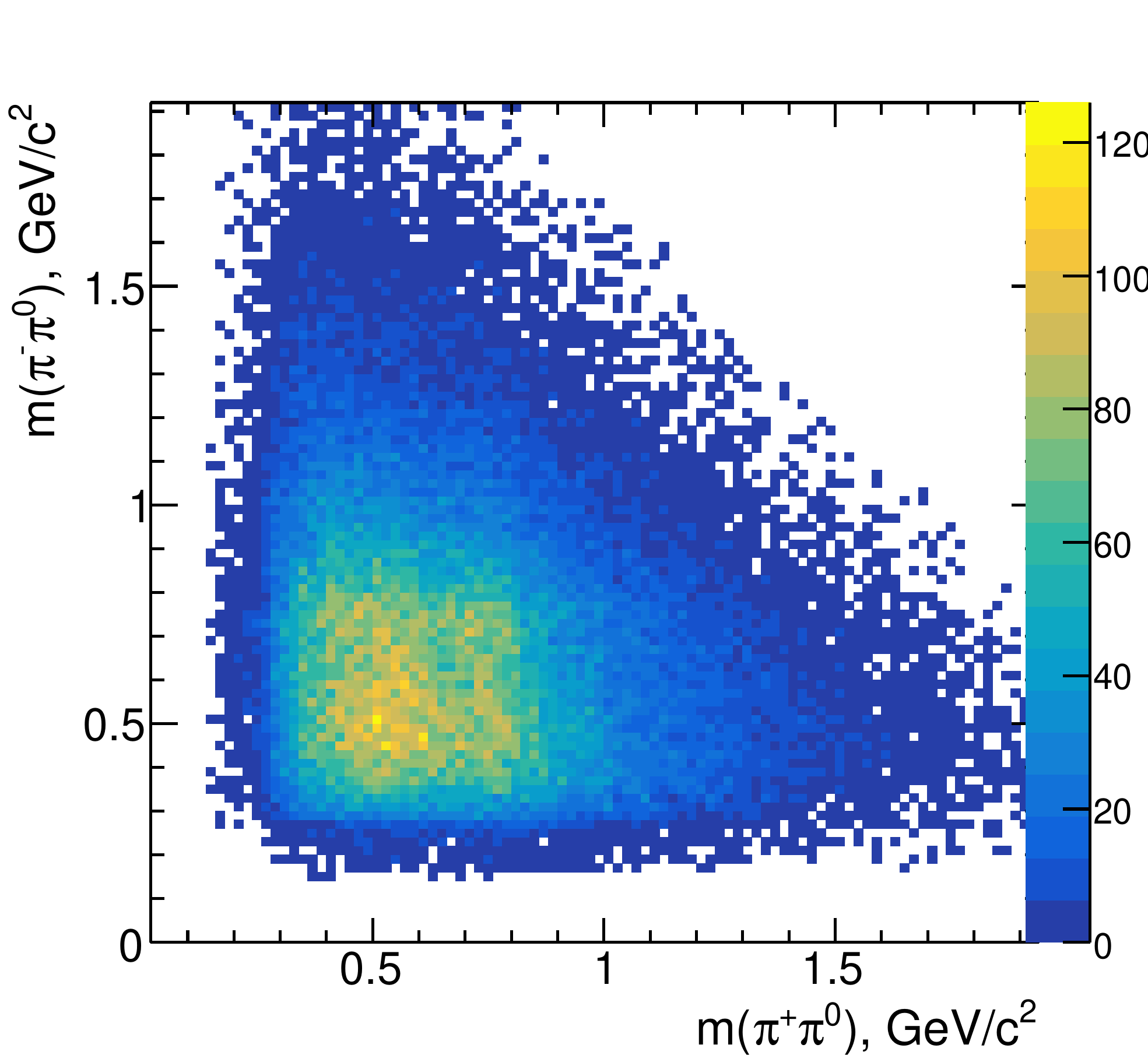}
\put(-50,120){\makebox(0,0)[lb]{\bf(b)}}
\includegraphics[width=0.315\linewidth]{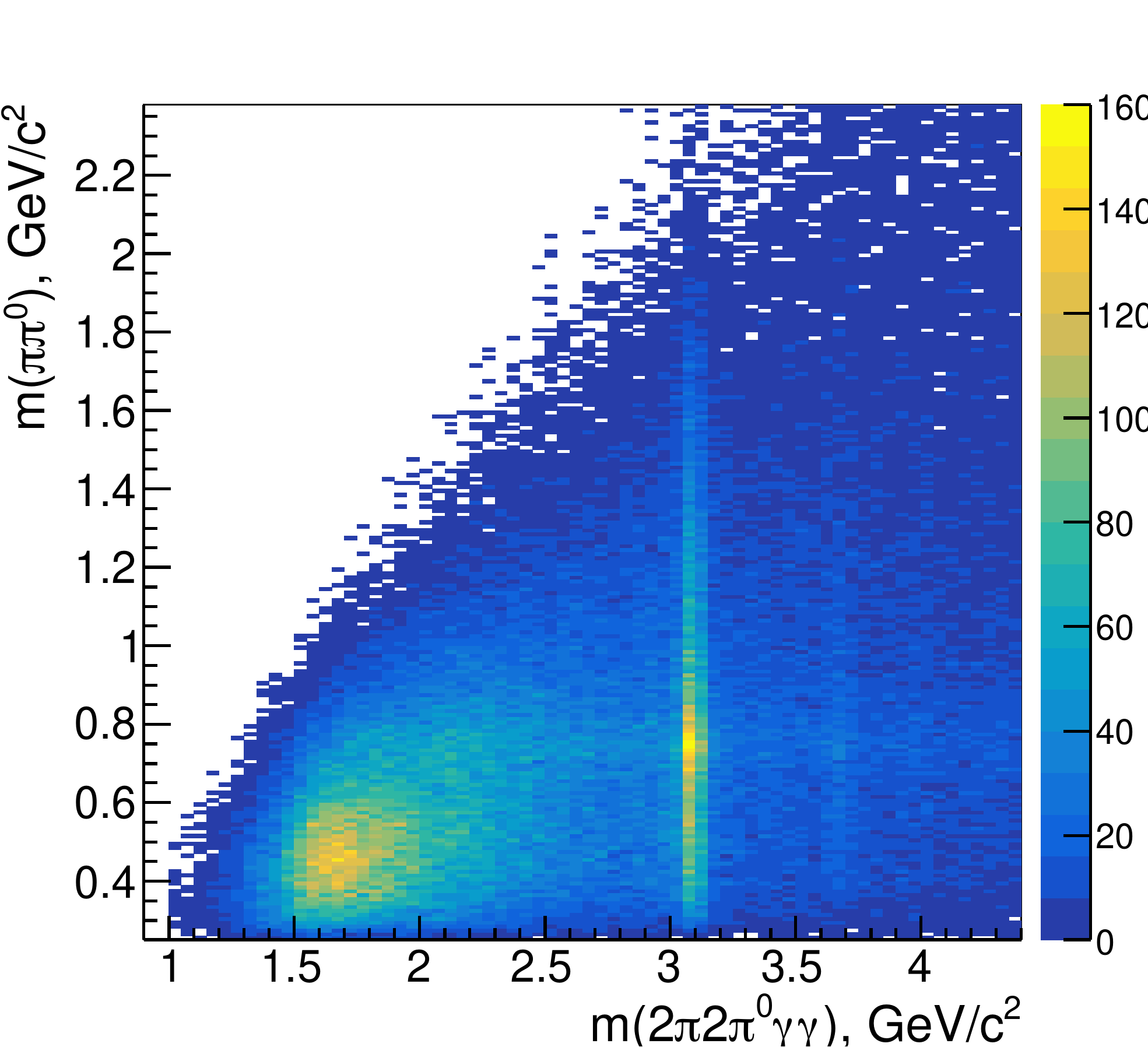}
\put(-130,120){\makebox(0,0)[lb]{\bf(c)}}
\caption{
(a) The $\pi^+\pi^0$ (solid) and $\pi^-\piz$ (dashed)  invariant masses
(three combinations per event). 
(b) The $\pi^-\piz$ vs the $\pi^+\piz$ invariant mass.
(c)  The $\pi^{\pm}\piz$  invariant mass vs the five-pion invariant mass.
}
\label{pipi0vs5pi}  
\end{center}
\end{figure*}
\subsection{\boldmath Cross section for $\epem\to \pipi\ppz\piz$}
\label{2pi3pi0}

The $\epem\to\pipi\ppz\piz$ Born cross section is  determined from
\begin{equation}
  \sigma(2\pi3\piz)(\Ecm)
  = \frac{dN_{5\pi\gamma}(\Ecm)}
         {d{\cal L}(E_{\rm c.m.})\epsilon_{5\pi}^{\rm corr}
          \epsilon_{5\pi}^{\rm MC}(E_{\rm c.m.})(1+\delta_R)}\ ,
\label{xseq}
\end{equation}
where $\Ecm$ is the invariant mass of
the five-pion system; $dN_{5\pi\gamma}$ is the background-subtracted number of selected
five-pion events in the interval $dE_{\rm c.m.}$,  and
$\epsilon_{5\pi}^{\rm MC}(E_{\rm c.m.})$ is the corresponding detection
efficiency from simulation. The factor $\epsilon_{5\pi}^{\rm corr}$ 
accounts for the difference between data and
simulation in the tracking
(1.0$\pm$1.0\%/per track)~\cite{isr4pi} and $\piz$
(3.0$\pm$1.0\% per pion)~\cite{isr2pi2pi0} reconstruction efficiencies.  
The ISR differential luminosity, $d{\cal L}$, is calculated using the 
total integrated \babar~ luminosity of 469 fb$^{-1}$~\cite{isr3pi}.
The initial- and final-state soft-photon emission is accounted for
by the radiative correction factor $(1+\delta_R)$, which is
close to unity for our selection criteria.
The cross section results contain the effect of
 vacuum polarization because this effect is not accounted for in
the luminosity calculation.

Our results for the $\epem\to\pipi\ppz\piz$ cross section
are shown in Fig.~\ref{2pi3pi0_ee_babar}.  The cross section exhibits a
structure around 1.7~\gev with  a peak value of about 2.5~nb, 
followed by a monotonic decrease toward higher energies. 
Because we present our data in bins of width 0.050~\gevcc, compatible   
with the experimental resolution, we do not apply an unfolding procedure to the data.
Numerical values for the cross section are presented in Table~\ref{2pi3pi0_tab}.
The $J/\psi$ region is discussed later.

\begin{figure}[tbh]
\begin{center}
\includegraphics[width=0.47\linewidth]{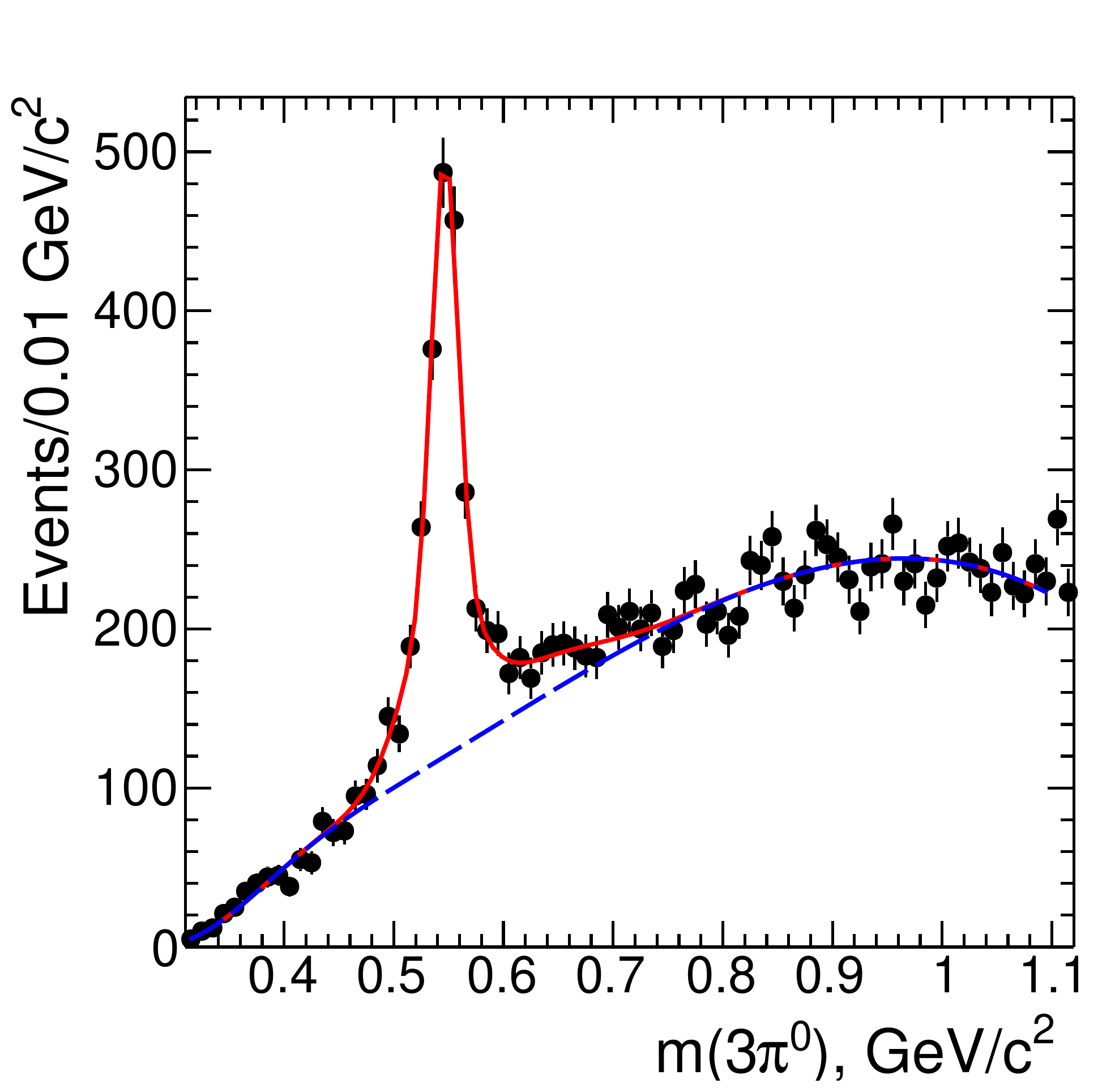}
\put(-40,90){\makebox(0,0)[lb]{\bf(a)}}
\includegraphics[width=0.50\linewidth]{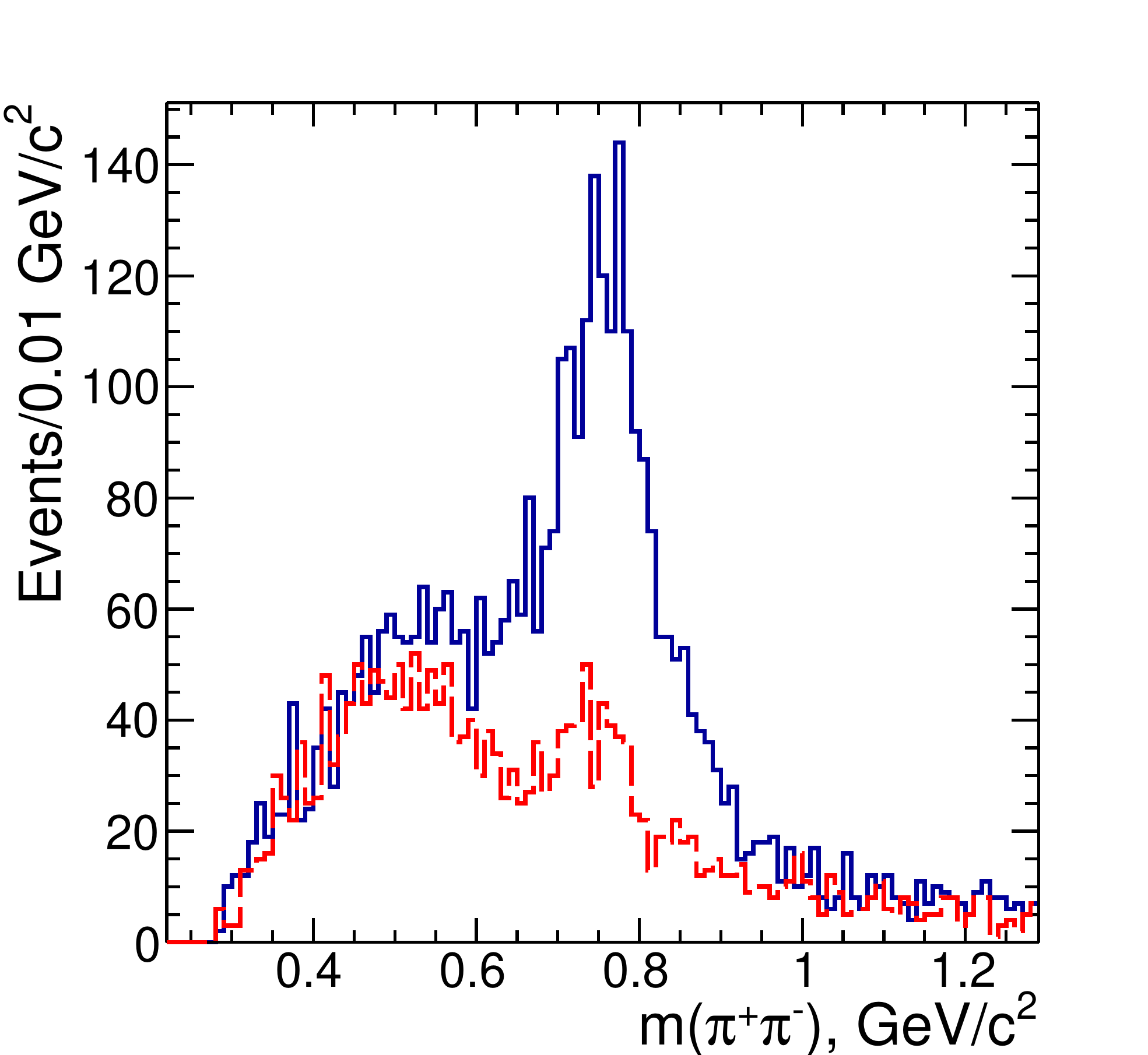}
\put(-40,90){\makebox(0,0)[lb]{\bf(b)}}
\caption{(a) The $3\piz$ invariant mass for data. 
  The curves show the fit functions.  The solid curve shows 
the $\eta$ peak (based on MC simulation) plus the non-$\eta$ continuum
background (dashed).
(b) The $\pipi$ invariant mass for events selected in the $\eta$
peak region. 
The dashed histogram shows the continuum events in the
$\eta$-peak sidebands.
}
\label{3pi0slices}
\end{center}
\end{figure}
\begin{figure}[tbh]
\begin{center}
\includegraphics[width=0.8\linewidth]{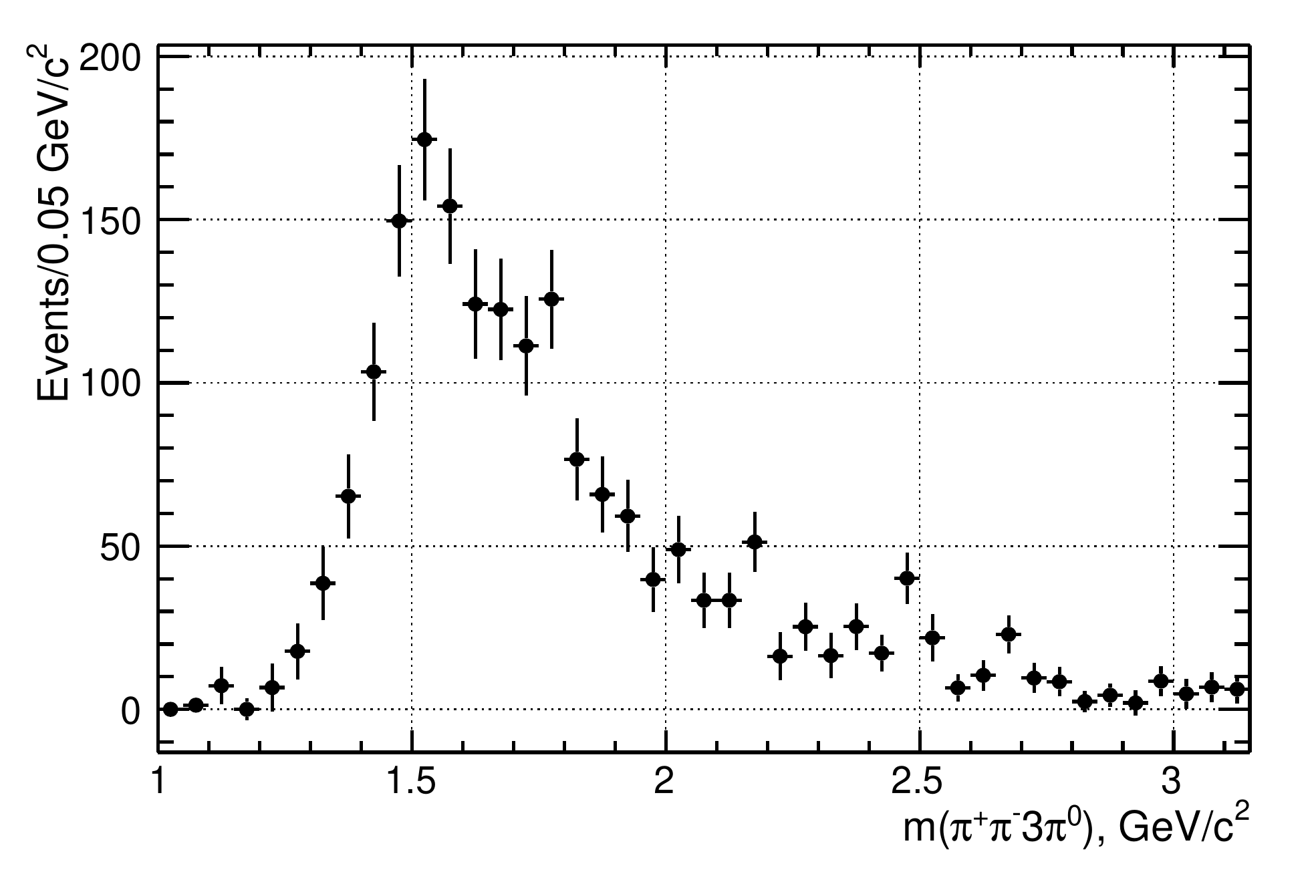}
\caption{ The $m(\pipi3\piz)$ invariant mass dependence of the selected data events
for $\epem\to\eta\pipi, \eta\to 3\piz$.
}
\label{neveta2pi}
\end{center}
\end{figure}
\subsection{\boldmath Summary of the systematic studies}
\label{sec:Systematics}
The systematic 
uncertainties, presented in the previous sections, are summarized in
Table~\ref{error_tab},  along 
with the corrections that are applied to the measurements.
\begin{table}[tbh]
\caption{
Summary of the systematic uncertainties in the $\epem\to
\pipi\ppz\pi^0$ cross section measurement.
}
\label{error_tab}
\begin{tabular}{l c c} 
Source & Correction & Uncertainty\\
\hline
Luminosity  &  --  &  $1\%$ \\
MC-data difference ISR\\ Photon efficiency & +1.5\%  & $1\%$\\ 
\chisq cut uncertainty & -- & $3\%$\\
Fit and background subtraction & -- &  $7\%$ \\
~~~~~ $\Ecm>2.5~\gev$                  & --  &  $20\%$ \\
~~~~~ $\Ecm>3.5~\gev$                 & --  &  $50\%$  \\
MC-data difference in track losses & $+2\%$ & $2\%$ \\
MC-data difference in $\pi^0$ losses & $+9\%$ & $3\%$ \\
Radiative corrections accuracy & -- & $1\%$ \\
Acceptance from MC \\(model-dependent) & -- & $5\%$  \\
\hline
Total (assuming no correlations)    &  $+12.5\%$   & $10\%$  \\
  ~~~~~$\Ecm>2.5~\gev$        &           &  $21\%$  \\
  ~~~~~$\Ecm>3.5~\gev$      &          & $50\%$\\
\end{tabular}
\end{table}

The three corrections applied to the cross sections sum
up to 12.5\%. The systematic uncertainties
vary from 10\% for \Ecm $<$ 2.5 GeV  to 50\% for \Ecm $>$ 3.5 GeV.
The largest systematic uncertainty arises from the fitting
and background subtraction procedures.
It is estimated by varying the background levels and the parameters of the functions used.

\subsection{Overview of the intermediate structures}
The $\epem\to\pipi\ppz\piz$ process has a rich internal
substructure. 
To study this substructure, we restrict events to $\mgg <
0.35$~\gevcc, eliminating the region populated
by $\epem\to\pipi\ppz\eta$.  We then assume that
the $m(\pipi2\piz\gamma\gamma)$ invariant mass can be taken to
represent $m(\pipi3\piz)$.

Figure~\ref{3pi0vs5pi}(a) shows the distribution of the $\ppz\piz$
invariant mass.  The distribution is seen to exhibit a  
prominent $\eta$ peak, which is due to the
$\epem\to\eta\pipi$ reaction.
Figure~\ref{3pi0vs5pi}(b) presents a scatter plot of the $\pipi$  vs
the $3\piz$ invariant mass.
From this plot, the $\rho(770)\eta$ intermediate state is seen to dominate.
Figure~\ref{3pi0vs5pi}(c)  presents a scatter plot of the $3\piz$
invariant mass versus $m(\pipi\ppz\gamma\gamma)$.

The distribution of the $\pipi\piz$ invariant mass (three entries per event)
is shown in ~\ref{3pivs5pi}(a).  A prominent $\omega$ peak from
$\epem\to\omega\ppz$ is seen.
Some indications of $\phi$ and $J/\psi$ peaks are also present.
The scatter plot in Fig.~\ref{3pivs5pi}(b) shows the $\ppz$  vs
the $\pipi\piz$ invariant mass. 
A scatter plot of the $\pipi\piz$ vs the $\pipi\ppz\gamma\gamma$
 mass is shown in Fig.~\ref{3pivs5pi}(c).
A clear signal for a $J/\psi$ peak is seen.

Figure~\ref{pipi0vs5pi}(a) shows the
$\pi^+\piz$(dotted) and $\pi^-\piz$(solid) invariant masses (three entries per
event). A prominent $\rho(770)$ peak, corresponding to  
$\epem\to3\pi\rho$, is visible.
The scatter plot in Fig.~\ref{pipi0vs5pi}(b) shows the $\pi^-\piz$  vs
the $\pi^+\piz$ invariant mass.  An indication of the $\rho^+\rho^-\piz$
intermediate state is visible. Figure~\ref{pipi0vs5pi}(c) shows
the $\pi\piz$ invariant mass  vs the five-pion invariant mass: a clear signal for
the $J/\psi$ and an indication of the $\psi(2S)$  are seen. 
\begin{figure}[tbh]
\begin{center}
\includegraphics[width=0.9\linewidth]{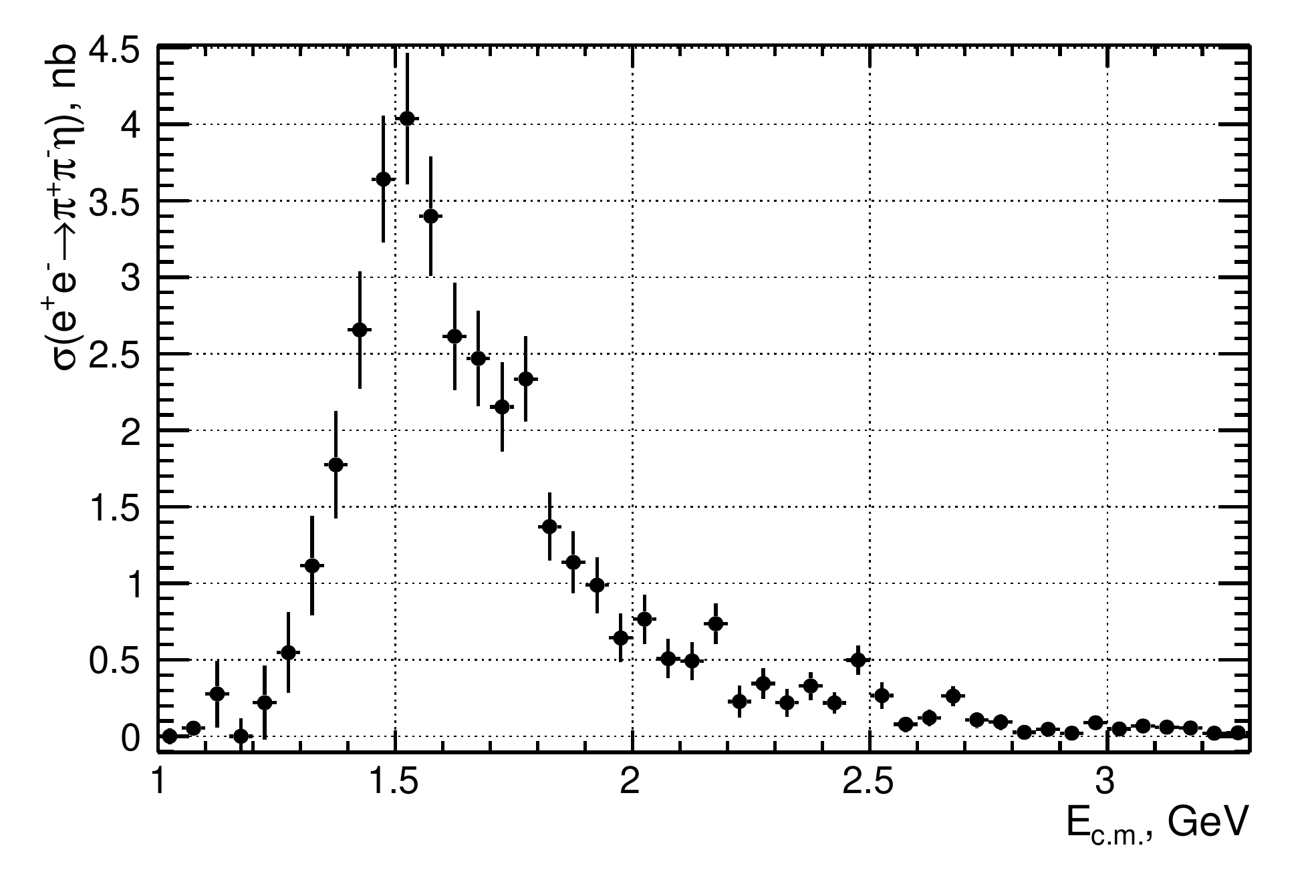}
\put(-50,120){\makebox(0,0)[lb]{\bf(a)}}\\
\includegraphics[width=1.0\linewidth]{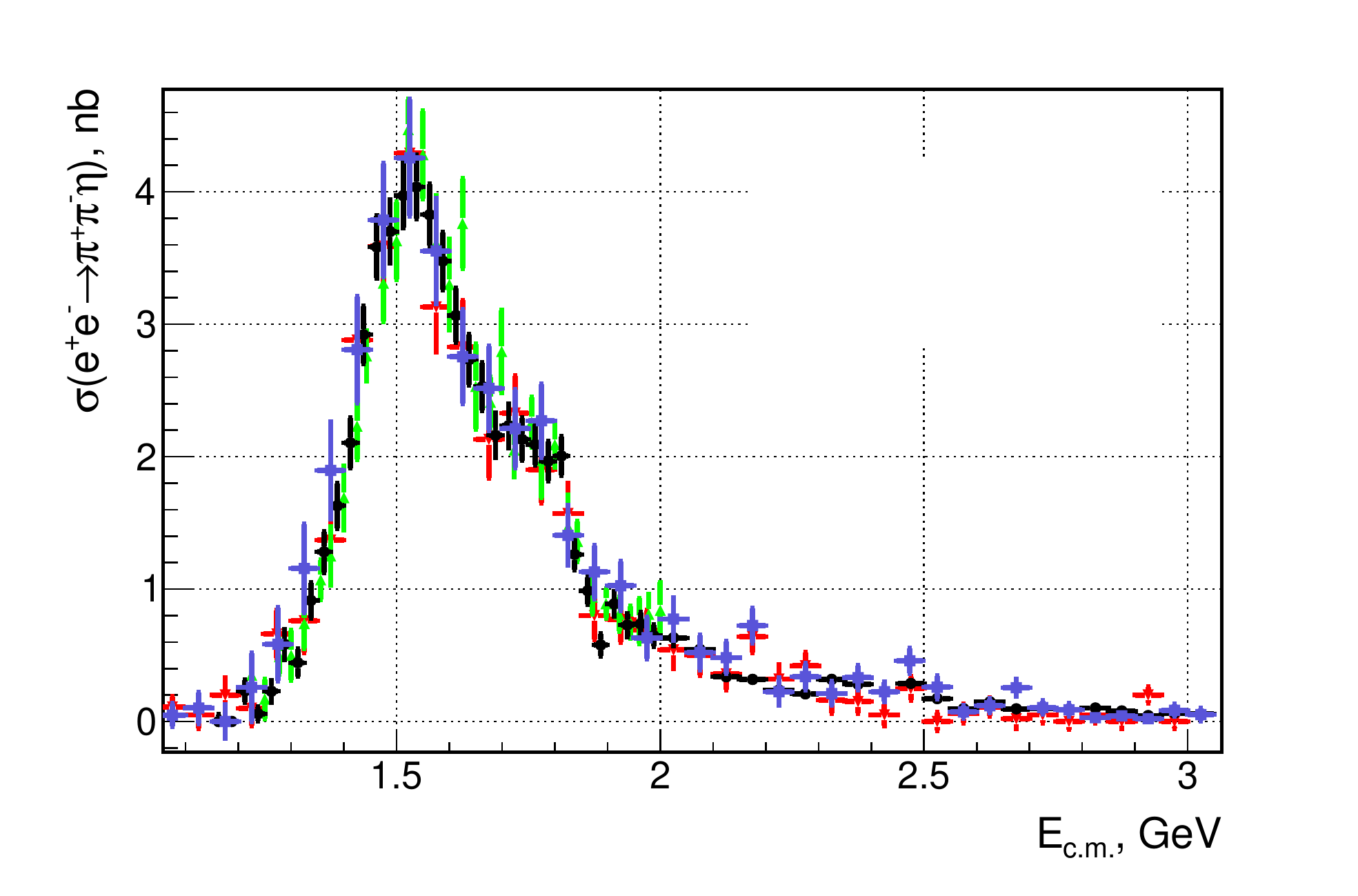}
\put(-60,120){\makebox(0,0)[lb]{\bf(b)}}
\caption{(a) The energy dependent $\epem\to\eta\pipi$ cross
  section obtained  in the $2\pi3\piz$ mode.
(b) 
Comparison of the current results (squares) with previous measurements
from \babar~ in the $\eta\to\pipi\piz$ (upside-down triangles)~\cite{isr5pi} and
$\eta\to\gamma\gamma$ modes (circles)~\cite{isretapipi}. Results from the
SND experiment~\cite{eta2pisnd} are shown by triangles.
}
\label{xs_2pieta}
\end{center}
\end{figure}
\subsection{\bf\boldmath The $\eta\pipi$ intermediate state}
\label{Sec:etapipi}
To determine the contribution of the $\eta\pipi$
 intermediate state, we fit the events of Fig.~\ref{3pi0vs5pi}(a)
using a triple-Gaussian function to describe the
signal peak, as in Fig.~\ref{m3pi_omega_eta_mc}(a), and a polynomial to
describe the background.
The result of the fit is shown in Fig.~\ref{3pi0slices}(a).
We obtain $2102\pm112$ $\eta\pipi$ events. 
The number of $\eta\pipi$ events as a function of the five-pion invariant mass
is determined by performing an analogous
fit of events in Fig.~\ref{3pi0vs5pi}(c) in each 0.05~\gevcc interval of $m(\pipi3\piz)$.  
The resulting distribution is shown in Fig.~\ref{neveta2pi}.

The $\pipi$ invariant mass distribution for events within
$\pm$0.7~\gevcc of the $\eta$ peak in Fig.~\ref{3pi0slices}(a) is
shown in Fig.~\ref{3pi0slices}(b).
A clear signal from $\rho(770)$ is observed, supporting the statement that the reaction
is dominated by the $\rho(770)\eta$ intermediate state. 
The distribution of events from $\eta$-peak
sidebands is shown by the dashed histogram.

Using Eq.~(\ref{xseq}), we determine the cross section for the
$\epem\to\eta\pipi$ process. Our simulation takes into account all
$\eta$ decays, so the cross section results, shown in Fig.~\ref{xs_2pieta}(a)
and listed in Table~\ref{2pieta_table}, correspond to all $\eta$ decays.
Systematic uncertainties in this measurement are the same as those listed in Table~\ref{error_tab}.
Figure~\ref{xs_2pieta}(b) shows our measurement
in comparison to our previous results~\cite{isr5pi,isretapipi}  and to
those from the SND experiment~\cite{eta2pisnd}.
These previous results are based on different
$\eta$ decay modes than that considered here.
The different results are seen to agree within the
uncertainties.  Including the results of the present study,
we have thus now measured the  $\epem\to\eta\pipi$ cross
section in three different $\eta$ decay modes.

\begin{figure}[tbh]
\begin{center}
\includegraphics[width=0.79\linewidth]{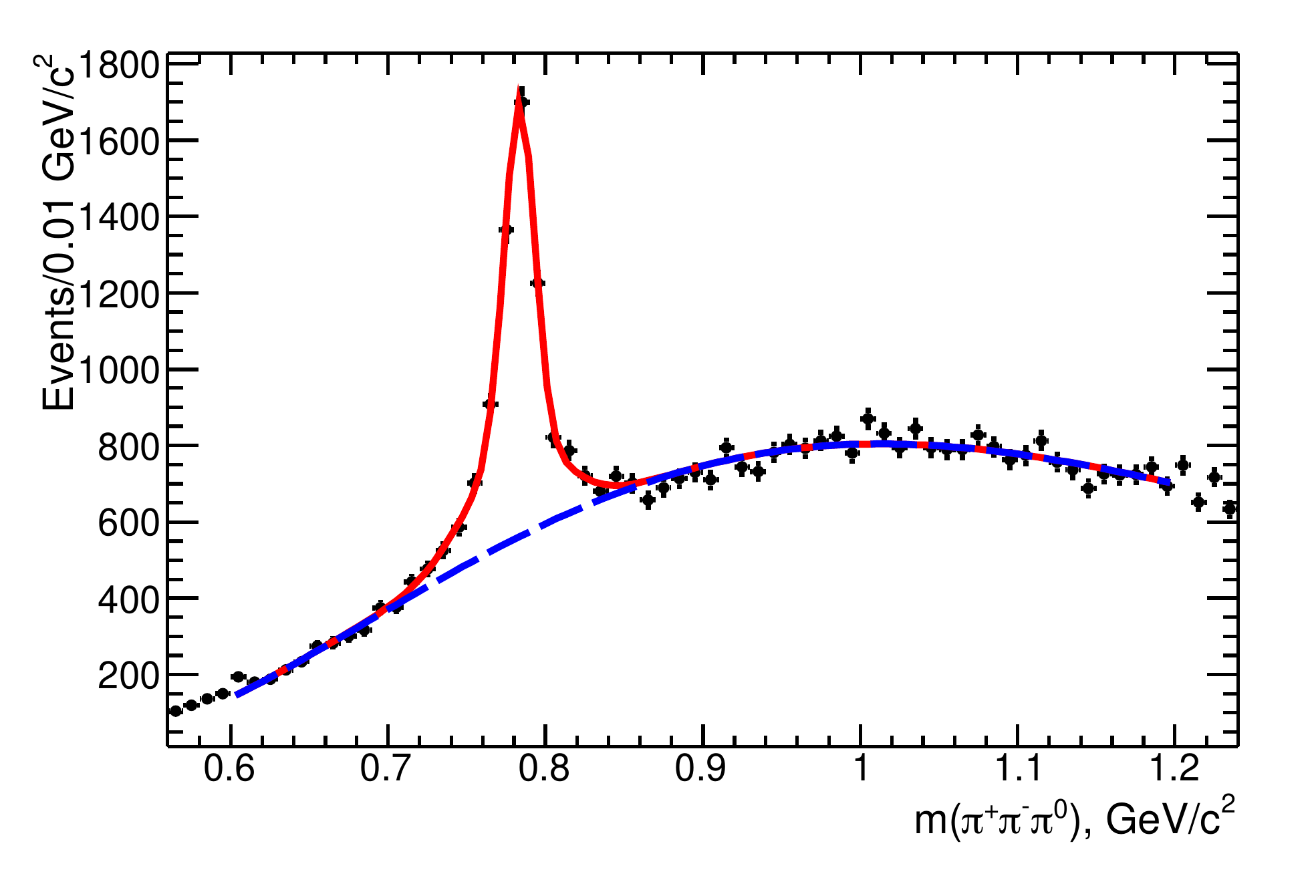}
\put(-50,100){\makebox(0,0)[lb]{\bf(a)}}\\
\includegraphics[width=0.79\linewidth]{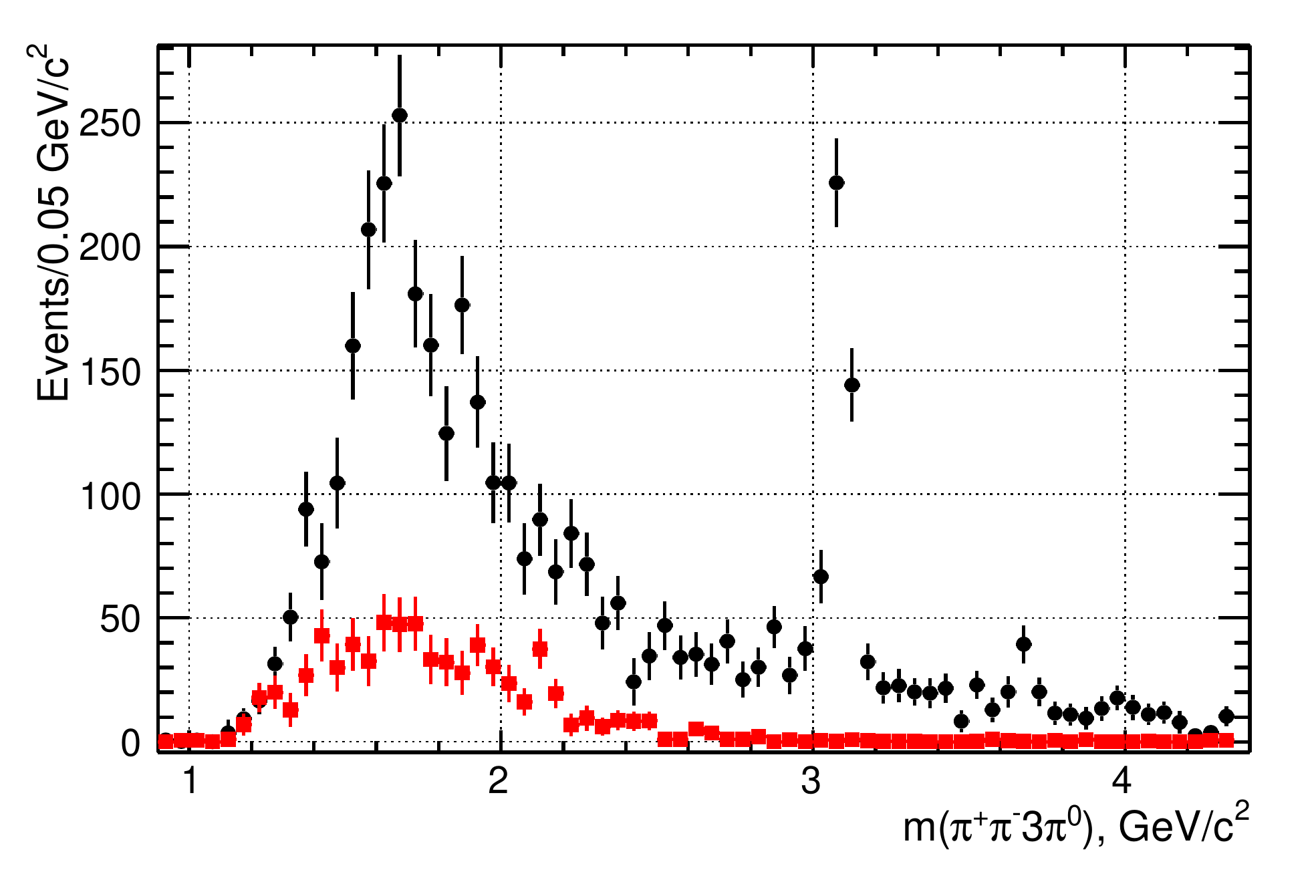}
\put(-50,100){\makebox(0,0)[lb]{\bf(b)}}
\caption{(a) The $\pipi\piz$ invariant mass for data. The solid curve
  shows the fit function for signal (based on MC-simulation) plus
 the combinatorial background (dashed curve).
(b) The mass distribution of the $\pipi\ 3\piz$ events in the 
$\omega$ peak (circles) and estimated contribution from the
$\omega\piz$ background (squares).
}
\label{nevomega2pi0}
\end{center}
\end{figure}

\subsection{\bf\boldmath The $\omega\ppz$ intermediate state}
\label{Sec:omegapipi}
%
To determine the contribution of the $\omega\ppz$
 intermediate state, we fit the events of Fig.~\ref{3pivs5pi}(a)
 using a BW function to model the signal and a polynomial to model the background. 
The BW function is convoluted with a Gaussian distribution that accounts for the detector
resolution, as described for the fit of Fig.~\ref{m3pi_omega_eta_mc}(b).
The result of the fit is shown in Fig.~\ref{nevomega2pi0}(a).
We obtain $3960\pm146$ $\omega\ppz$ events. 
The number of the $\omega\ppz$ events as a function of the five-pion invariant mass
is determined by performing an analogous
fit of events in Fig.~\ref{3pivs5pi}(c) in each 0.05~\gevcc interval of $m(\pipi3\piz)$.  
The resulting distribution is shown by the circle symbols in Fig.~\ref{nevomega2pi0}(b).
We do not observe a clear $f_0(980)\to\ppz$
signal in the $\ppz$ invariant mass, perhaps because of a large combinatorial background.
In contrast, in our previous study of the
 $\epem\to\omega\pipi\to\pipi\pipi\piz$ process~\cite{isr5pi},
 a clear $f_0(980)\to\pipi$ signal was seen.

For the $\epem\to\omega\ppz$ channel, there is a peaking background from
$\epem\to\omega\piz\to\pipi\ppz$. 
A simulation of this reaction
with proper normalization leads to the peaking-background
estimation  shown by the square symbols in Fig.~\ref{nevomega2pi0}(b).
This background is subtracted from the $\omega\ppz$
signal candidate distribution.

\begin{figure}[tbh]
\begin{center}
\includegraphics[width=0.9\linewidth]{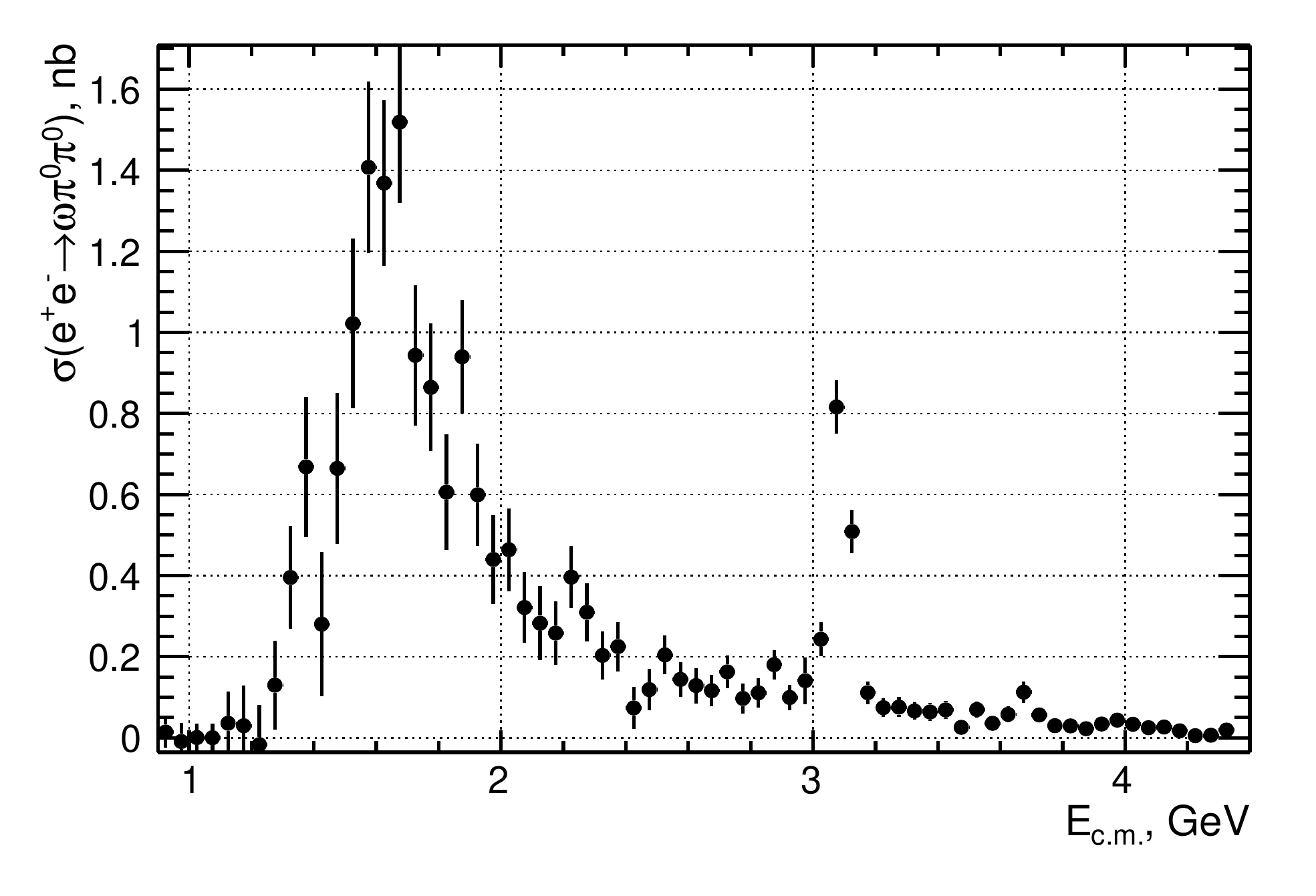}
\caption{The energy dependent $\epem\to\omega\ppz$ cross
  section  in the $\pipi3\piz$ mode.
}
\label{xs_2pi0omega}
\end{center}
\end{figure}

The  $\epem\to\omega\ppz$ cross section, corrected
for the $\omega\to\pipi\piz$ branching fraction,
is shown in Fig.~\ref{xs_2pi0omega} and tabulated in Table~\ref{omega2pi0_table}.
The uncertainties are statistical only.
The systematic uncertainties are about 10\% for
\Ecm $<$ 2.4 GeV, as discussed in Sec.~\ref{sec:Systematics}.
No previous measurement exists for this process.
The cross section exhibits a rise at threshold,
a decrease at large \Ecm, and a clear resonance
at around 1.6 GeV, possibly from
the $\omega(1650)$.  The measured $\epem\to\omega\ppz$
cross section is around a factor of two smaller than
that we observed for $\epem\to\omega\pipi$~\cite{isr5pi}, as is expected
from isospin symmetry.
\begin{figure*}[tbh]
\begin{center}
\includegraphics[width=0.29\linewidth]{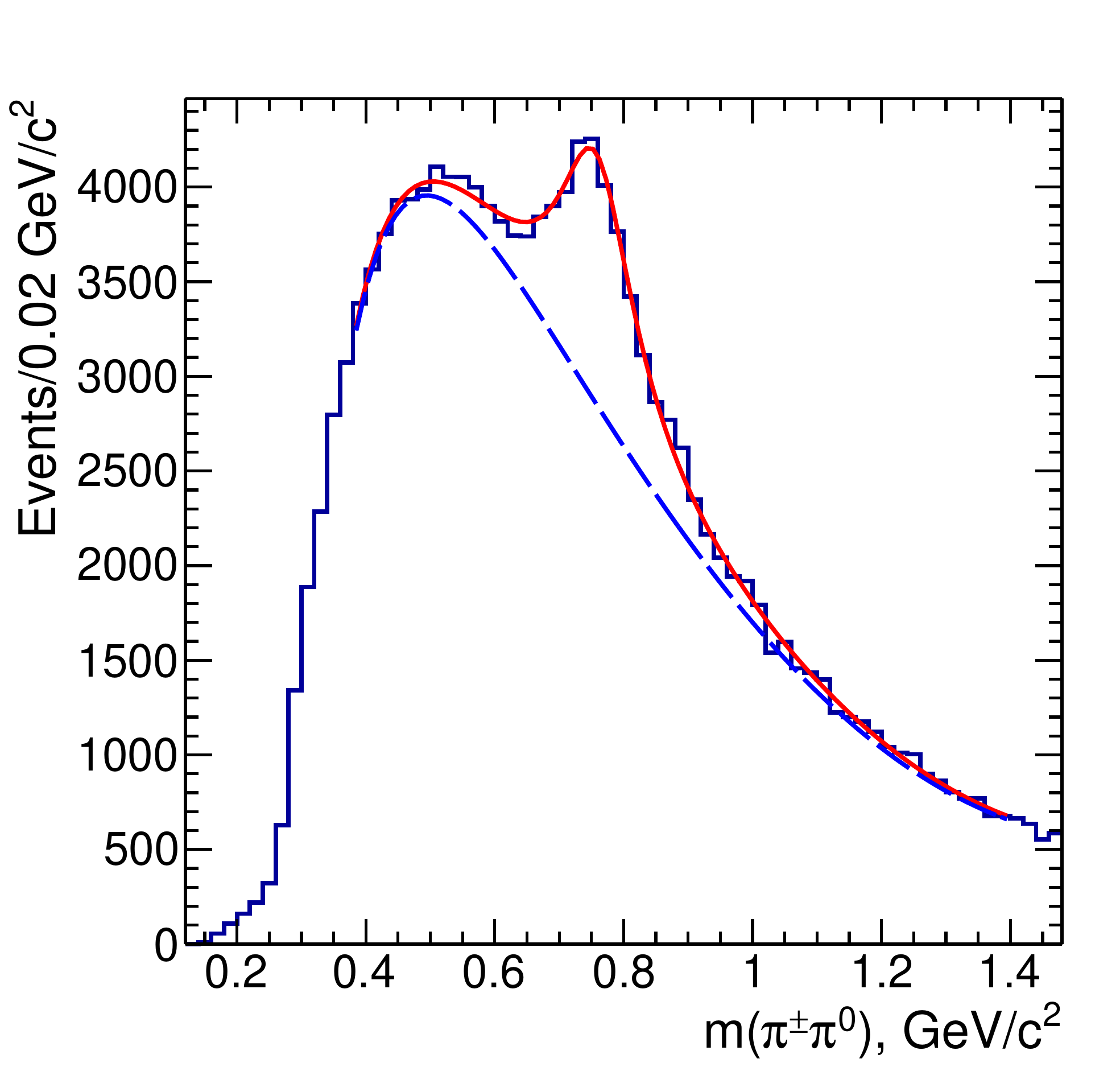}
\put(-50,100){\makebox(0,0)[lb]{\bf(a)}}
\includegraphics[width=0.29\linewidth]{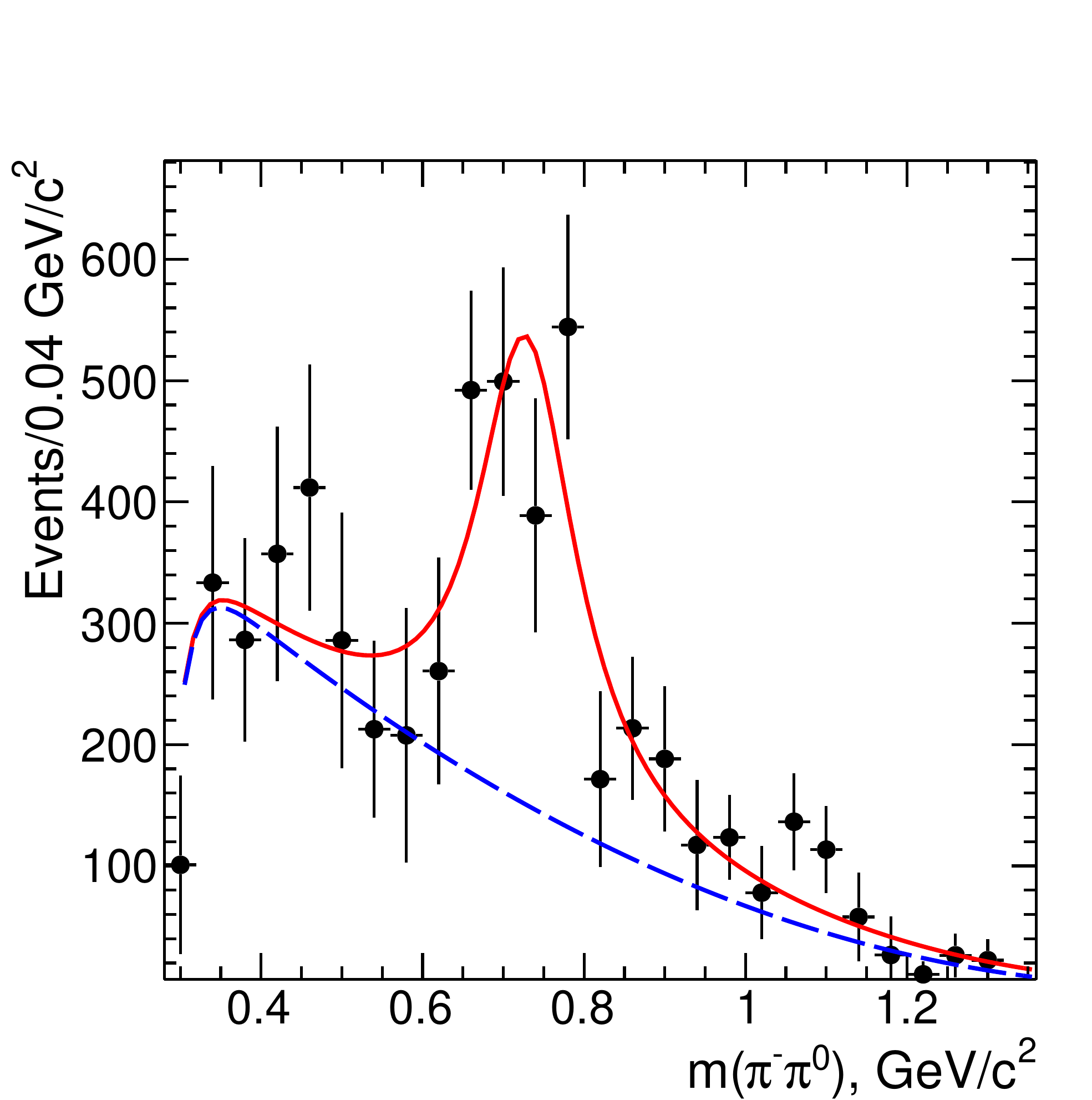}
\put(-50,100){\makebox(0,0)[lb]{\bf(b)}}
\includegraphics[width=0.30\linewidth]{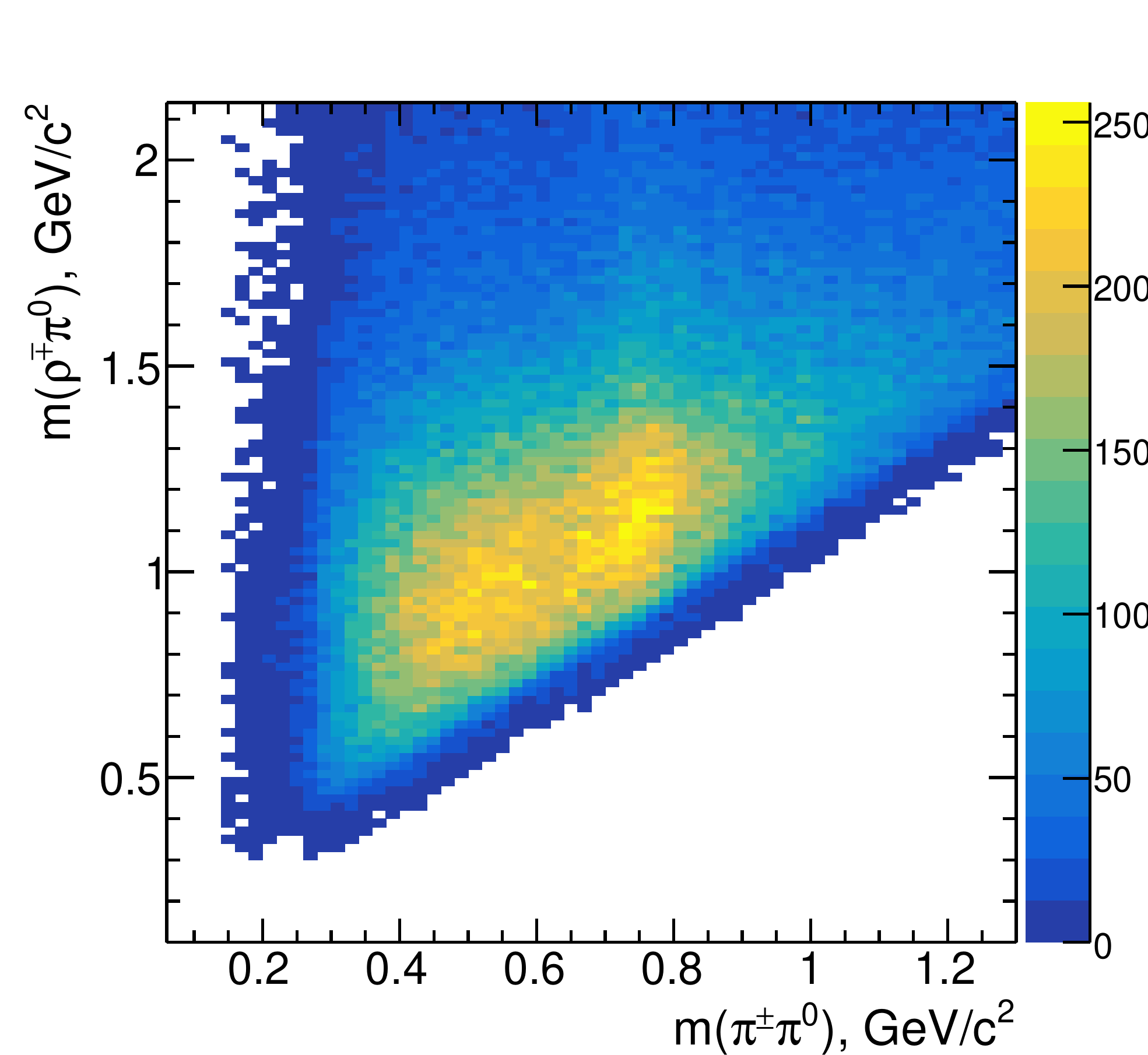}
\put(-50,40){\makebox(0,0)[lb]{\bf(c)}}
\vspace{-0.3cm}
\caption{(a) The $\pi^{\pm}\piz$ invariant mass for data. 
The dashed curve shows the fit to the combinatorial background. The solid curve 
is the sum of the background curve and the BW function for the
$\rho^{\pm}$.
(b) The result of the $\rho^+$ fit in bins of 0.04~\gevcc
in the $\rho^-$ mass.
(c) Scatter plot of the $\rho^{\pm}\piz$ invariant mass vs
the $\pi^{\mp}\piz$ invariant mass. 
}
\label{pipi0slices}
\end{center}
\end{figure*}
\begin{figure}[h]
\begin{center}
\includegraphics[width=0.9\linewidth]{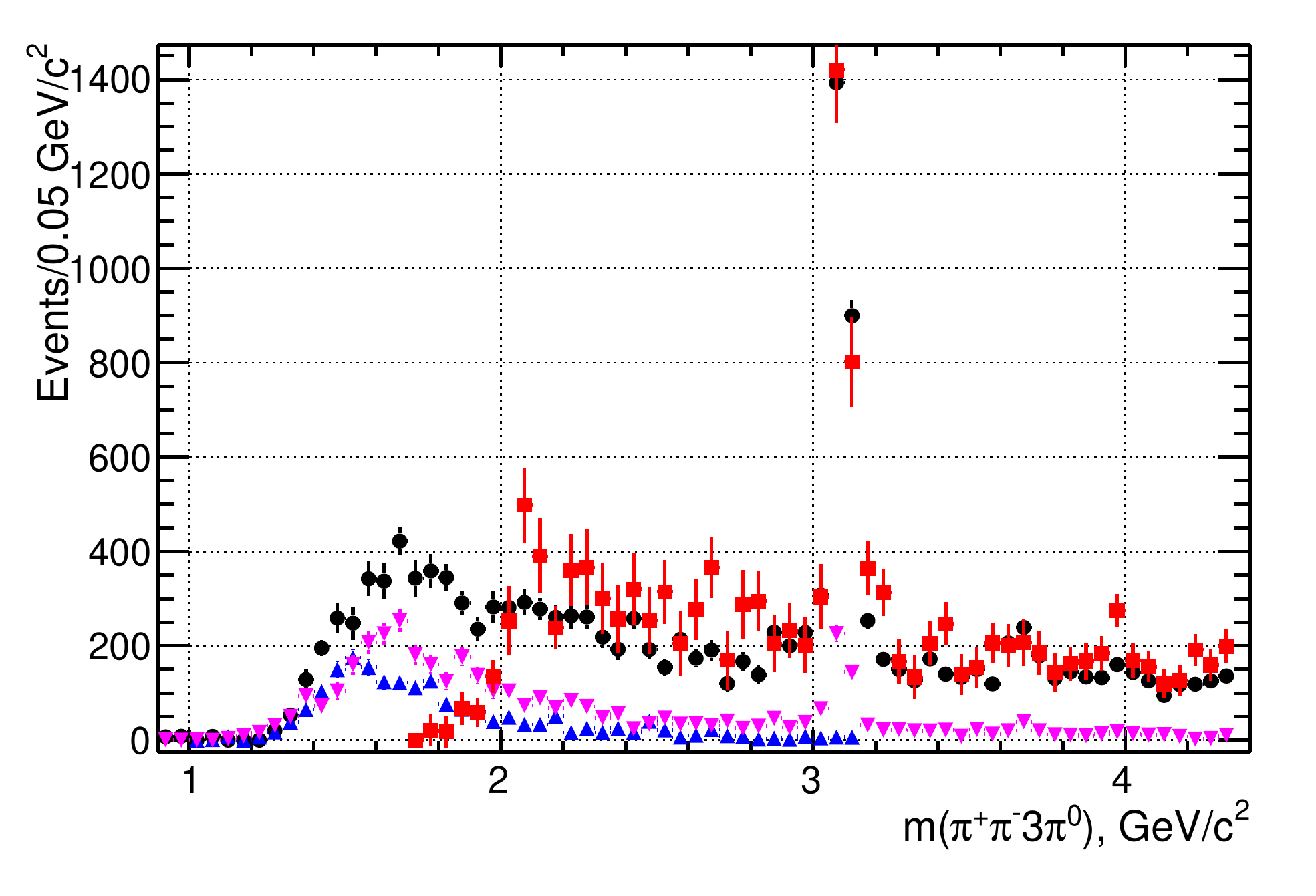}
\put(-50,120){\makebox(0,0)[lb]{\bf(a)}}\\
\includegraphics[width=0.9\linewidth]{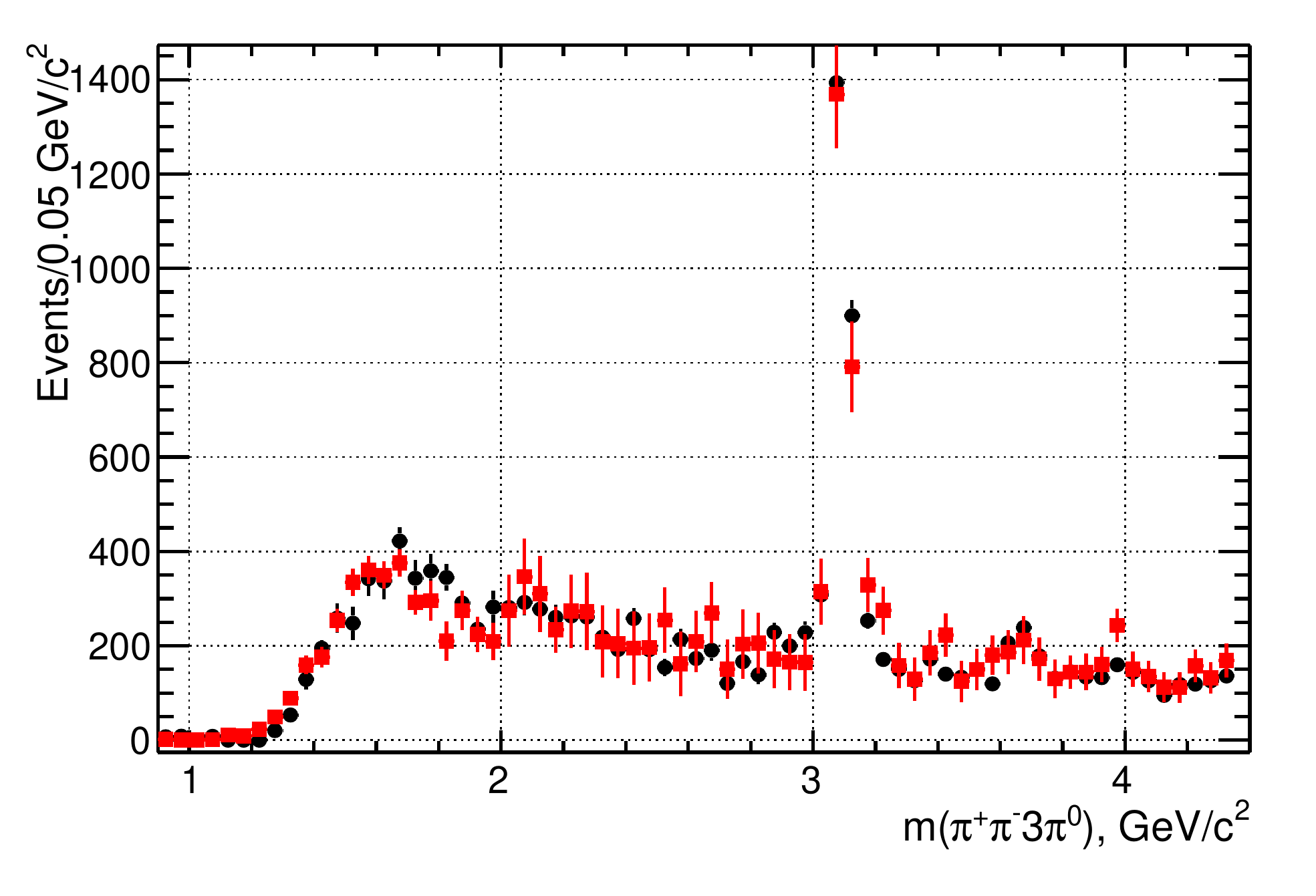}
\put(-50,120){\makebox(0,0)[lb]{\bf(b)}}
\vspace{-0.5cm}
\caption{(a) Number of events in bins of \Ecm from the $\eta\pipi$
  (triangles), $\omega\ppz$ (upside-down triangles), and  
$\rho\to\pi\piz$ (squares) intermediate states. The circles show the total
event numbers obtained  from the fit to the $\piz$ peak. 
(b) The circles as are described for  (a). The squares show the 
sums of event numbers with $\eta$, $\omega$ and the $\rho$
contribution for correlated $\rho^+\rho^-$ production.
}
\label{sumallev}
\end{center}
\end{figure}
\subsection{\bf\boldmath The $\rho(770)^{\pm}\pi^{\mp}\ppz$ intermediate state}
\label{sec:rhoselectpi0}
A similar approach is followed to study events
with a $\rho^{\pm}$ meson in the intermediate state.
Because the $\rho$ meson is broad, a BW
function is used to describe the signal shape.
There are six $\rho^\pm$ entries per event, leading to
a large combinatoric background.
To extract the contribution of the $\rho^{\pm}\pi^{\mp}\ppz$ intermediate state we
fit the events in Fig.~\ref{pipi0vs5pi}(a) with a BW function
to describe the signal and a polynomial to describe the background.
The parameters of the $\rho$ resonance are taken from Ref.~\cite{PDG}.
The result of the fit  is shown in Fig.~\ref{pipi0slices}(a). 
We obtain $14\,894\pm501$ $\rho^{\pm}\pi^{\mp}\ppz$ events. The
distribution of these events vs the five-pion invariant mass is shown
by the square symbols in Fig.~\ref{sumallev}(a).

The circle symbols in Fig.~\ref{sumallev}(a)  show the
total number of $\pipi3\piz$ events,
repeated from Fig.~\ref{nev_2pi3pi0_data}.
It is seen that the number of events with
a $\rho^{\pm}$ exceeds the total number of
$\pipi3\piz$ events, implying that there is
more than one $\rho^{\pm}$ per event,
namely a significant production
 of $\epem\to\rho^+\rho^-\piz$.
To determine the rate of $\rho^+\rho^-\piz$ events,
we perform a fit to determine the number
of $\rho^+$ in intervals of 0.04~\gevcc
in the $\pi^-\piz$ distribution of Fig.~\ref{pipi0vs5pi}(b).
The result is shown in Fig.~\ref{pipi0slices}(b).
Indeed, a significant $\rho^+$ peak is observed.
 
The number of $\epem\to\rho^+\rho^-\piz$ events is
determined by fitting the data of    Fig.~\ref{pipi0slices}(b)                        
with the sum of a BW function and a polynomial.
The sample is divided into three mass intervals:
$m(\pipi3\piz)<2.5$~\gevcc,  $2.5<m(\pipi3\piz)<3.0$~\gevcc, and
$m(\pipi3\piz)>3.0$~\gevcc. For each mass interval we determine
the number of $\rho^+$ events.
We find that the fraction of correlated
$\rho^+\rho^-$ events, relative to the total number of
$\pipi3\piz$ events with a $\rho^{\pm}$,
decreases with the mass interval as  0.49$\pm$0.05,
0.37$\pm$0.07, and 0.23$\pm0.10$, respectively,
where the uncertainties are statistical.
Thus, the $\rho^+\rho^-\piz$ intermediate state dominates at threshold.

Intermediate states with either one or two $\rho(770)$
 are expected to be produced, at least in part, 
 through  $\epem\to\rho(1400,1700)^0\piz\to
a_1(1260)^{\pm}\pi^{\mp}\piz\to\rho^{\pm}\pi^{\mp}\ppz$ and 
$\epem\to\rho^{\pm}a_1^{\mp}\to\rho^+\rho^-\piz$, respectively.
Figure~\ref{pipi0slices}(c) shows a scatter plot of the $\rho^{\pm}\piz$ invariant mass vs
the $\pi^{\mp}\piz$ invariant mass. An indication of the $a_1(1260)$ is
seen, but it is not statistically significant.

\subsection{\bf\boldmath The sum of intermediate states}
\label{sec:sum}

Figure ~\ref{sumallev}(a) shows the number of $\eta\pipi$ (upside-down triangles),
$\omega\ppz$ (triangles), and $\rho^{\pm}\pi^{\mp}\ppz$ (square) intermediate state
events, found as described in the previous sections, in comparison to the
total number of $\pipi3\piz$ events (circles) found from the fit to the
$\piz$ mass peak.  The results for the $\eta$ and $\omega$
are repeated from Figs.~\ref{neveta2pi} and~\ref{nevomega2pi0},
respectively. 
As noted above, a significant excess of events with a $\rho$ is observed. 
Based on the results of our study of correlated $\rho^+\rho^-$
production, we scale the number of events found from the
fit to the rho peak so that it corresponds to the number
of events with either a single $\rho^{\pm}$ or with a $\rho^+\rho^-$ pair.
We then sum this latter result with the eta and omega
curves in Fig.~\ref{sumallev}(a).  The result of this sum is shown
by the square symbols in Fig.~\ref{sumallev}(b).  This summed curve
is seen to be in agreement with the total number of
 $\pipi3\piz$ events, shown by the circular symbols.

Note that below
\Ecm=2 GeV,  the number of events is completely dominated by the $\eta\pipi$ and
$\omega\ppz$ channels, so the cross section of the intermediate states with
a $\rho$ can be estimated as the difference between the total
$\epem\to\pipi\ppz\piz$ cross section and the sum of the $\eta\pipi$ and 
$\omega\ppz$ contributions. 

\section{\bf\boldmath The $\pipi2\piz\eta$ final state}
\subsection{Determination of  the number of events }

\begin{figure}[tbh]
\begin{center}
\includegraphics[width=0.9\linewidth]{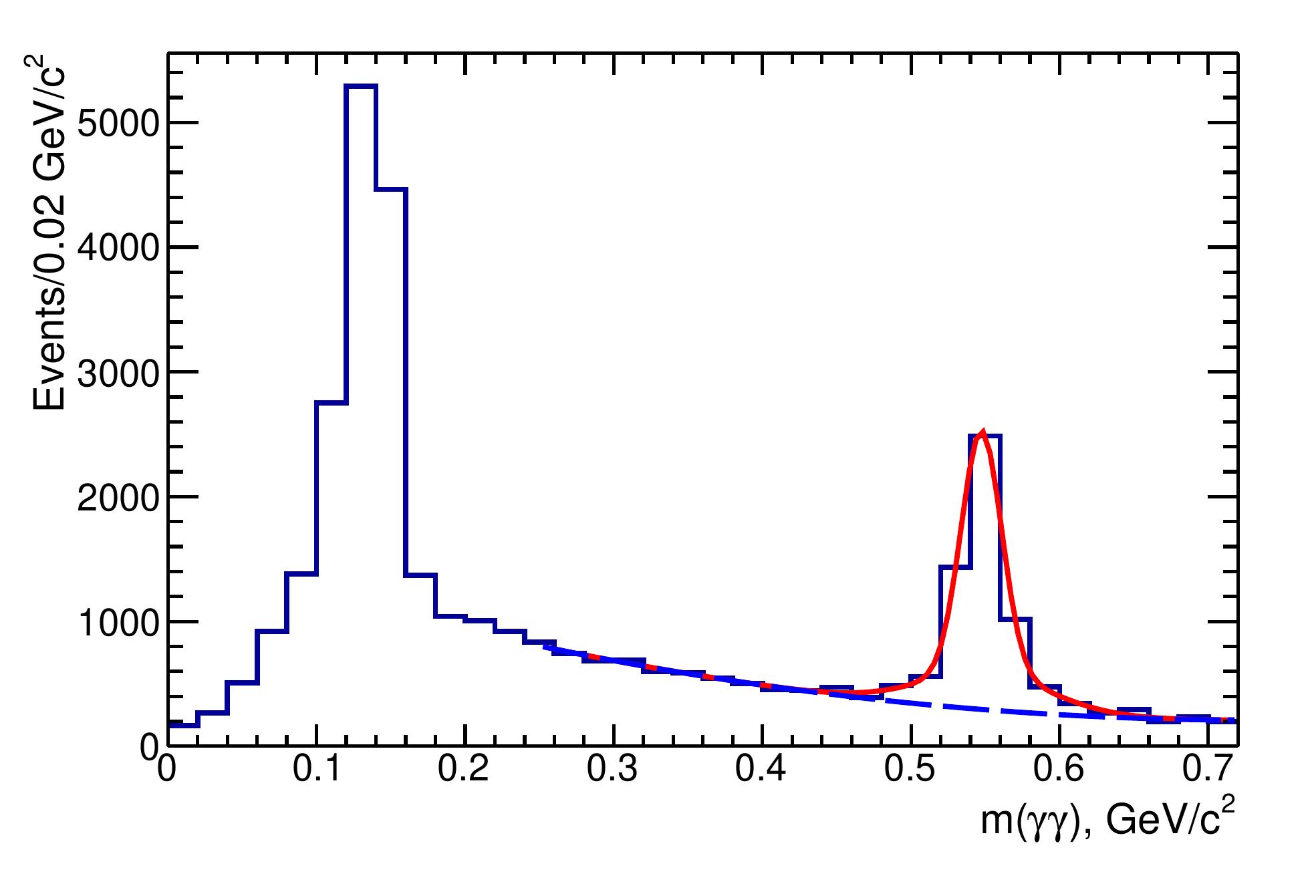}
\put(-50,120){\makebox(0,0)[lb]{\bf(a)}}\\
\includegraphics[width=0.9\linewidth]{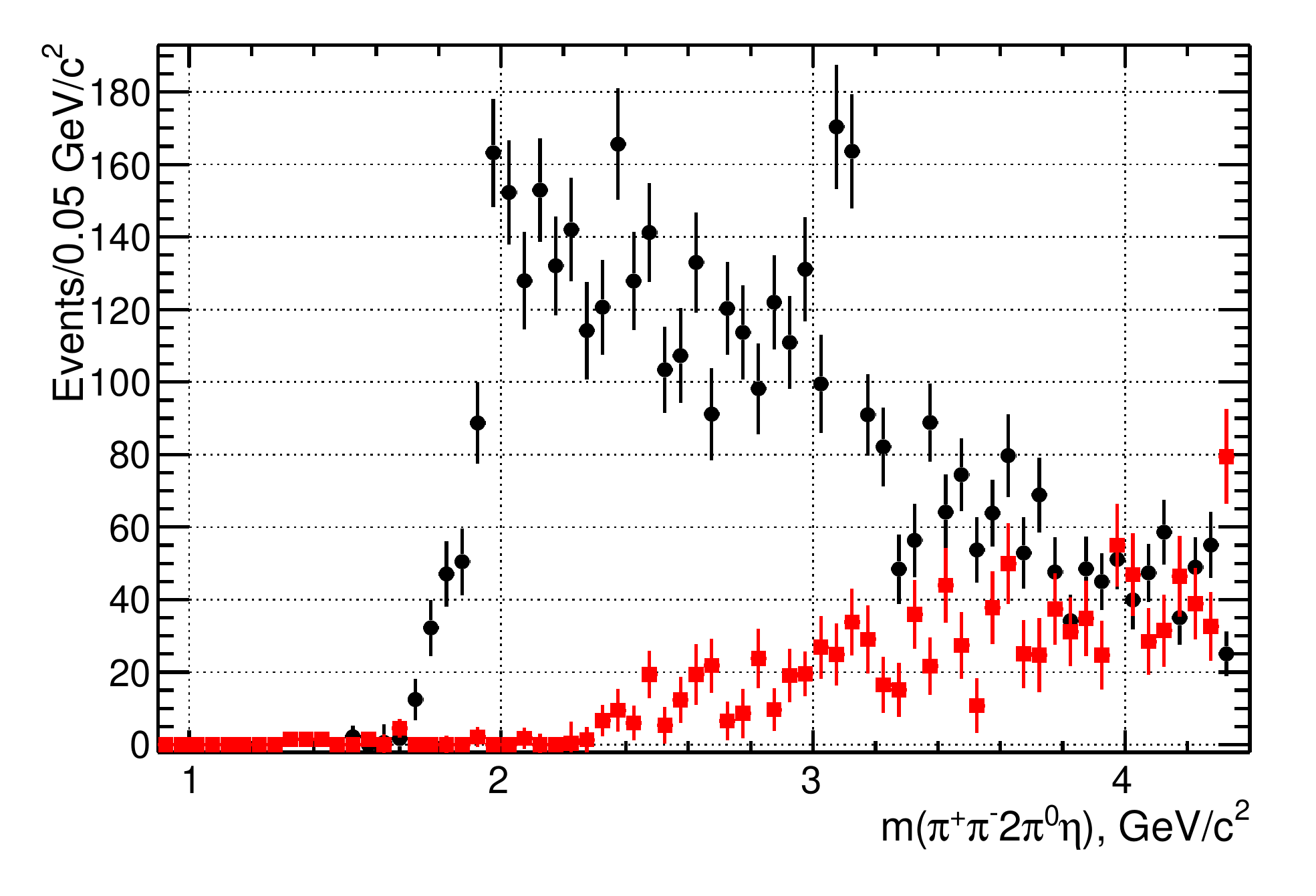}
\put(-50,120){\makebox(0,0)[lb]{\bf(b)}}
\vspace{-0.5cm}
\caption{(a) The third-photon-pair invariant mass for data.
  The dashed curve shows the fitted background. 
  The solid curve
  shows the sum of background and the two-Gaussian fit function used to obtain the number of events
  with an $\eta$.
(b) The invariant mass distribution for the
$\pipi2\piz\eta$ events obtained from the $\eta$ signal fit. The
contribution of the $uds$ background events is shown by the squares.
}
\label{meta_data_fit}
\end{center}
\end{figure}

The analogous approach to that described above for
$\epem\to\pipi\ppz\piz$ events is used to study
$\epem\to\pipi\ppz\eta$ events.
We fit the $\eta$ signal in the third-photon-pair
invariant mass distribution (cf., Fig.~\ref{2pi3pi0_chi2_all}) with
the sum of
two Gaussians with a common mean, while the relatively smooth 
background is described by a second-order polynomial function, as shown
in Fig.~\ref{meta_data_fit}(a).  We obtain $4700\pm84$ events.
Figure~\ref{meta_data_fit}(b) shows the mass distribution of these events.
\subsection{Peaking background}\label{sec:udsbkg2}

The major background producing an $\eta$ peak is the
non-ISR background, in particular $\epem\to\pipi\ppz\piz\eta$
when one of the neutral pions decays asymmetrically, producing a 
photon interpreted as ISR.  The $\eta$ peak from the $uds$ simulation
is visible in Fig.~\ref{udsbkg}.

To normalize the $uds$ simulation,
we form the diphoton invariant mass distribution of
 the ISR candidate with all the remaining photons in the event.
Comparing the number of events in the $\piz$ peaks
in data and $uds$ simulation, we assign
a scale factor of $1.5\pm0.2$ to the simulation.  We fit the $\eta$
peak in the $uds$ simulation in intervals of 0.05~\gevcc in
$m(\pipi\ppz\gamma\gamma)$. The results are shown by the squares in Fig.~\ref{meta_data_fit} (b).

\begin{figure}[tbh]
\begin{center}
\includegraphics[width=0.49\linewidth]{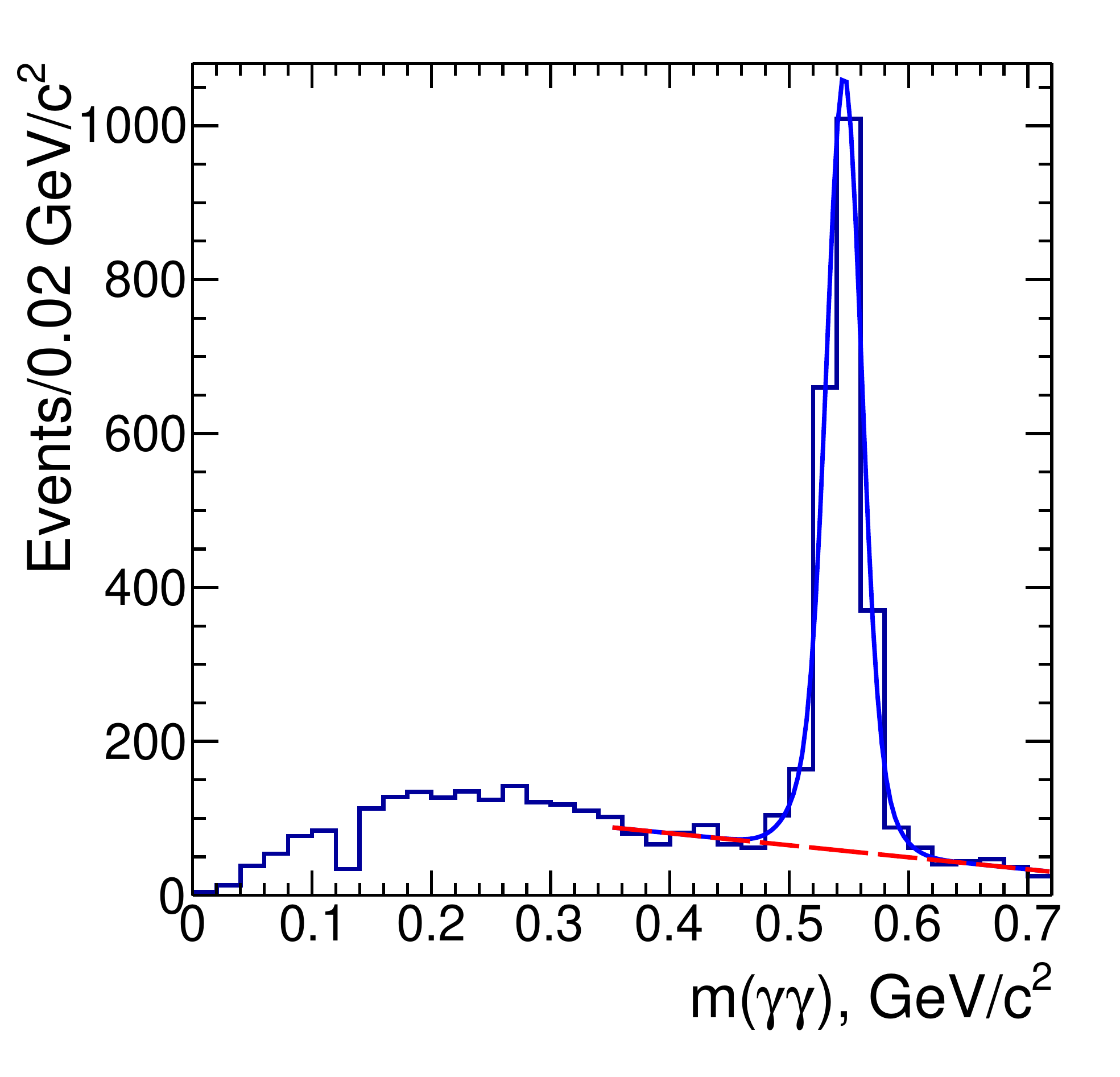}
\put(-80,80){\makebox(0,0)[lb]{\bf(a)}}
\includegraphics[width=0.49\linewidth]{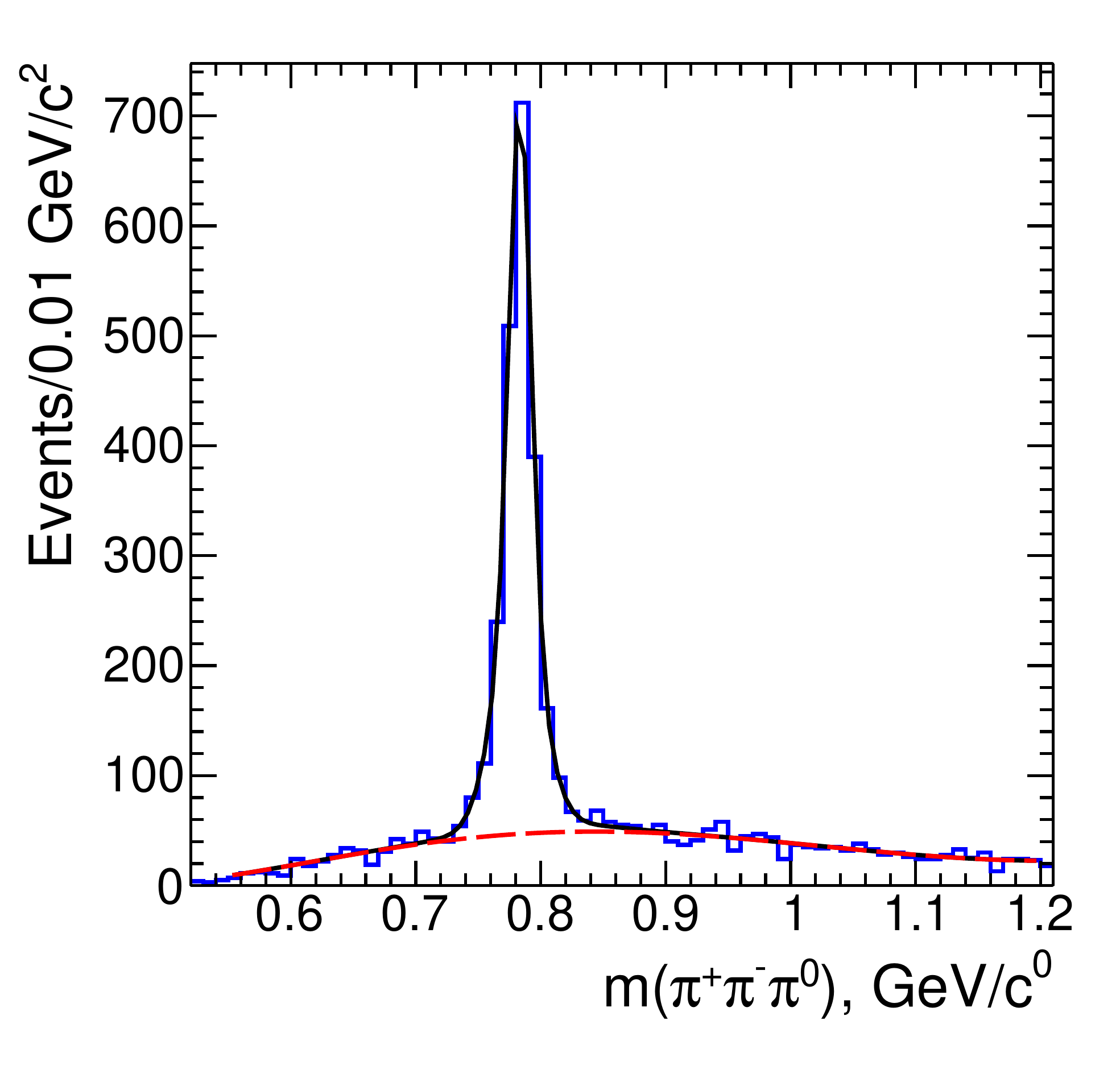}
\put(-50,80){\makebox(0,0)[lb]{\bf(b)}}
\vspace{-0.3cm}
\caption{(a) The third-photon-pair invariant mass for simulation of
  the $\epem\to\pipi\ppz\eta\gamma$ process.
   The dashed curve shows the fitted background. The solid curve
  shows the sum of background and the two-Gaussian fit function used to obtain the number of events
  with an $\eta$.
(b) The $\pipi\piz$ invariant mass for simulation. The solid curve shows a two-Gaussian
fit function for the $\omega$ signal plus the combinatorial
background (dashed).
}
\label{meta_mc_fit}
\end{center}
\end{figure}

\subsection{Detection efficiency}
\label{sec:eff_2pi2pi0eta}
We use simulated  $\epem\to\pipi\ppz\eta\gamma$ events from the
phase space model and with the $\omega\piz\eta$ intermediate state to
determine the efficiency. 
As for the data, we fit to find the $\eta$ signal in the third  
photon pair in intervals of 0.05~\gevcc in $m(\pipi\ppz\gamma\gamma)$.
The fit is illustrated in Fig.~\ref{meta_mc_fit}(a) using all $\pipi\ppz\gamma\gamma$
candidates. The efficiency is determined as the ratio of the
number of fitted events in each interval to the number generated
in that interval. For the $\omega\piz\eta$ intermediate channel,
we also determine the efficiency using an alternative method,
 by fitting the $\omega$ peak
in the $\pipi\piz$ invariant mass distribution, shown in
Fig.~\ref{meta_mc_fit}(b).

\begin{figure}[tbh]
\begin{center}
\includegraphics[width=0.9\linewidth]{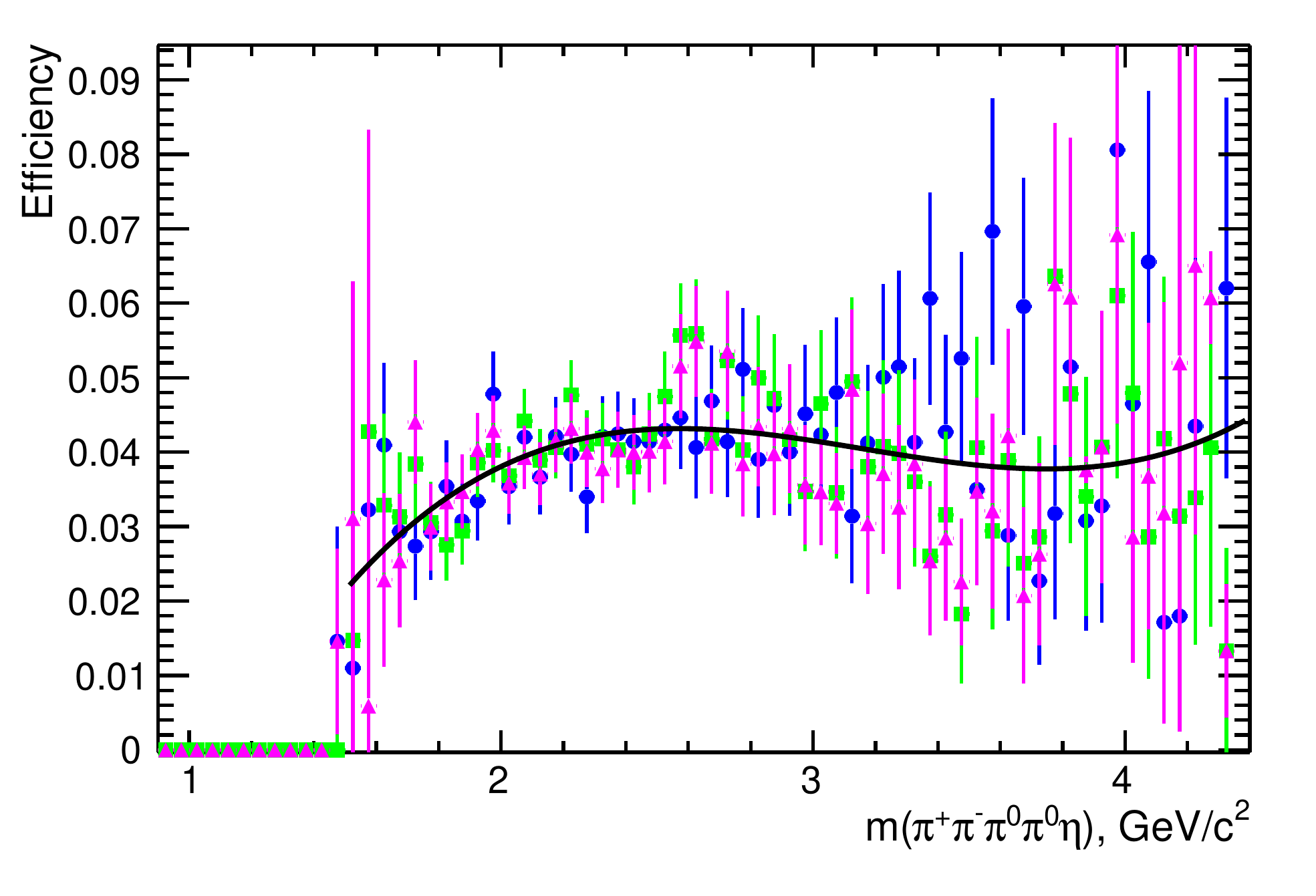}
\vspace{-0.5cm}
\caption{ The energy dependent detection efficiency, determined in
  three different ways: see text. 
The curve shows the fit to the average of the three and 
is used in the cross section determination. 
}
\label{eff_2pi2pi0eta}
\end{center}
\end{figure}

The efficiencies obtained for the three methods are
shown in Fig.~\ref{eff_2pi2pi0eta}.  The circles and squares show the
results from the fit to the $\eta$ peak for the phase space
and $\omega\piz\eta$ channels, respectively.  The triangles
show the results for the fit to the $\omega$ peak.  The efficiencies
are calculated assuming the $\eta\to\gamma\gamma$ mode only.
The obtained efficiencies are around 4\%, similar to what is
found for $\pipi3\piz$  (Fig.~\ref{mc_acc}).
The results from the three methods are consistent with each other,
and are averaged.
The average is fit with a third-order polynomial,
shown by the curve in Fig.~\ref{eff_2pi2pi0eta}.
The result of the fit is used for the cross section determination.

We estimate the systematic uncertainty in the efficiency due to the fit procedure and
the model dependence to be not more than 10\%.

\begin{figure}[tbh]
\begin{center}
\includegraphics[width=0.9\linewidth]{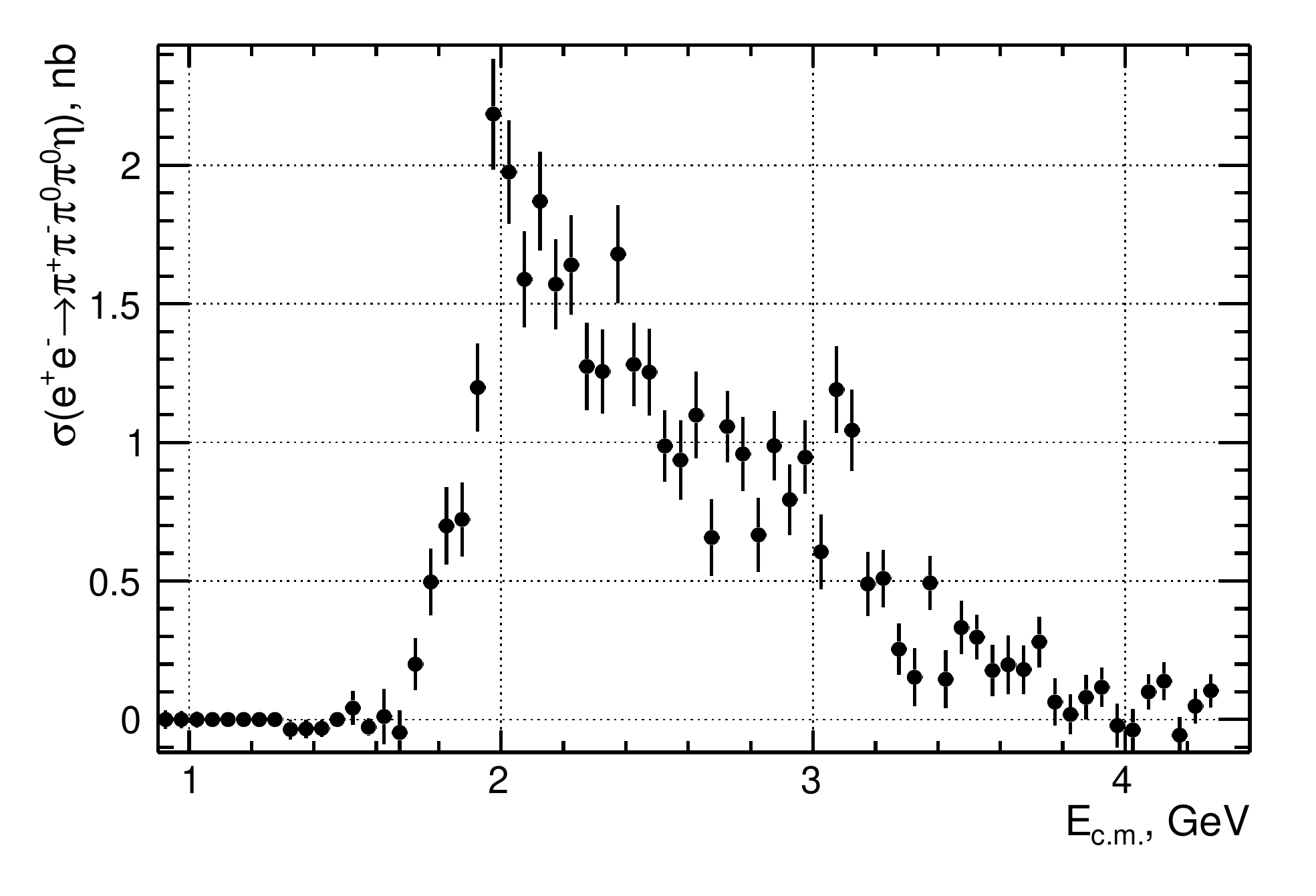}
\vspace{-0.5cm}
\caption{
Energy dependent  cross section
for   $\epem\to\pipi\ppz\eta$.
The uncertainties are statistical only.
}
\label{2pi2pi0eta_ee_babar}
\end{center}
\end{figure} 
\subsection{\boldmath Cross section for $\epem\to \pipi\ppz\eta$}
\label{2pi2pi0eta}
The cross section for $\epem\to\pipi\ppz\eta$ is determined
using Eq.~(\ref{xseq}).  The results are shown in
Fig.~\ref{2pi2pi0eta_ee_babar} and
listed in Table~\ref{4pieta_table}. These are the first results for
this process. 
The systematic uncertainties and corrections
are the same as those presented in Table~\ref{error_tab} except
there is an increase in the uncertainty in the detection
efficiency.  The total systematic uncertainty for
\Ecm $< 2.5$ GeV is 13\%.

\begin{figure*}[tbh]
\begin{center}
\includegraphics[width=0.34\linewidth]{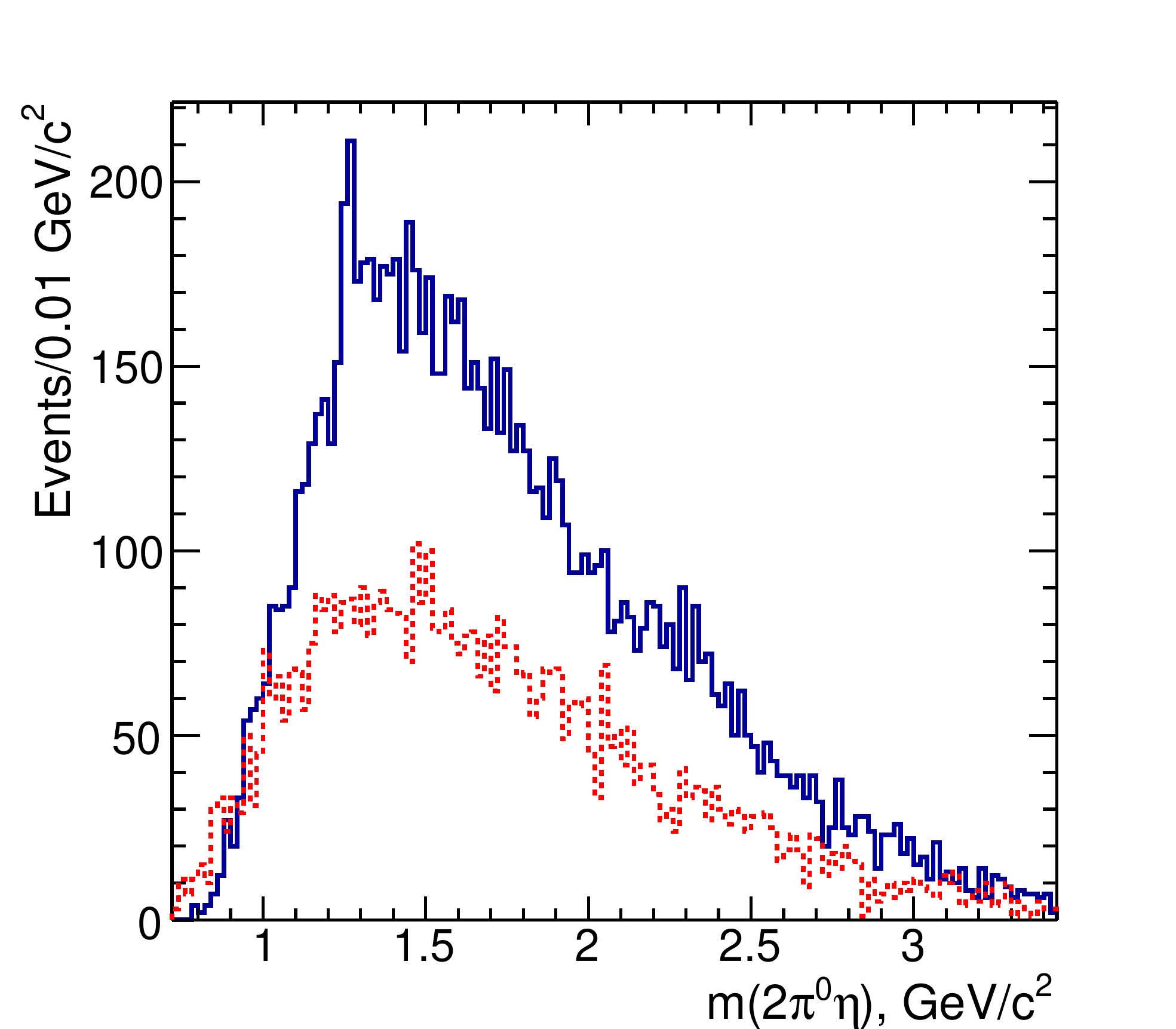}
\put(-50,120){\makebox(0,0)[lb]{\bf(a)}}
\includegraphics[width=0.34\linewidth]{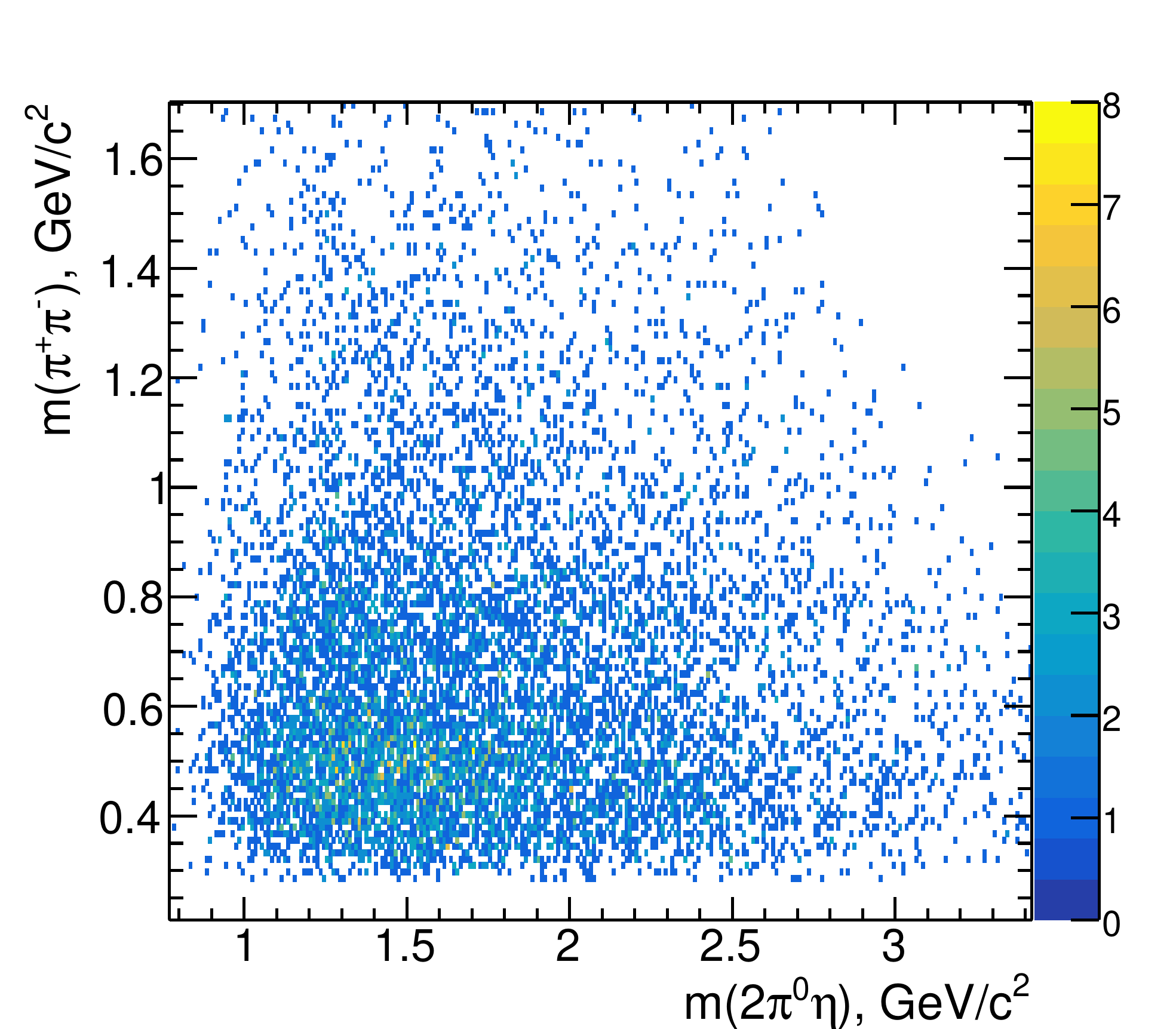}
\put(-50,120){\makebox(0,0)[lb]{\bf(b)}}
\caption{
(a) The $2\piz\eta$
invariant mass of the selected $\pipi 2\piz\eta$ events (solid
histogram), and  the background determined from the \chisq sideband  (dotted histogram).
(b) The $\pipi$ vs the $2\piz\eta$ mass for the selected events.
}
\label{2pi0etavs4pieta}  
\end{center}
%
\begin{center}
\includegraphics[width=0.34\linewidth]{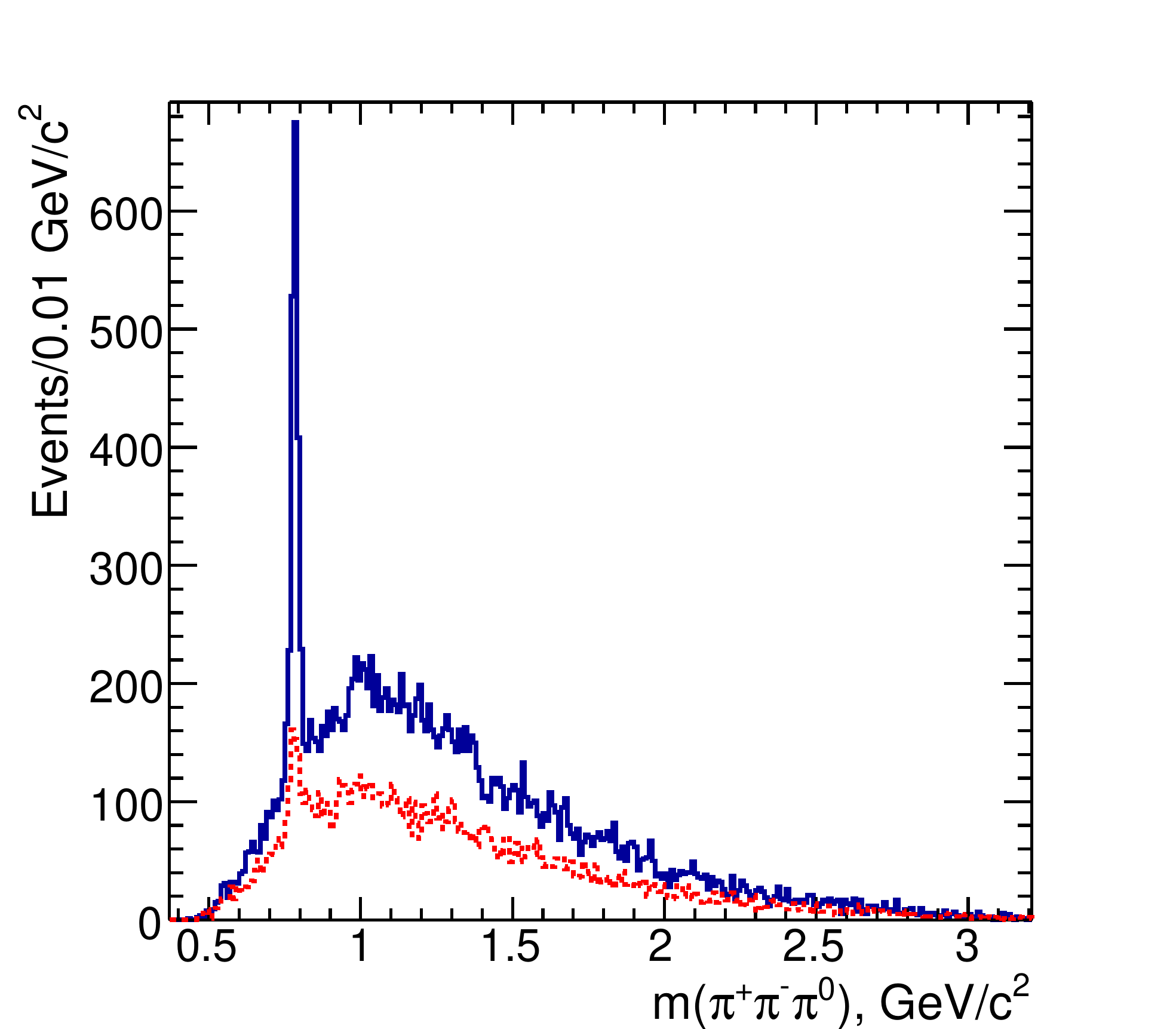}
\put(-50,120){\makebox(0,0)[lb]{\bf(a)}}
\includegraphics[width=0.34\linewidth]{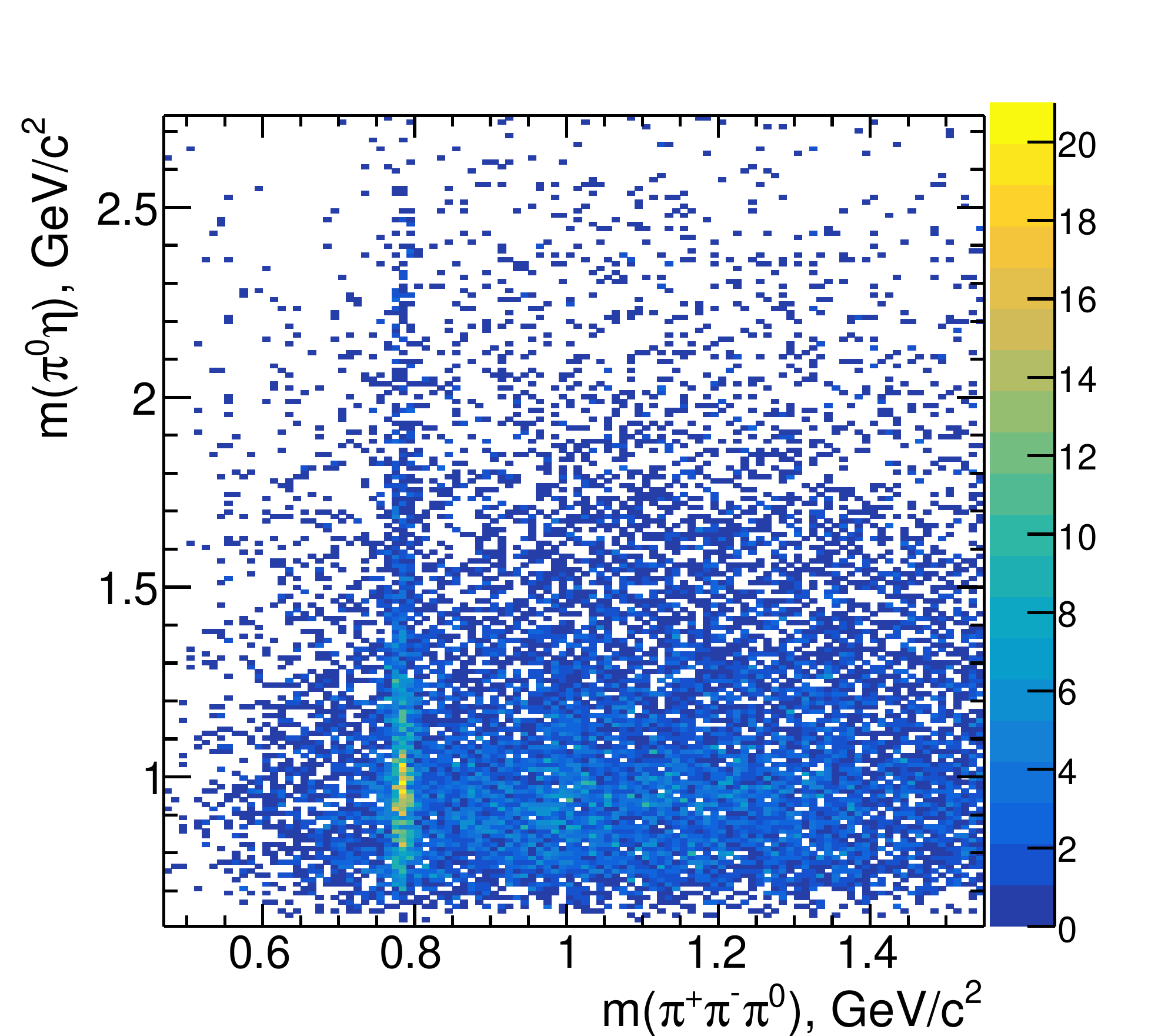}
\put(-140,120){\makebox(0,0)[lb]{\bf(b)}}
\includegraphics[width=0.34\linewidth]{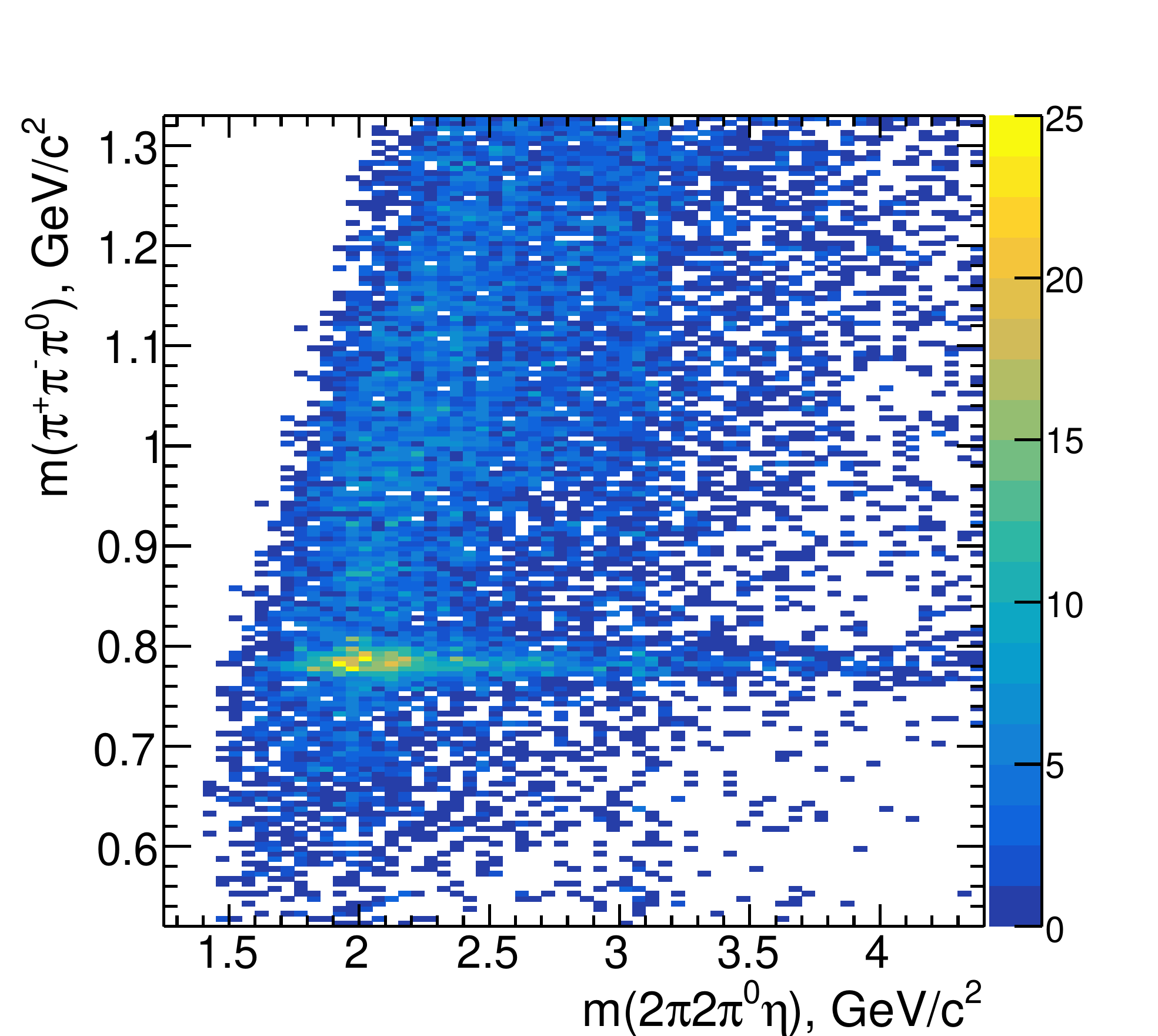}
\put(-140,120){\makebox(0,0)[lb]{\bf(c)}}
\vspace{-0.3cm}
\caption{
(a) The $\pipi\piz$  invariant mass with two entries per event (solid
histogram) and the background estimate from the $\eta$ sideband
(dotted histogram).
(b) The $\piz\eta$ vs the $\pipi\pi^0$ invariant mass.
(c)  The $\pipi\piz$  invariant mass vs the $\pipi2\piz\eta$ invariant mass.
}
\label{3pivs4pieta}  
\end{center}
%
\begin{center}
\includegraphics[width=0.34\linewidth]{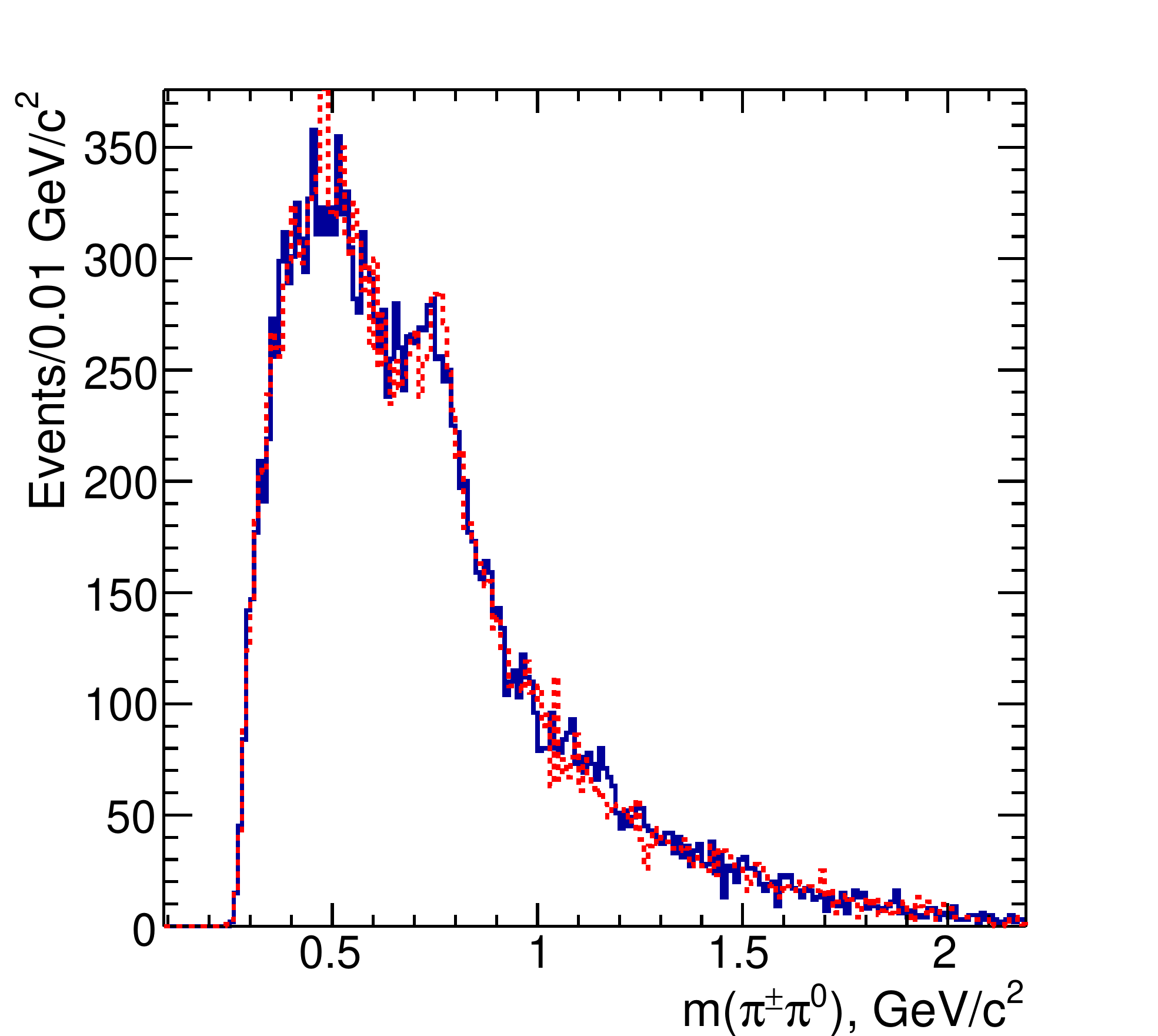}
\put(-50,125){\makebox(0,0)[lb]{\bf(a)}}
\includegraphics[width=0.34\linewidth]{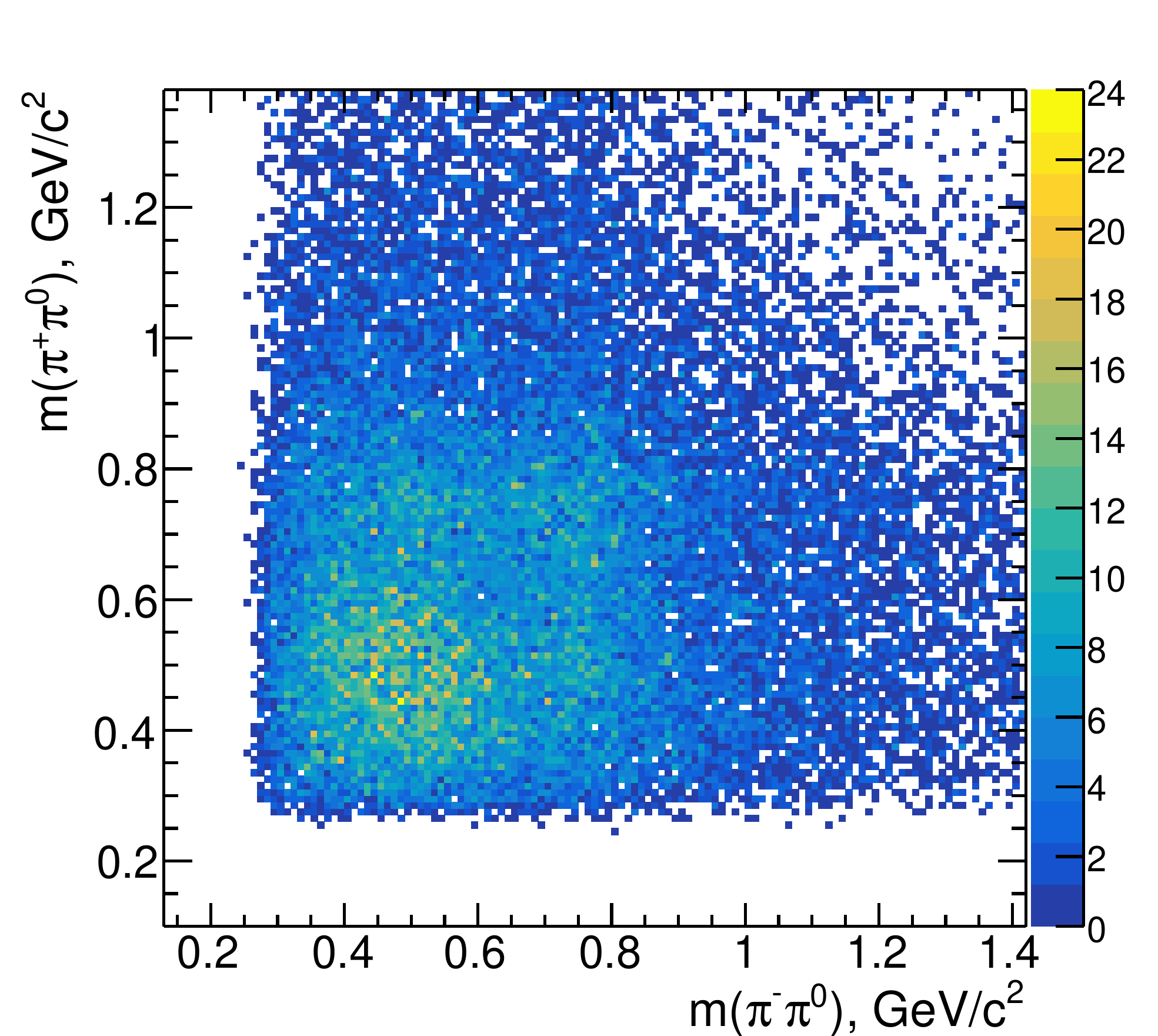}
\put(-147,125){\makebox(0,0)[lb]{\bf(b)}}
\includegraphics[width=0.34\linewidth]{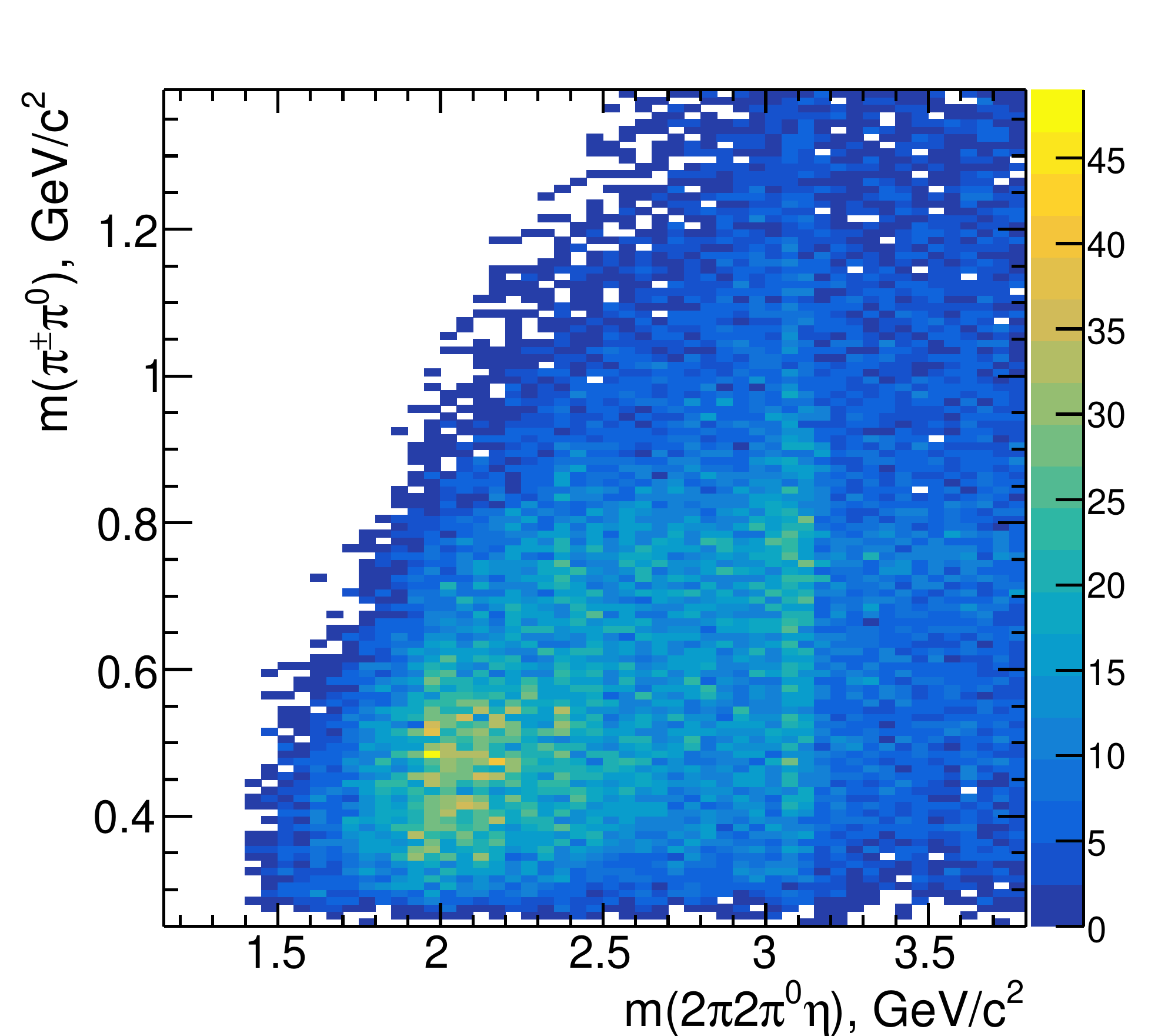}
\put(-120,125){\makebox(0,0)[lb]{\bf(c)}}
\vspace{-0.3cm}
\caption{
(a) the $\pi^+\piz$ (solid) and $\pi^-\piz$
(dotted) invariant mass for the selected $\pipi 2\piz\eta$
events (two entries per event).
(b) The $\pi^-\piz$ vs the $\pi^+\piz$ invariant mass for the selected
events.
(c)  The $\pi^{\pm}\piz$  invariant mass vs the $\pipi2\piz\eta$ invariant mass.
}
\label{pipi0vs4pieta}  
\end{center}
\end{figure*}
\subsection{Overview of the intermediate structures}
The $\pipi2\piz\eta$ final state, like that for $\pipi3\piz$, has a rich
substructure.  
Figure~\ref{2pi0etavs4pieta}(a) shows the
$2\piz\eta$ invariant mass distribution for  events selected by requiring
$|m(\gamma\gamma)-m(\eta)|<0.07$~\gevcc in Fig.~\ref{meta_data_fit}(a).
There is a small but clear signal for $\eta(1285)$ production.  The
dotted histogram shows the background 
distribution, determined using an $\eta$ sideband  control
region defined by $0.07<|m(\gamma\gamma)-m(\eta)|<0.14$~\gevcc.
Figure~\ref{2pi0etavs4pieta}(b)
shows a scatter plot of the $\pipi$ invariant mass vs the
$2\piz\eta$ invariant mass.
No structures are seen.

Figure~\ref{3pivs4pieta}(a) shows the $\pipi\piz$ mass distribution
(two entries per event). An $\omega$ signal is  clearly visible, as
well as a bump close to 1~\gevcc corresponding to  $\phi\to\pipi\piz$. 
The dotted histogram shows the estimate of the background,
evaluated using the $\eta$ sideband described above.
The scatter plot in  Fig.~\ref{3pivs4pieta}(b) shows the $\piz\eta$
vs the $\pipi\piz$ invariant mass. A clear correlation  of $\omega$ and
$a_0(980)\to\piz\eta$ production is seen. Figure~\ref{3pivs4pieta}(c) shows
how $\omega\piz\eta$ events are distributed over the $\pipi2\piz\eta$
invariant mass.

Figure~\ref{pipi0vs4pieta}(a) presents the $\pi^+\piz$ (solid) and
$\pi^-\piz$ (dotted) mass combinations (two entries per event) for the
selected $\pipi2\piz\eta$ events. 
Signals from the $\rho^{\pm}$ are clearly visible, but they
can also come from events with a $\rho^+\rho^-$ pair.   
The fraction of $\rho^+\rho^-$ events is extracted from the
distribution in Fig.~\ref{pipi0vs4pieta}(b), where the $\pi^+\piz$ vs the $\pi^-\piz$
invariant mass is shown.
Figure~\ref{pipi0vs4pieta}(c) displays the $\pi^{\pm}\piz$
vs the $\pipi2\piz\eta$ invariant mass.
\begin{figure}[tbh]
\begin{center}
\includegraphics[width=0.49\linewidth]{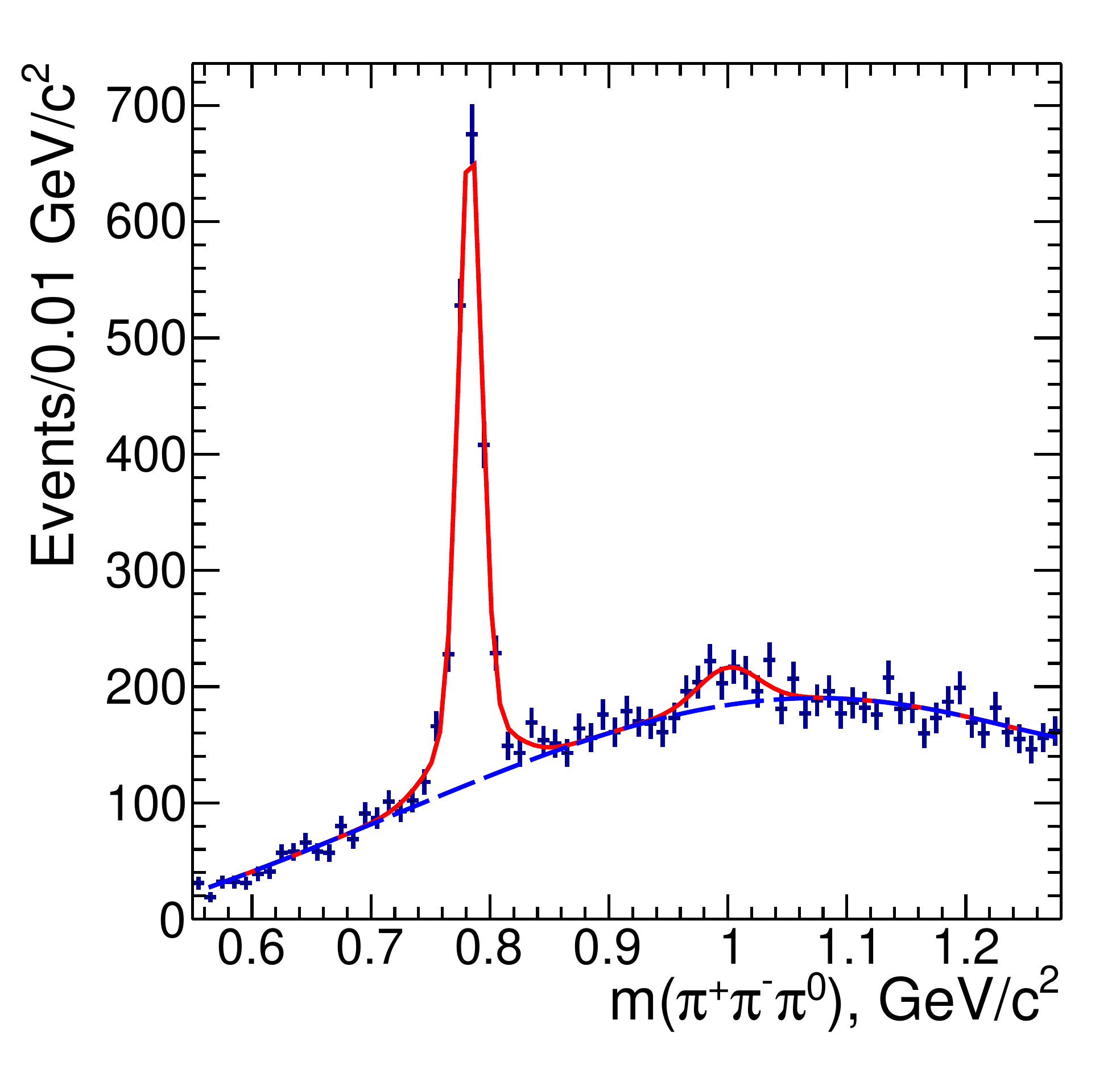}
\put(-50,80){\makebox(0,0)[lb]{\bf(a)}}
\includegraphics[width=0.53\linewidth]{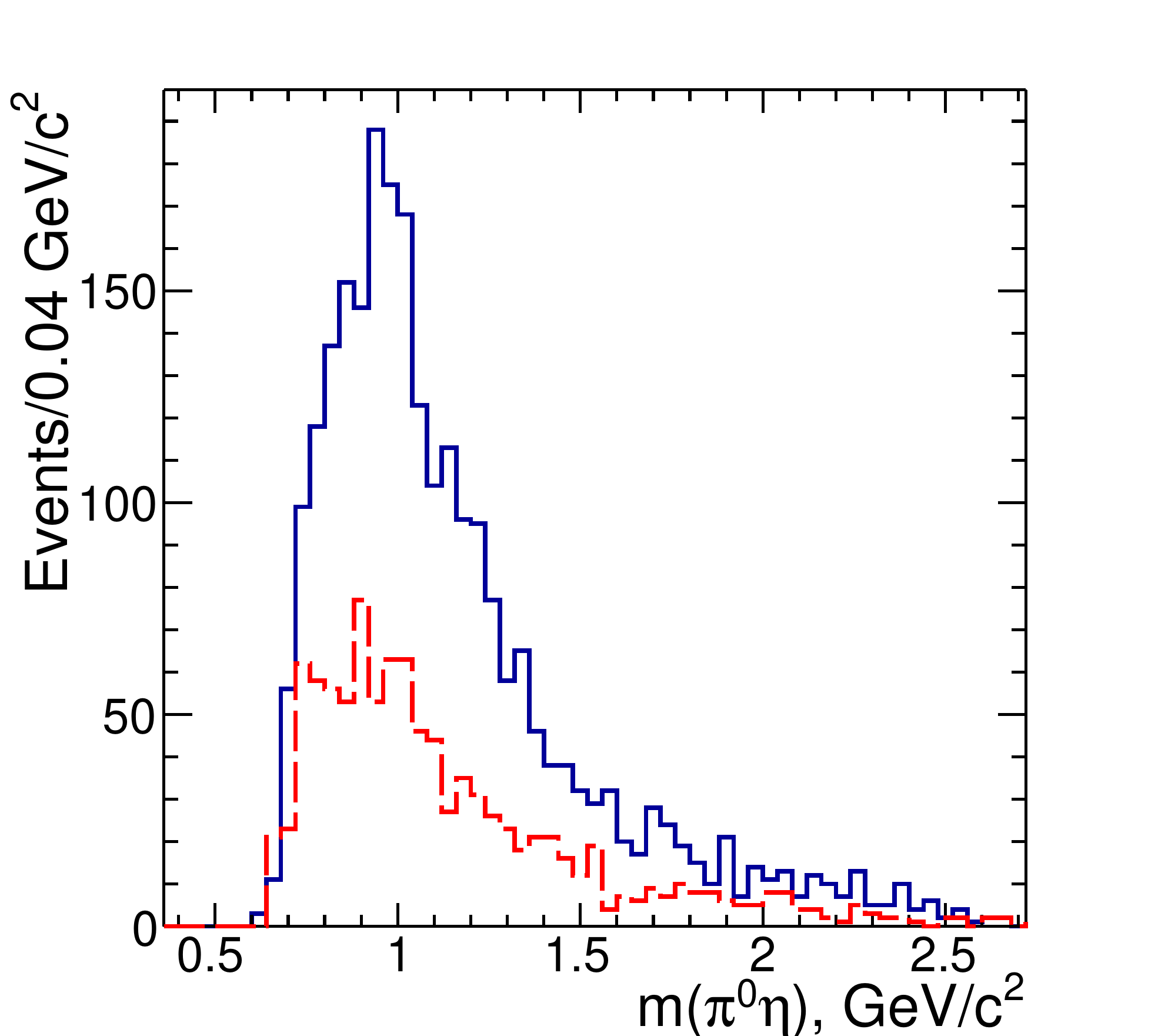}
\put(-50,80){\makebox(0,0)[lb]{\bf(b)}}
\vspace{-0.3cm}
\caption{(a) 
The $\pipi\piz$ invariant mass for data. The dashed curve describes the non-resonant
background. The solid curve shows the sum of   the background and 
the fit functions for the $\omega$ and $\phi$ contributions,  described 
in the text. 
(b) The $\piz\eta$ invariant mass distribution for the events selected
in the $\omega$ peak (solid). The dashed histogram shows the distribution from the $\omega$-peak side band.
}
\label{3pislices_eta}
\end{center}
\end{figure}
\subsection{\bf\boldmath The $\omega\piz\eta$ and $\phi\piz\eta$ intermediate states}
\label{Sec:omegapi0eta}
%
To determine the contribution of the $\omega\piz\eta$ and $\phi\piz\eta$ intermediate states, we
fit the events in Fig.~\ref{3pivs4pieta}(a) with two Gaussian
functions, one to describe the
$\omega$ peak and the other the $\phi$ peak, and a polynomial function,
which describes the background.
 The results of
the fit are shown  in Fig.~\ref{3pislices_eta}(a). We obtain 
$1676\pm22$ and $269\pm68$ events for the $\omega$ and $\phi$,
respectively. The number of events as a function of the
$\pipi2\piz\eta$ invariant mass is determined by performing an
analogous fit of events in Fig.~\ref{3pivs4pieta}(c) in intervals of 0.05~\gevcc in $m(\pipi2\piz\eta)$.

We select events within $\pm 0.7$~\gevcc 
of the $\omega$ peak in Fig.~\ref{3pislices_eta}(a) and display the
resulting $\piz\eta$
invariant mass in Fig.~\ref{3pislices_eta}(b). A very clear
signal from the $a_0(980)$ is observed, while no signal is seen in an
$\omega$ sideband defined by $0.07<|m(\pipi\piz)-m(\omega)|<0.14$~\gevcc.
\begin{figure}[tbh]
\begin{center}
\includegraphics[width=0.9\linewidth]{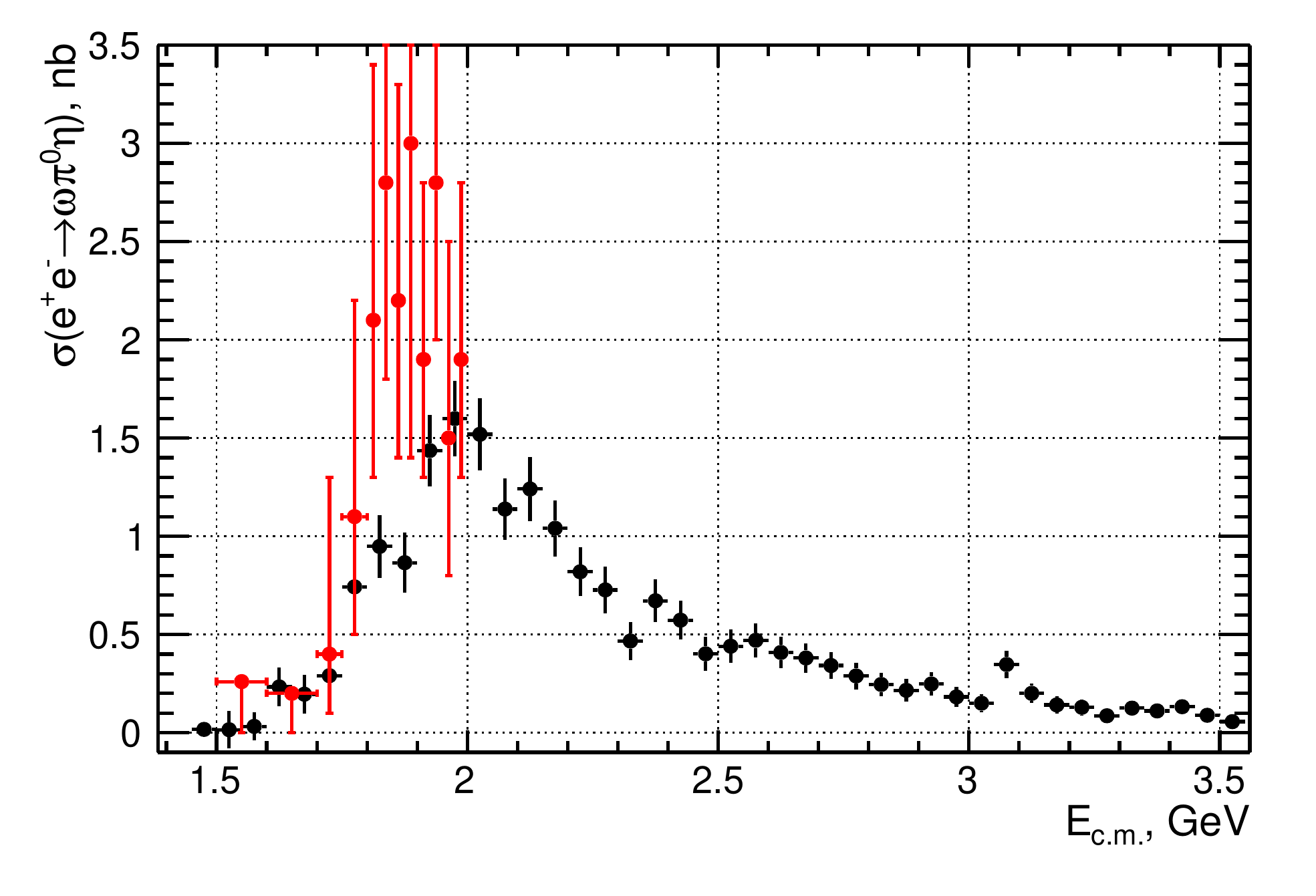}
\caption{
The \Ecm~ dependence of the $\epem\to\omega\piz\eta$ cross
  section (circles) in
comparison with the SND results~\cite{SNDompi0eta} (squares).
}
\label{xs_omegapi0eta}
\end{center}
\end{figure}
The obtained $\epem\to\omega\piz\eta$ cross section, corrected for the $\omega\to\pipi\piz$
branching fraction, is shown in Fig.~\ref{xs_omegapi0eta} in
comparison to previous results from SND~\cite{SNDompi0eta}.
The SND results, which are available only for energies below
2 GeV, are seen to lie systematically above our data.
All systematic uncertainties discussed in
section~\ref{sec:Systematics} are applied to the measured
$\epem\to\omega\piz\eta$ cross section, 
resulting in a total systematic uncertainty of 13\% below 2.4 GeV. 
The results are presented in Table~\ref{ompi0eta_table} (statistical uncertainties only)
in bin widths of 0.05 GeV.  Above 3.5 GeV, the cross section measurements
are consistent with zero within the experimental accuracy.
\begin{figure}[tbh]
\begin{center}
\includegraphics[width=0.49\linewidth]{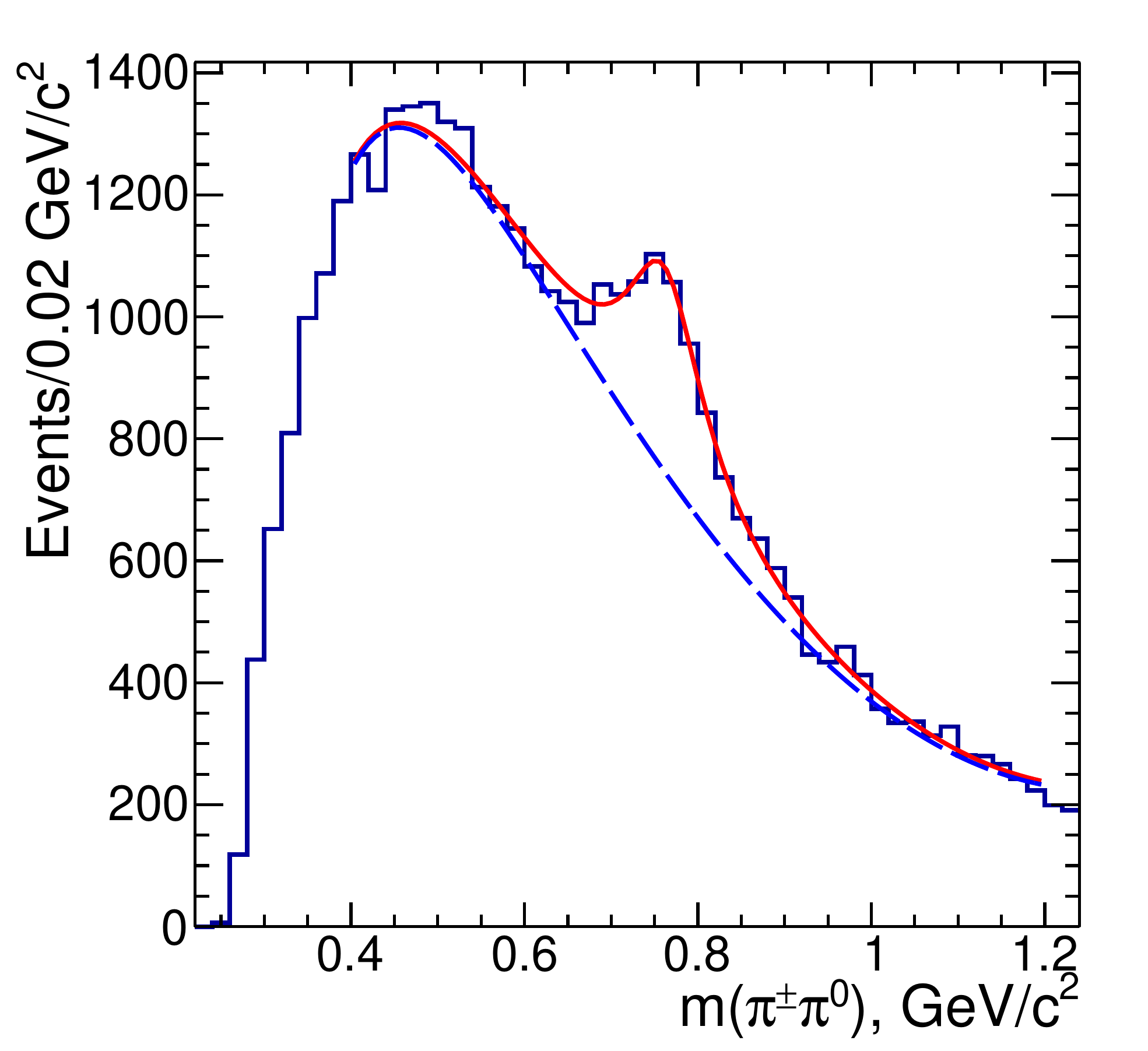}
\put(-40,85){\makebox(0,0)[lb]{\bf(a)}}
\includegraphics[width=0.51\linewidth]{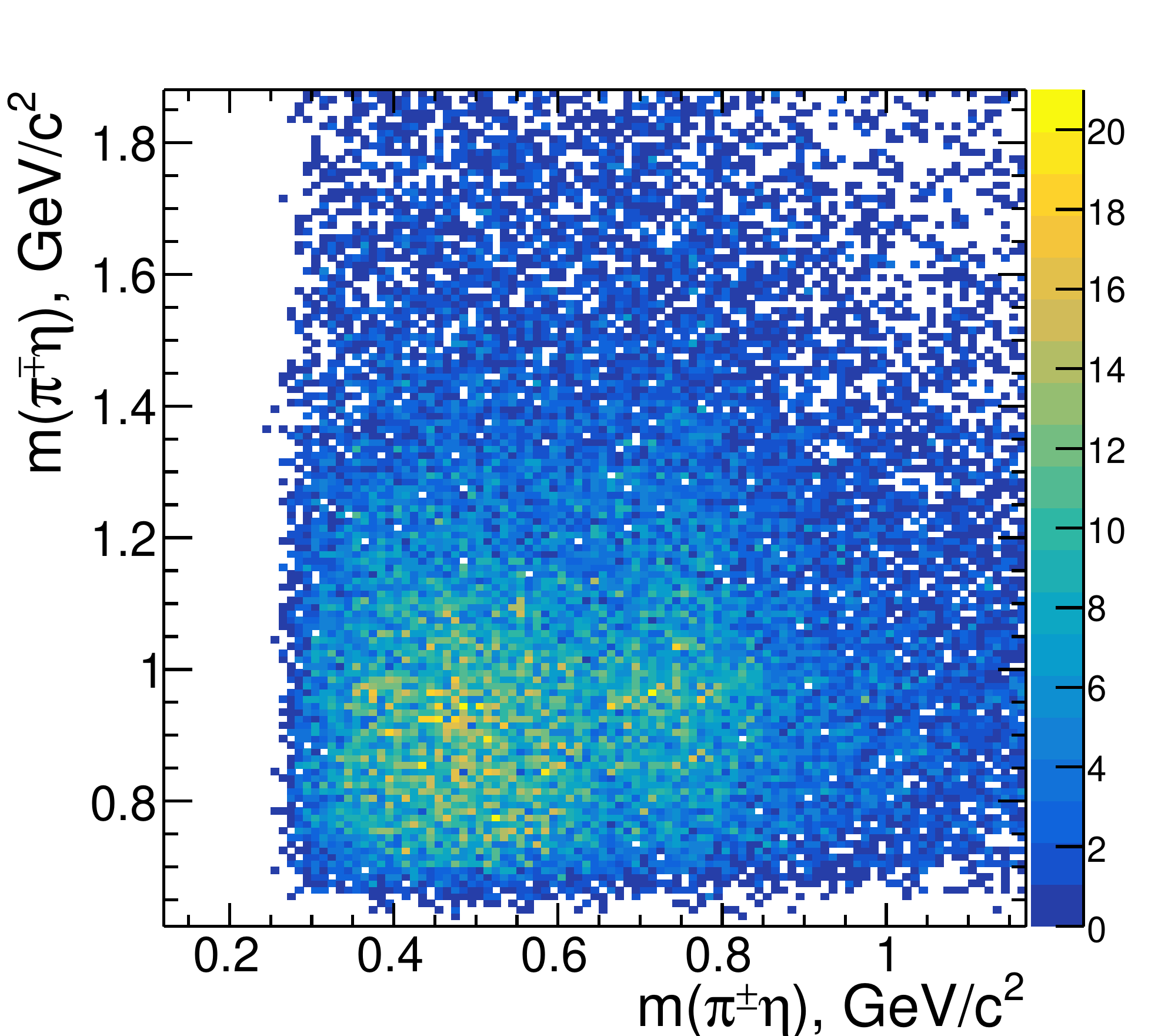}
\put(-105,85){\makebox(0,0)[lb]{\bf(b)}}
\vspace{-0.3cm}
\caption{(a) The $\pi^{\pm}\piz$ invariant mass for data. The curves
  show the fit functions, described in the text.
(b) The $\pi^{\pm}\eta$ vs the $\pi^{\mp}\piz$ invariant mass.
}
\label{rhoslices_eta}
\end{center}
\end{figure}
\begin{figure}[tbh]
\begin{center}
\includegraphics[width=0.9\linewidth]{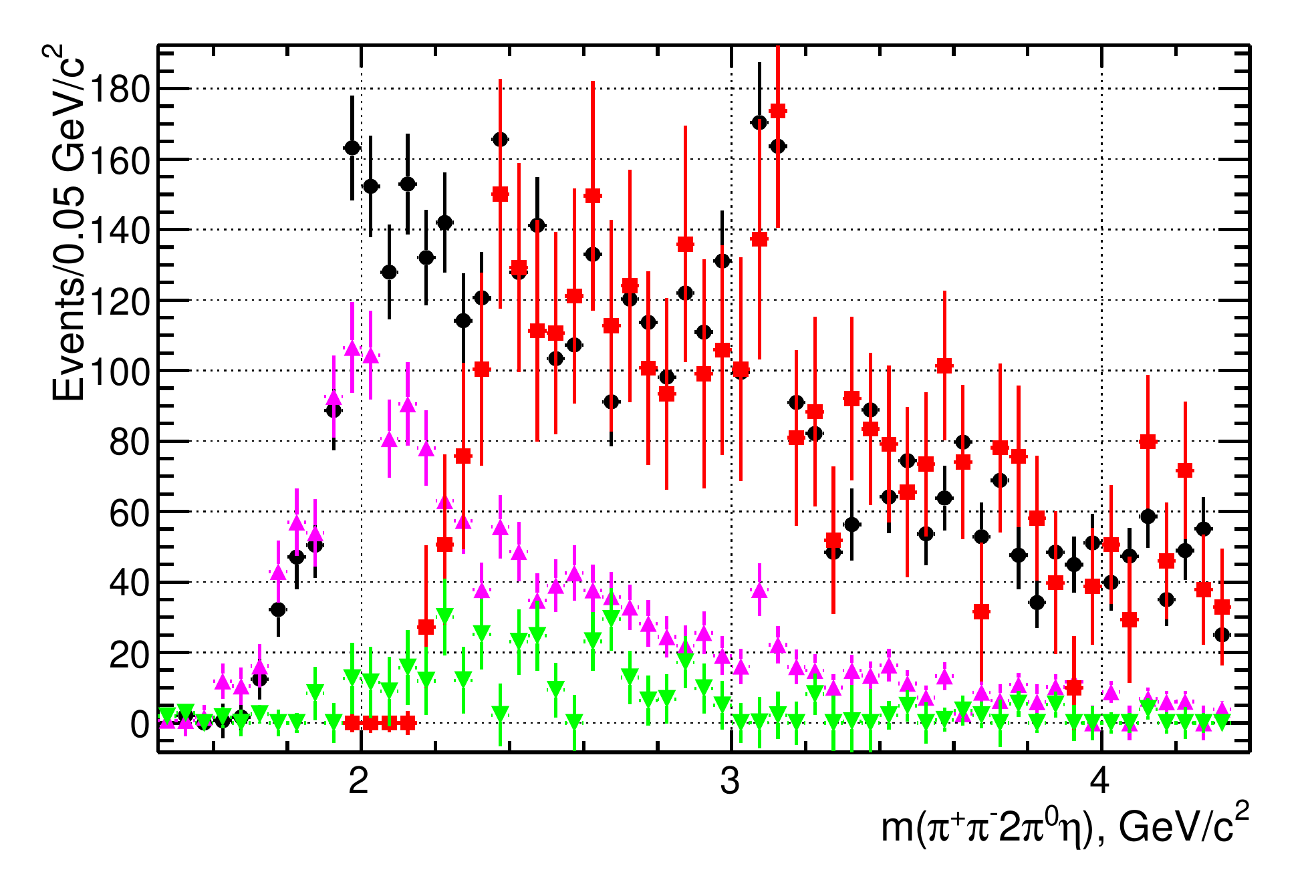}
\caption{
Number of events in bins of \Ecm for inclusive $\pipi2\piz\eta$ events (circles)
and for the $\omega\piz\eta$ (triangles), $\phi\piz\eta$ (upside-down triangles),
and $\rho^{\pm}\pi^{\mp}\piz\eta$ (squares) intermediate states.
}
\label{sumallev_eta}
\end{center}
\end{figure}
\subsection{\bf\boldmath The $\rho(770)^{\pm}\pi^{\mp}\piz\eta$ intermediate state}
\label{sec:rhoselect}
The  approach described in Sec.~\ref{sec:rhoselectpi0} is used 
to study events with a $\rho^{\pm}$ meson in the
intermediate state. We fit 
the events in Fig.~\ref{pipi0vs4pieta}(a) 
using a BW function to describe the $\rho$ signal and a polynomial
function to describe the background
(four entries per event). The fit yields  2908$\pm$202
$\rho^{\pm}\pi^{\mp}\piz\eta$ events. The result of the fit is shown
in Fig.~\ref{rhoslices_eta}(a). The distribution of these events vs the $\pipi2\piz\eta$ invariant mass is shown
by the squares in Fig.~\ref{sumallev_eta}.

The size of our data sample is not sufficient to justify
 a sophisticated amplitude analysis, as would be needed
to extract detailed information on all the intermediate states.
 We can deduce that an intermediate $a_0(980)\rho\pi$ state is present:
a correlated bump at the
$a_0(980)$ and $\rho$ invariant masses is seen in the scatter plot of
Fig.~\ref{rhoslices_eta}(b),  where the $\pi^{\pm}\eta$ invariant mass
is plotted vs the $\pi^{\mp}\piz$ mass.
Also, there is a contribution from  $\rho^+\rho^-\eta$: a scatter plot
of the  $\pi^{\pm}\piz$ vs the $\pi^{\mp}\piz$ invariant mass is presented in 
 Fig.~\ref{pipi0vs4pieta}(b), from which an enhancement corresponding to correlated   
$\rho^+\rho^-$ production is visible.

\subsection{\bf\boldmath The sum of intermediate states}
\label{sec:sumeta}

Figure~\ref{sumallev_eta} displays the number of events obtained from the 
fits described above to the $\omega$ (triangles),
$\phi$ (upside-down triangles), and $\rho$ (square) peaks.
The results are shown in comparison to the total
number of $\pipi2\piz\eta$ events (circles) 
obtained from the fit to the third photon pair invariant mass
distribution.
The sum of events from the intermediate states is seen to
agree within the uncertainties with the total number of
$\pipi2\piz\eta$  events, except in the region around 2 GeV.

\begin{table*}
\caption{Summary of the $\epem\to\eta\pipi$ 
cross section measurement. The uncertainties are statistical only.}
\label{2pieta_table}
\begin{tabular}{c c c c c c c c c c}
$E_{\rm c.m.}$, GeV & $\sigma$, nb  
& $E_{\rm c.m.}$, GeV & $\sigma$, nb 
& $E_{\rm c.m.}$, GeV & $\sigma$, nb 
& $E_{\rm c.m.}$, GeV & $\sigma$, nb  
& $E_{\rm c.m.}$, GeV & $\sigma$, nb  
\\
\hline
1.075 & 0.06 $\pm$ 0.03 &1.475 & 3.74 $\pm$ 0.43 &1.875 & 1.16 $\pm$ 0.21 &2.275 & 0.35 $\pm$ 0.10 &2.675 & 0.27 $\pm$ 0.07 \\ 
1.125 & 0.29 $\pm$ 0.23 &1.525 & 4.14 $\pm$ 0.44 &1.925 & 1.00 $\pm$ 0.19 &2.325 & 0.22 $\pm$ 0.09 &2.725 & 0.11 $\pm$ 0.05 \\ 
1.175 & 0.00 $\pm$ 0.12 &1.575 & 3.48 $\pm$ 0.40 &1.975 & 0.65 $\pm$ 0.16 &2.375 & 0.33 $\pm$ 0.09 &2.775 & 0.09 $\pm$ 0.05 \\ 
1.225 & 0.23 $\pm$ 0.25 &1.625 & 2.67 $\pm$ 0.36 &2.025 & 0.78 $\pm$ 0.16 &2.425 & 0.22 $\pm$ 0.07 &2.825 & 0.03 $\pm$ 0.04 \\ 
1.275 & 0.57 $\pm$ 0.27 &1.675 & 2.52 $\pm$ 0.32 &2.075 & 0.51 $\pm$ 0.13 &2.475 & 0.51 $\pm$ 0.10 &2.875 & 0.05 $\pm$ 0.04 \\ 
1.325 & 1.15 $\pm$ 0.34 &1.725 & 2.20 $\pm$ 0.30 &2.125 & 0.50 $\pm$ 0.13 &2.525 & 0.27 $\pm$ 0.09 &2.925 & 0.02 $\pm$ 0.04 \\ 
1.375 & 1.83 $\pm$ 0.36 &1.775 & 2.38 $\pm$ 0.29 &2.175 & 0.75 $\pm$ 0.13 &2.575 & 0.08 $\pm$ 0.05 &2.975 & 0.09 $\pm$ 0.05 \\ 
1.425 & 2.74 $\pm$ 0.40 &1.825 & 1.39 $\pm$ 0.23 &2.225 & 0.23 $\pm$ 0.11 &2.625 & 0.12 $\pm$ 0.06 &3.025 & 0.05 $\pm$ 0.05 \\ 
\hline
\end{tabular}
\end{table*}

\begin{table*}
\caption{Summary of the $\epem\to\omega\ppz$ 
cross section measurement. The uncertainties are statistical only.}
\label{omega2pi0_table}
\begin{tabular}{c c c c c c c c c c}
$E_{\rm c.m.}$, GeV & $\sigma$, nb  
& $E_{\rm c.m.}$, GeV & $\sigma$, nb 
& $E_{\rm c.m.}$, GeV & $\sigma$, nb 
& $E_{\rm c.m.}$, GeV & $\sigma$, nb  
& $E_{\rm c.m.}$, GeV & $\sigma$, nb  
\\
\hline
1.125 & 0.04 $\pm$ 0.08 &1.775 & 0.88 $\pm$ 0.16 &2.425 & 0.07 $\pm$ 0.05 &3.075 & 0.83 $\pm$ 0.07 &3.725 & 0.06 $\pm$ 0.02 \\ 
1.175 & 0.03 $\pm$ 0.10 &1.825 & 0.62 $\pm$ 0.14 &2.475 & 0.12 $\pm$ 0.05 &3.125 & 0.52 $\pm$ 0.05 &3.775 & 0.03 $\pm$ 0.02 \\ 
1.225 & --0.02 $\pm$ 0.10 &1.875 & 0.96 $\pm$ 0.14 &2.525 & 0.21 $\pm$ 0.05 &3.175 & 0.11 $\pm$ 0.03 &3.825 & 0.03 $\pm$ 0.01 \\ 
1.275 & 0.13 $\pm$ 0.11 &1.925 & 0.61 $\pm$ 0.13 &2.575 & 0.15 $\pm$ 0.04 &3.225 & 0.08 $\pm$ 0.02 &3.875 & 0.02 $\pm$ 0.01 \\ 
1.325 & 0.41 $\pm$ 0.13 &1.975 & 0.45 $\pm$ 0.11 &2.625 & 0.13 $\pm$ 0.04 &3.275 & 0.08 $\pm$ 0.02 &3.925 & 0.03 $\pm$ 0.02 \\ 
1.375 & 0.69 $\pm$ 0.18 &2.025 & 0.47 $\pm$ 0.10 &2.675 & 0.12 $\pm$ 0.04 &3.325 & 0.07 $\pm$ 0.02 &3.975 & 0.04 $\pm$ 0.01 \\ 
1.425 & 0.29 $\pm$ 0.18 &2.075 & 0.33 $\pm$ 0.09 &2.725 & 0.17 $\pm$ 0.04 &3.375 & 0.06 $\pm$ 0.02 &4.025 & 0.03 $\pm$ 0.01 \\ 
1.475 & 0.68 $\pm$ 0.19 &2.125 & 0.29 $\pm$ 0.09 &2.775 & 0.10 $\pm$ 0.04 &3.425 & 0.07 $\pm$ 0.02 &4.075 & 0.02 $\pm$ 0.01 \\ 
1.525 & 1.05 $\pm$ 0.21 &2.175 & 0.26 $\pm$ 0.08 &2.825 & 0.11 $\pm$ 0.04 &3.475 & 0.03 $\pm$ 0.02 &4.125 & 0.03 $\pm$ 0.01 \\ 
1.575 & 1.44 $\pm$ 0.22 &2.225 & 0.40 $\pm$ 0.08 &2.875 & 0.18 $\pm$ 0.04 &3.525 & 0.07 $\pm$ 0.02 &4.175 & 0.02 $\pm$ 0.01 \\ 
1.625 & 1.40 $\pm$ 0.21 &2.275 & 0.31 $\pm$ 0.07 &2.925 & 0.10 $\pm$ 0.03 &3.575 & 0.04 $\pm$ 0.02 &4.225 & 0.01 $\pm$ 0.01 \\ 
1.675 & 1.55 $\pm$ 0.20 &2.325 & 0.21 $\pm$ 0.06 &2.975 & 0.14 $\pm$ 0.06 &3.625 & 0.06 $\pm$ 0.02 &4.275 & 0.01 $\pm$ 0.01 \\ 
1.725 & 0.96 $\pm$ 0.18 &2.375 & 0.23 $\pm$ 0.06 &3.025 & 0.25 $\pm$ 0.04 &3.675 & 0.11 $\pm$ 0.03 &4.325 & 0.02 $\pm$ 0.01 \\ 
\hline
\end{tabular}
\end{table*}

\begin{table*}
\caption{Summary of the $\epem\to\pipi\ppz\eta$ 
cross section measurement. The uncertainties are statistical only.}
\label{4pieta_table}
\begin{tabular}{c c c c c c c c c c}
$E_{\rm c.m.}$, GeV & $\sigma$, nb  
& $E_{\rm c.m.}$, GeV & $\sigma$, nb 
& $E_{\rm c.m.}$, GeV & $\sigma$, nb 
& $E_{\rm c.m.}$, GeV & $\sigma$, nb  
& $E_{\rm c.m.}$, GeV & $\sigma$, nb  
\\
\hline
1.625 & 0.01 $\pm$ 0.10 &2.175 & 1.59 $\pm$ 0.16 &2.725 & 1.07 $\pm$ 0.13 &3.275 & 0.26 $\pm$ 0.09 &3.825 & 0.02 $\pm$ 0.07 \\ 
1.675 & --0.05 $\pm$ 0.08 &2.225 & 1.66 $\pm$ 0.18 &2.775 & 0.97 $\pm$ 0.14 &3.325 & 0.15 $\pm$ 0.11 &3.875 & 0.08 $\pm$ 0.08 \\ 
1.725 & 0.20 $\pm$ 0.10 &2.275 & 1.29 $\pm$ 0.16 &2.825 & 0.68 $\pm$ 0.14 &3.375 & 0.50 $\pm$ 0.10 &3.925 & 0.12 $\pm$ 0.07 \\ 
1.775 & 0.51 $\pm$ 0.12 &2.325 & 1.27 $\pm$ 0.15 &2.875 & 1.00 $\pm$ 0.13 &3.425 & 0.15 $\pm$ 0.11 &3.975 & --0.02 $\pm$ 0.08 \\ 
1.825 & 0.71 $\pm$ 0.14 &2.375 & 1.70 $\pm$ 0.18 &2.925 & 0.81 $\pm$ 0.13 &3.475 & 0.34 $\pm$ 0.10 &4.025 & --0.04 $\pm$ 0.08 \\ 
1.875 & 0.73 $\pm$ 0.14 &2.425 & 1.30 $\pm$ 0.15 &2.975 & 0.96 $\pm$ 0.13 &3.525 & 0.30 $\pm$ 0.08 &4.075 & 0.10 $\pm$ 0.06 \\ 
1.925 & 1.22 $\pm$ 0.16 &2.475 & 1.27 $\pm$ 0.16 &3.025 & 0.61 $\pm$ 0.14 &3.575 & 0.18 $\pm$ 0.09 &4.125 & 0.14 $\pm$ 0.07 \\ 
1.975 & 2.22 $\pm$ 0.20 &2.525 & 1.00 $\pm$ 0.13 &3.075 & 1.21 $\pm$ 0.16 &3.625 & 0.20 $\pm$ 0.11 &4.175 & --0.06 $\pm$ 0.07 \\ 
2.025 & 2.01 $\pm$ 0.19 &2.575 & 0.95 $\pm$ 0.15 &3.125 & 1.06 $\pm$ 0.15 &3.675 & 0.18 $\pm$ 0.09 &4.225 & 0.05 $\pm$ 0.06 \\ 
2.075 & 1.61 $\pm$ 0.18 &2.625 & 1.11 $\pm$ 0.16 &3.175 & 0.50 $\pm$ 0.12 &3.725 & 0.28 $\pm$ 0.09 &4.275 & 0.10 $\pm$ 0.06 \\ 
2.125 & 1.90 $\pm$ 0.18 &2.675 & 0.67 $\pm$ 0.14 &3.225 & 0.52 $\pm$ 0.11 &3.775 & 0.06 $\pm$ 0.09 &4.325 & 0.04 $\pm$ 0.06 \\ 
\hline
\end{tabular}
\end{table*}

\begin{table*}
\caption{Summary of the $\epem\to\omega\piz\eta$ 
cross section measurement. The uncertainties are statistical only.}
\label{ompi0eta_table}
\begin{tabular}{c c c c c c c c c c}
$E_{\rm c.m.}$, GeV & $\sigma$, nb  
& $E_{\rm c.m.}$, GeV & $\sigma$, nb 
& $E_{\rm c.m.}$, GeV & $\sigma$, nb 
& $E_{\rm c.m.}$, GeV & $\sigma$, nb  
& $E_{\rm c.m.}$, GeV & $\sigma$, nb  
\\
\hline
1.525 & 0.02 $\pm$ 0.10 &2.125 & 1.26 $\pm$ 0.17 &2.725 & 0.35 $\pm$ 0.07 &3.325 & 0.13 $\pm$ 0.04 &3.925 & 0.08 $\pm$ 0.03 \\ 
1.575 & 0.03 $\pm$ 0.07 &2.175 & 1.06 $\pm$ 0.14 &2.775 & 0.29 $\pm$ 0.07 &3.375 & 0.11 $\pm$ 0.03 &3.975 & 0.00 $\pm$ 0.03 \\ 
1.625 & 0.24 $\pm$ 0.10 &2.225 & 0.83 $\pm$ 0.13 &2.825 & 0.25 $\pm$ 0.06 &3.425 & 0.13 $\pm$ 0.04 &4.025 & 0.05 $\pm$ 0.02 \\ 
1.675 & 0.20 $\pm$ 0.10 &2.275 & 0.74 $\pm$ 0.12 &2.875 & 0.22 $\pm$ 0.06 &3.475 & 0.09 $\pm$ 0.03 &4.075 & 0.00 $\pm$ 0.03 \\ 
1.725 & 0.30 $\pm$ 0.11 &2.325 & 0.47 $\pm$ 0.10 &2.925 & 0.25 $\pm$ 0.06 &3.525 & 0.06 $\pm$ 0.03 &4.125 & 0.04 $\pm$ 0.02 \\ 
1.775 & 0.76 $\pm$ 0.15 &2.375 & 0.68 $\pm$ 0.11 &2.975 & 0.18 $\pm$ 0.05 &3.575 & 0.10 $\pm$ 0.03 &4.175 & 0.03 $\pm$ 0.02 \\ 
1.825 & 0.96 $\pm$ 0.16 &2.425 & 0.58 $\pm$ 0.10 &3.025 & 0.15 $\pm$ 0.05 &3.625 & 0.02 $\pm$ 0.02 &4.225 & 0.03 $\pm$ 0.02 \\ 
1.875 & 0.88 $\pm$ 0.16 &2.475 & 0.41 $\pm$ 0.09 &3.075 & 0.35 $\pm$ 0.07 &3.675 & 0.06 $\pm$ 0.03 &4.275 & 0.00 $\pm$ 0.03 \\ 
1.925 & 1.46 $\pm$ 0.18 &2.525 & 0.45 $\pm$ 0.09 &3.125 & 0.20 $\pm$ 0.05 &3.725 & 0.05 $\pm$ 0.03 &4.325 & 0.02 $\pm$ 0.01 \\ 
1.975 & 1.62 $\pm$ 0.20 &2.575 & 0.48 $\pm$ 0.09 &3.175 & 0.14 $\pm$ 0.04 &3.775 & 0.08 $\pm$ 0.02 &          &   \\ 
2.025 & 1.54 $\pm$ 0.19 &2.625 & 0.41 $\pm$ 0.08 &3.225 & 0.13 $\pm$ 0.04 &3.825 & 0.04 $\pm$ 0.03 &          &   \\ 
2.075 & 1.16 $\pm$ 0.16 &2.675 & 0.39 $\pm$ 0.08 &3.275 & 0.09 $\pm$ 0.03 &3.875 & 0.07 $\pm$ 0.02 &          &   \\ 
\hline
\end{tabular}
\end{table*}

\section{\bf\boldmath The $J/\psi$ region}
\subsection{\bf\boldmath The $\pipi3\piz$  final state}
\begin{figure}
\includegraphics[width=0.50\linewidth]{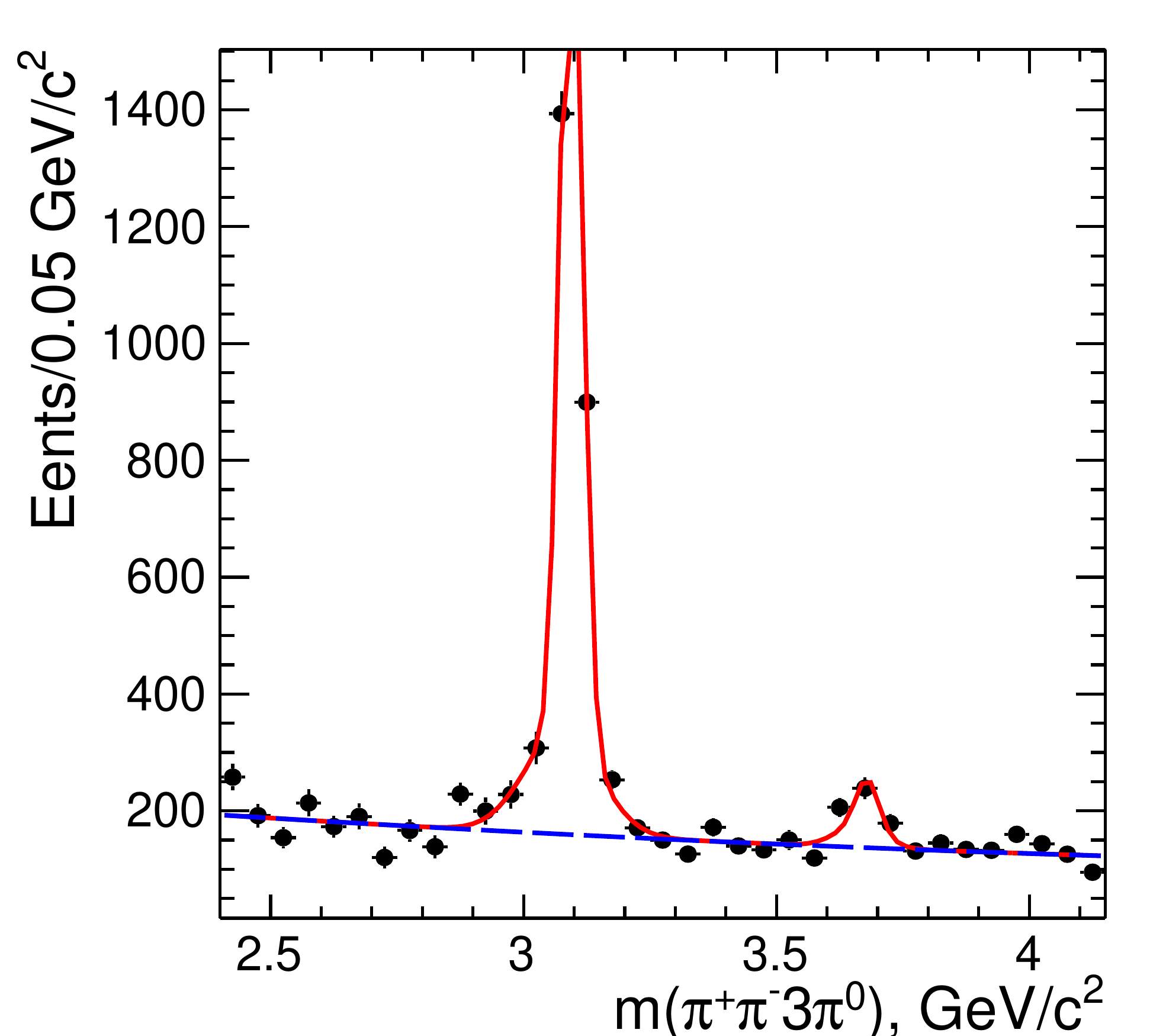}
\put(-40,80){\makebox(0,0)[lb]{\bf(a)}}
\includegraphics[width=0.49\linewidth]{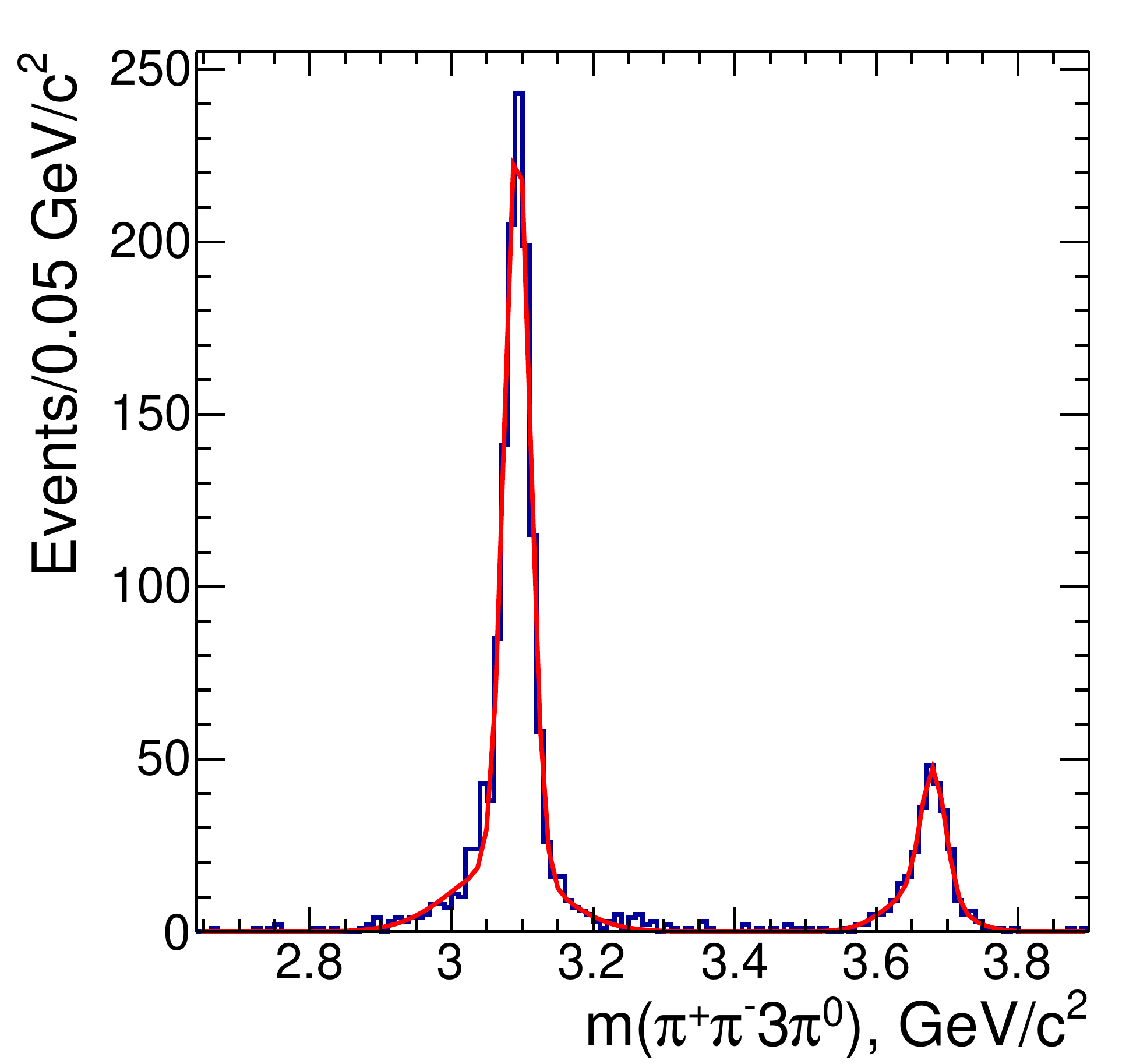}
\put(-40,80){\makebox(0,0)[lb]{\bf(b)}}
\caption{(a)
The $\pipi3\piz$ mass distribution for ISR-produced
$\epem\to\pipi\ppz\piz$ events in the $J/\psi$--$\psi(2S)$
        region.  
(b) The MC-simulated signals. The curves show the fit functions
described in the text.
}
\label{jpsi}
\end{figure}

Figure~\ref{jpsi}(a) shows an expanded view of the $J/\psi$ mass region 
from Fig.~\ref{nev_2pi3pi0_data} for the five-pion data sample. 
Signals from  $J/\psi\to\pipi\ppz\piz$ and $\psi(2S)\to 
\pipi\ppz\piz$  are clearly seen. 
The non-resonant background distribution is  flat in this region.

The observed peak shapes are not purely Gaussian because of radiation
effects and resolution, 
as is also seen in the simulated signal distributions
shown in Fig.~\ref{jpsi}(b).  The sum of two Gaussians with a common
mean is used to describe them.
We obtain $2389\pm63$  $J/\psi$ events and $177\pm27$
$\psi(2S)$ events.
Using the results for the number of events, the detection
efficiency, and the ISR luminosity,
we determine the product:
\begin{eqnarray}
  B_{J/\psi\to 5\pi}\cdot\Gamma^{J/\psi}_{ee}
  = \frac{N(J/\psi\to \pipi 3\piz)\cdot m_{J/\psi}^2}%
           {6\pi^2\cdot d{\cal L}/dE\cdot\epsilon^{\rm MC}\cdot\epsilon^{corr}\cdot C} \label{jpsieq}\\
  = (150\pm 4\pm  15)~\ev\ ,\nonumber
\end{eqnarray}
where $\Gamma^{J/\psi}_{ee}$ is
the electronic width, $d{\cal L}/dE =
  180~\invnb/\mev$ is the ISR luminosity 
at the $J/\psi$ mass $m_{J/\psi}$, $\epsilon^{\rm MC} = 0.041$ is the detection
efficiency from simulation with the corrections $\epsilon^{corr}=0.88$, discussed in 
Sec.~\ref{sec:Systematics},
and  $C = 3.894\times 10^{11}~\nb\mev^2$ is a
conversion constant ~\cite{PDG}. We estimate the systematic uncertainty for this
region to be 10\%, because no background subtraction is needed.  
The subscript ``$5\pi$'' for the branching fraction refers
to the $\pipi3\piz$ final state exclusively.

Using $\Gamma^{J/\psi}_{ee} =5.55\pm0.14~\kev$ ~\cite{PDG}, we obtain
$B_{J/\psi\to 5\pi} = (2.70\pm 0.07\pm 0.27)\times 10^{-2}$: no
other measurements for this channel exist.

\begin{figure}[tbh]
\includegraphics[width=0.50\linewidth]{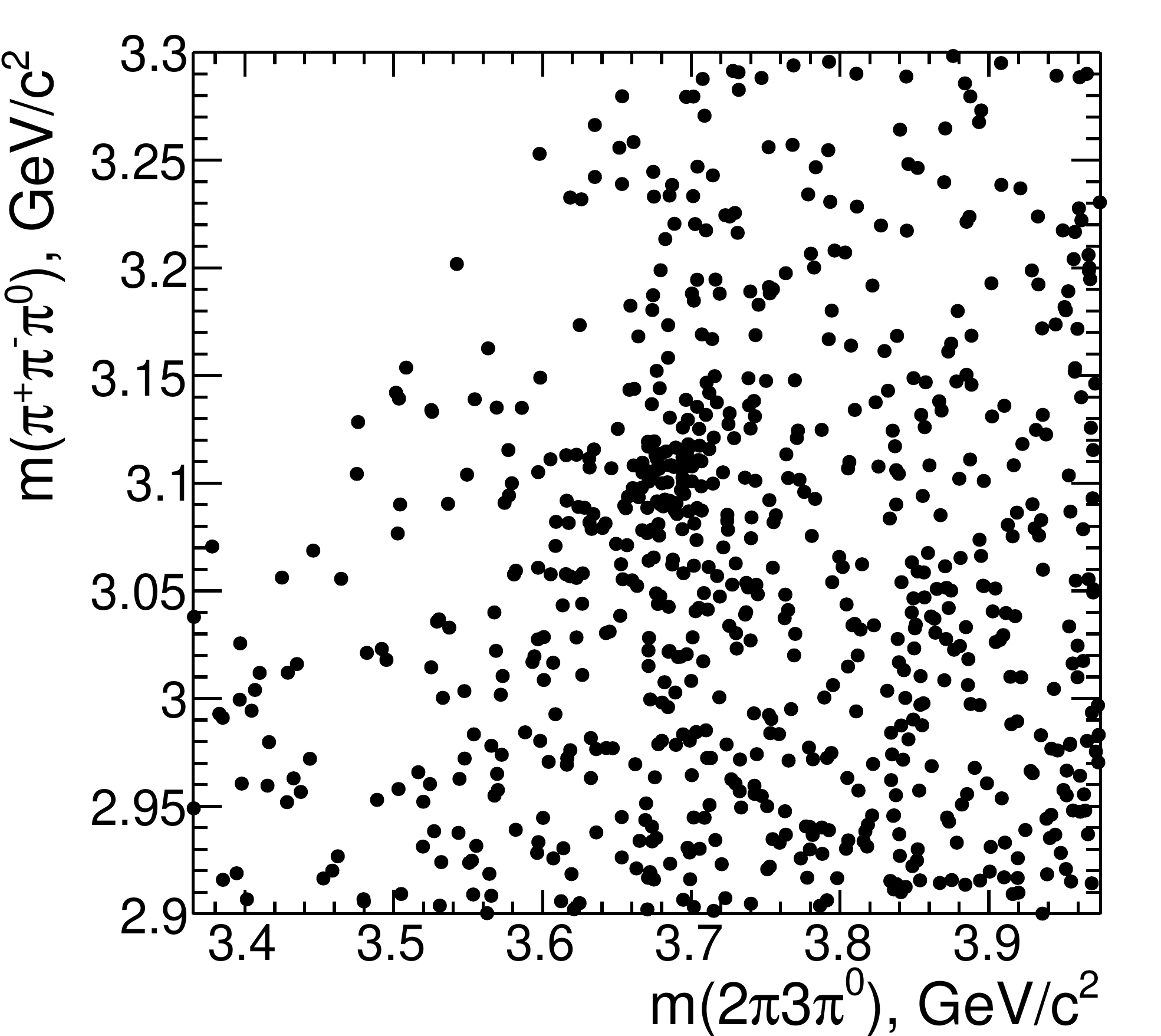}
\put(-80,80){\makebox(0,0)[lb]{\bf(a)}}
\includegraphics[width=0.48\linewidth]{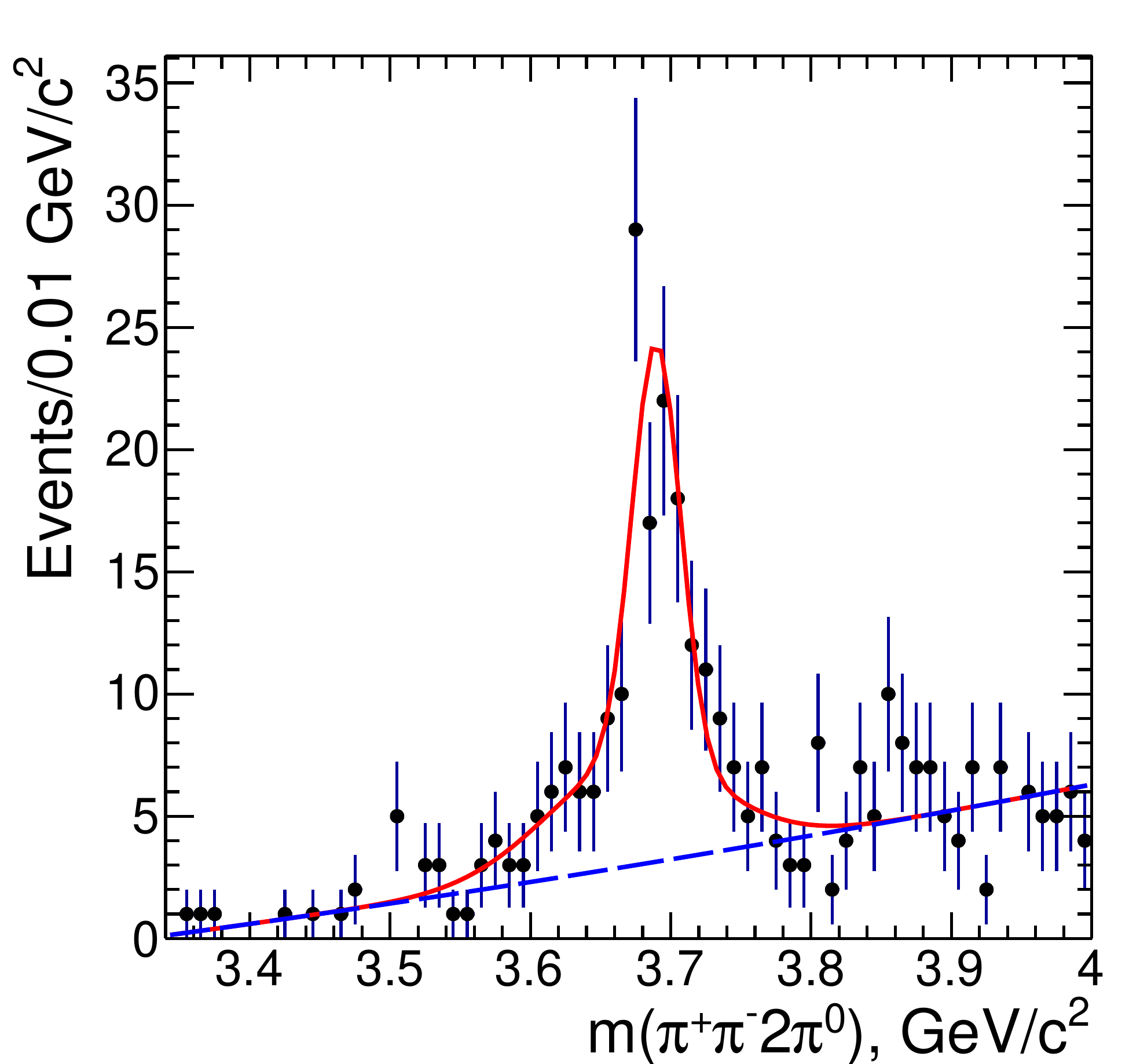}
\put(-80,80){\makebox(0,0)[lb]{\bf(b)}}
\caption{
(a) The three-pion combination closest to the $J/\psi$ mass vs
the five-pion mass.
(b) The five-pion mass for the 
events with the three-pion mass in the $\pm50$~\mevcc interval  around
the $J/\psi$ mass. The curves show the fit functions for all events
(solid) and the contribution of the background (dashed).
}
\label{psi2s_chain}
\end{figure}    
Using Eq.(\ref{jpsieq}) and the result $d{\cal L}/dE =
  228~\invnb/\mev$  at the $\psi(2S)$ mass, we obtain:
\begin{eqnarray*}
  B_{\psi(2S)\to 5\pi}\cdot\Gamma^{\psi(2S)}_{ee}
  &=& (12.4\pm1.9\pm1.2)~\ev\ .
\end{eqnarray*}
With $\Gamma^{\psi(2S)}_{ee} =2.34\pm0.06~\kev$ ~\cite{PDG}  we
find $B_{\psi(2S)\to 5\pi} = (5.2\pm 0.8\pm 0.5)\times
10^{-3}$. For this channel also, no previous result exists.

The $\psi(2S)$ peak partly corresponds to the decay chain
$\psi(2S)\to J/\psi\ppz\to\pipi\ppz\pi^0$, with $J/\psi$ decay to
three pions. We select the $\pipi\piz$
mass combination closest to the $J/\psi$ mass. 
Figure~\ref{psi2s_chain}(a) displays this $\pipi\piz$ mass vs the five-pion
invariant mass. A clear signal from the above decay chain is seen. 
We select events in a $\pm$0.05~\gevcc window
around the $J/\psi$ mass and project the results onto
$m(\pipi3\piz)$.  The results are shown in Fig.~\ref{psi2s_chain}(b).
Performing a fit to this distribution yields
$142\pm21$ $\psi(2S)\to J/\psi\ppz\to\pipi\ppz\piz$
events. In conjunction with the detection efficiency and
ISR luminosity, this yields:
\begin{eqnarray*}
  B_{\psi(2S)\to J/\psi\ppz}\cdot B_{J/\psi\to\pipi\pi^0}\cdot\Gamma^{\psi(2S)}_{ee}
  &=&\\ (10.1\pm1.5\pm1.1)~\ev\ .
\end{eqnarray*}
With $\Gamma^{\psi(2S)}_{ee}$as stated above  and $B_{\psi(2S)\to J/\psi\ppz} = 0.1817\pm
0.0031$~\cite{PDG}, we obtain $B_{J/\psi\to\pipi\pi^0} =
(2.29\pm 0.28\pm 0.23)\%$, in  agreement with our direct measurement
$B_{J/\psi\to\pipi\pi^0} = (2.18\pm 0.19)\%$~\cite{isr3pi} as well
as with the  PDG value $B_{J/\psi\to\pipi\pi^0} = (2.11\pm 0.07)\%$.
This gives us confidence that our normalization procedure is correct.

\begin{figure}[tbh]
\includegraphics[width=0.49\linewidth]{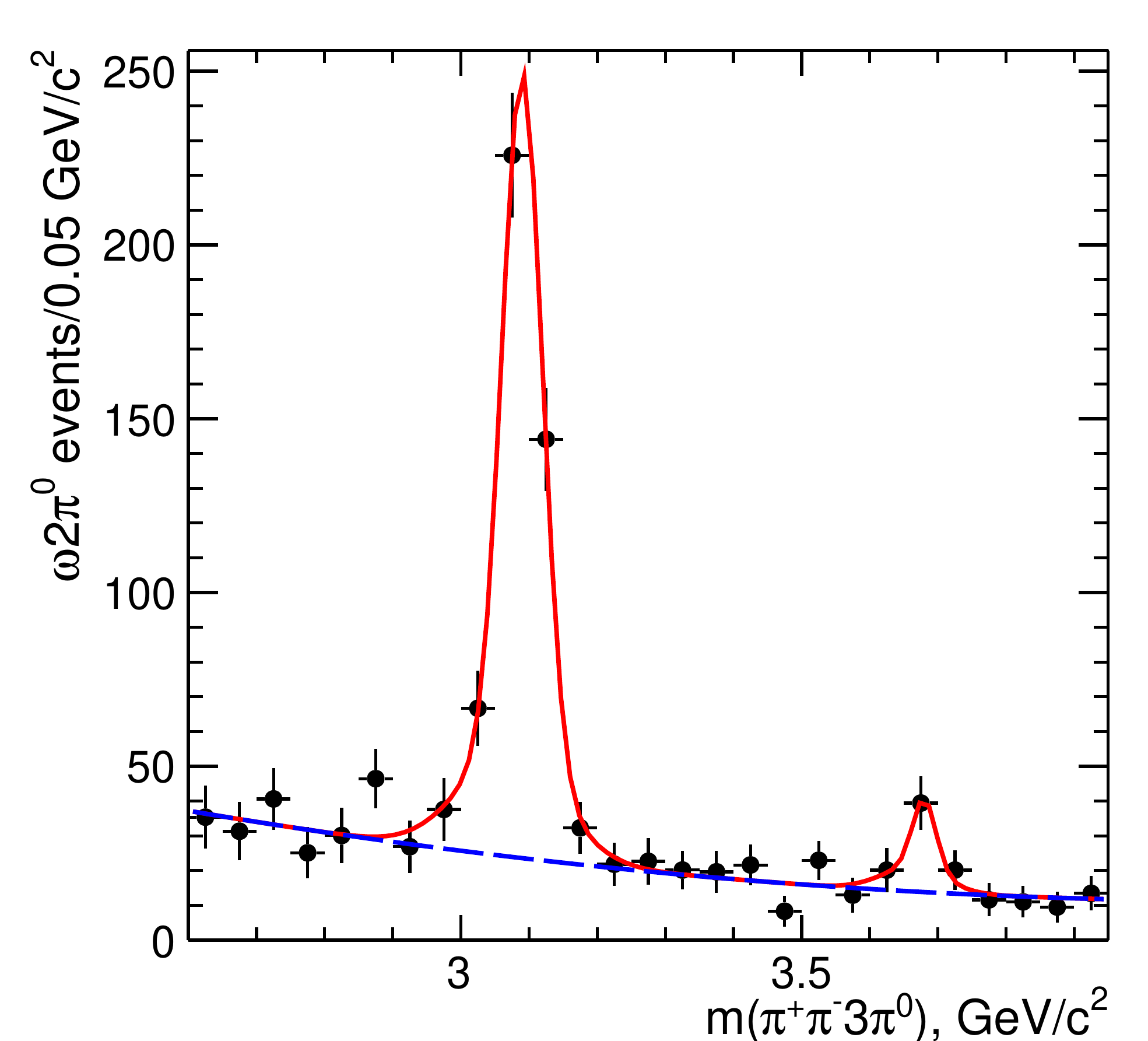}
\put(-40,80){\makebox(0,0)[lb]{\bf(a)}}
\includegraphics[width=0.49\linewidth]{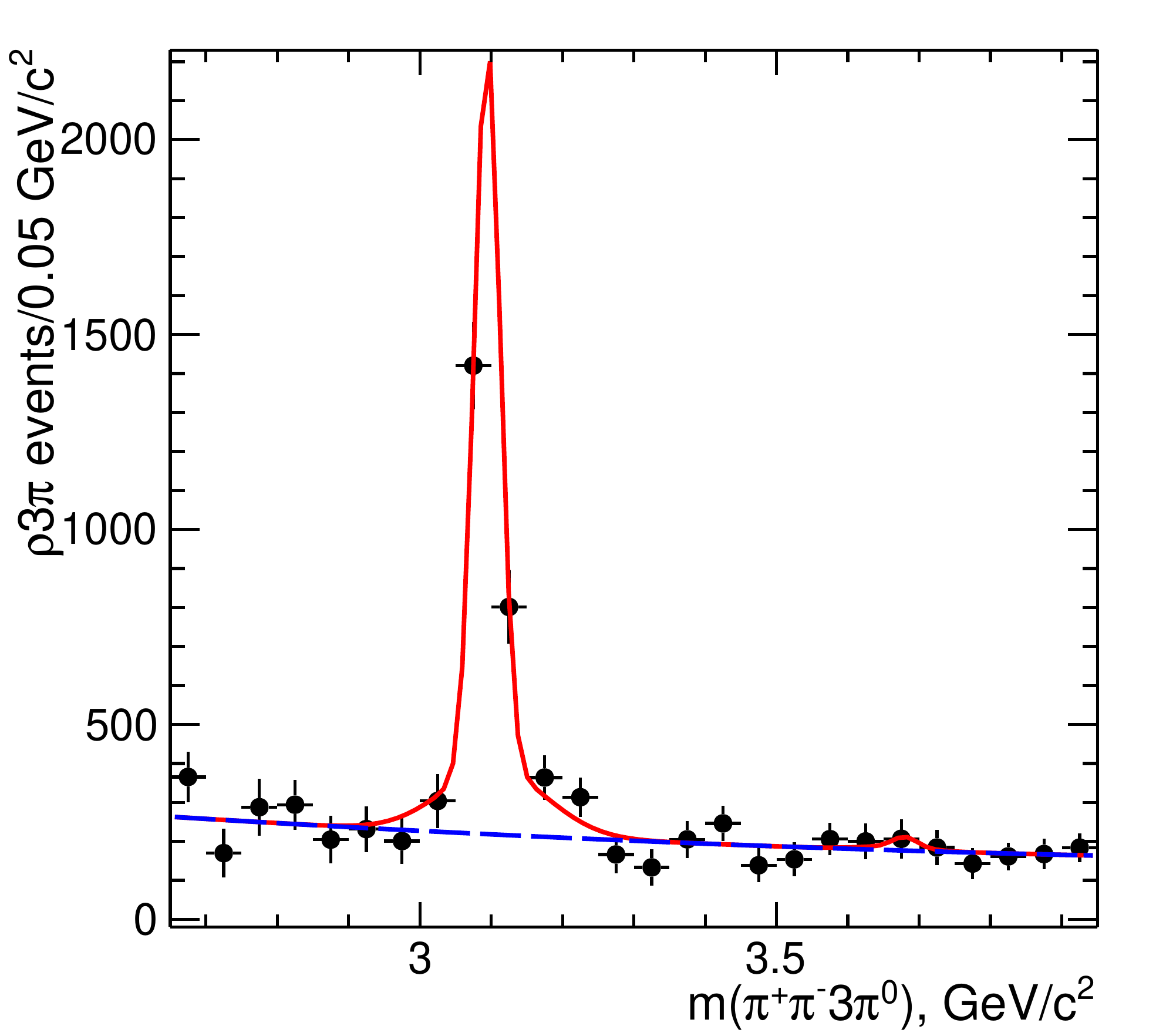}
\put(-40,80){\makebox(0,0)[lb]{\bf(b)}}
\caption{(a) The five-pion mass for
events with the three-pion combination in the $\omega(782)$ mass region.
(b) The five-pion mass for
events with $\pi^{\pm}\piz$ combination in the $\rho(770)$ mass
region. The curves show the fit functions
described in the text.
}
\label{etaomega2pi}
\end{figure}

\subsubsection{\bf\boldmath The $\omega\ppz$ intermediate state}

The $J/\psi\to\eta\pipi$ branching fraction is very small,  as we observed in our
previous publication~\cite{isretapipi}, 
and there is not a statistically significant
signal in our sample, shown in Fig.~\ref{neveta2pi}.
 We do not attempt to extract a $J/\psi$ branching
fraction for this channel.

Figure~\ref{etaomega2pi}(a) shows 
an expanded view of Fig.~\ref{nevomega2pi0} with the $\pipi3\piz$ mass
distribution for events obtained by a fit to the $\pipi\piz$
mass distribution.
The two-Gaussian fit, implemented as discribed above, yields $398\pm29$ and $33\pm10$ events for
the $J/\psi$ and $\psi(2S)$, respectively. 
Using Eq.(\ref{jpsieq}) we obtain:
\begin{eqnarray*}
  B_{J/\psi\to\omega\ppz}\cdot B_{\omega\to\pipi\pi^0}\cdot\Gamma^{J/\psi}_{ee}
  &=&\\ (24.9\pm1.8\pm2.5)\ev\ ,\\
  B_{\psi(2S)\to\omega\pipi}\cdot B_{\omega\to\pipi\pi^0}\cdot\Gamma^{\psi(2S)}_{ee}
  &=& \\(2.3\pm0.7\pm0.2)\ev\ .
\end{eqnarray*} 
Using $B_{\omega\to\pipi\pi^0} = 0.891$ and the value of $\Gamma_{ee}$ from
Ref.~\cite{PDG}, we
obtain $B_{J/\psi\to\omega\ppz} = (5.04\pm 0.37\pm 0.50)\times 10^{-3}$
and $B_{\psi(2S)\to\omega\ppz} = (1.1\pm 0.3\pm 0.1)\times
10^{-3}$. 
The value of $B_{J/\psi\to\omega\ppz}$ listed in Ref.~\cite{PDG}, based
on the DM2~\cite{DM2} result, is $(3.4\pm 0.8)\times 10^{-3}$ . There is no
previous result for $B_{\psi(2S)\to\omega\ppz}$.
Note that our result for $B_{J/\psi\to\omega\ppz}$  is about a factor of two lower
than our result $B_{J/\psi\to\omega\pipi} = (9.7\pm 0.9)\times
10^{-3}$~\cite{isr5pi}, as expected from  isospin symmetry. 
\begin{figure}[tbh]
\includegraphics[width=0.49\linewidth]{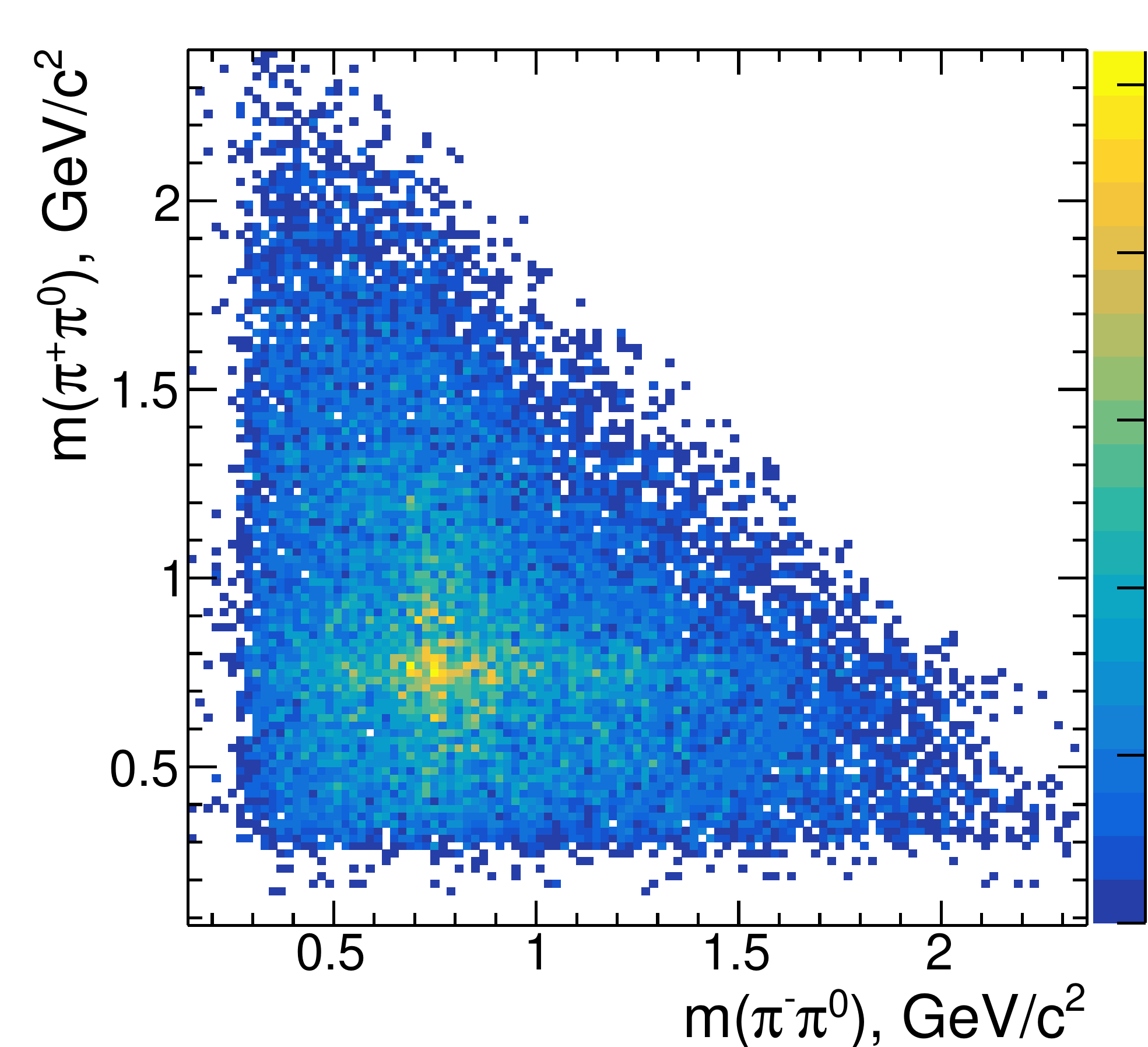}
\put(-40,80){\makebox(0,0)[lb]{\bf(a)}}
\includegraphics[width=0.49\linewidth]{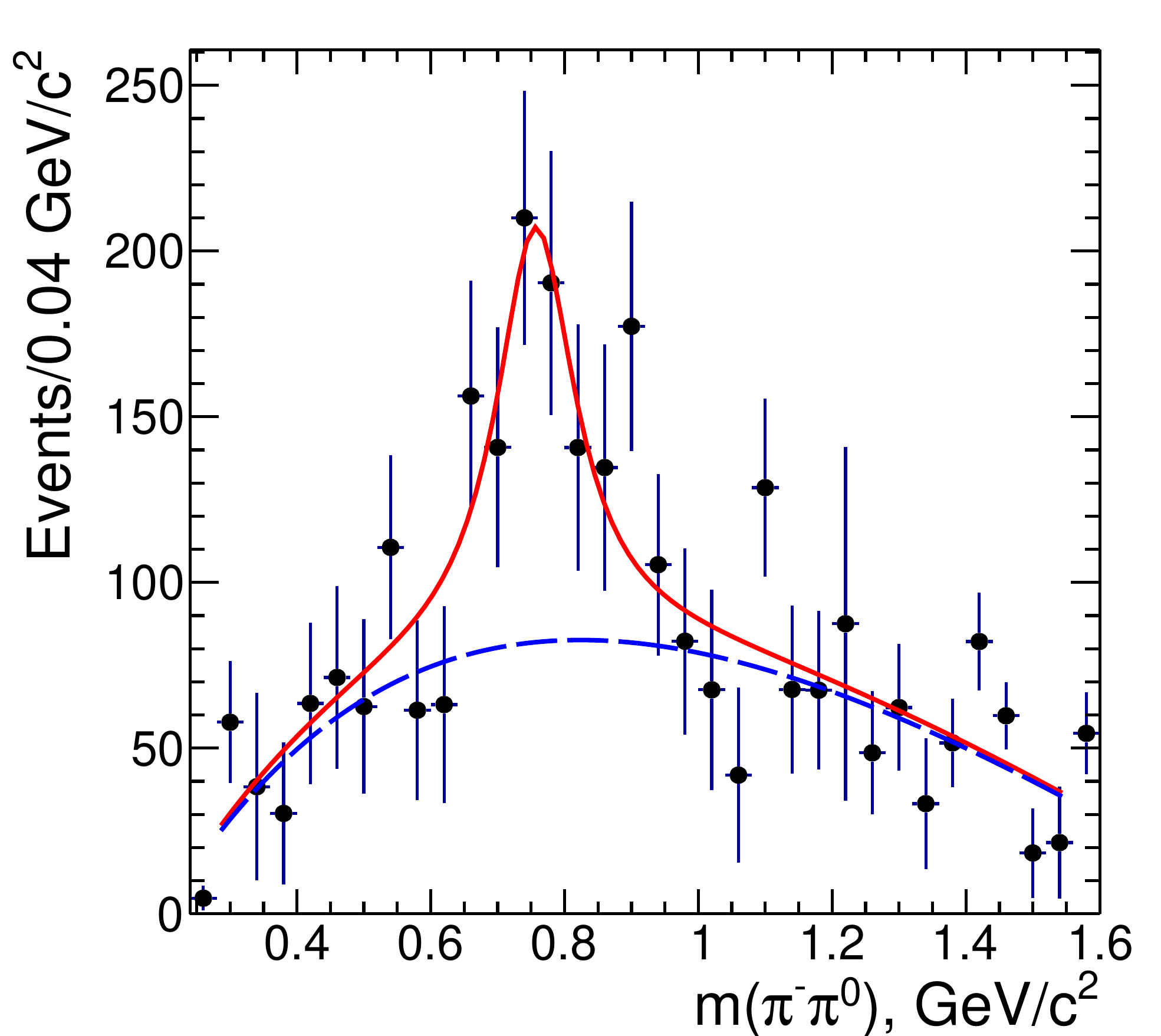}
\put(-40,80){\makebox(0,0)[lb]{\bf(b)}}
\caption{(a) Scatter plot of the $\pi^+\piz$ vs the $\pi^-\piz$ invariant
  mass for the $J/\psi$ region in Fig.~\ref{etaomega2pi}(b).
(b) Number of $\pi^+\piz$ events in bins of 0.04~\gevcc  
in the $\pi^-\piz$ mass. The curves show the fit functions for all events
(solid) and the contribution of the background (dashed).
}
\label{jpsirhorho}
\end{figure}    
\begin{figure*}[tbh]
\includegraphics[width=0.32\linewidth]{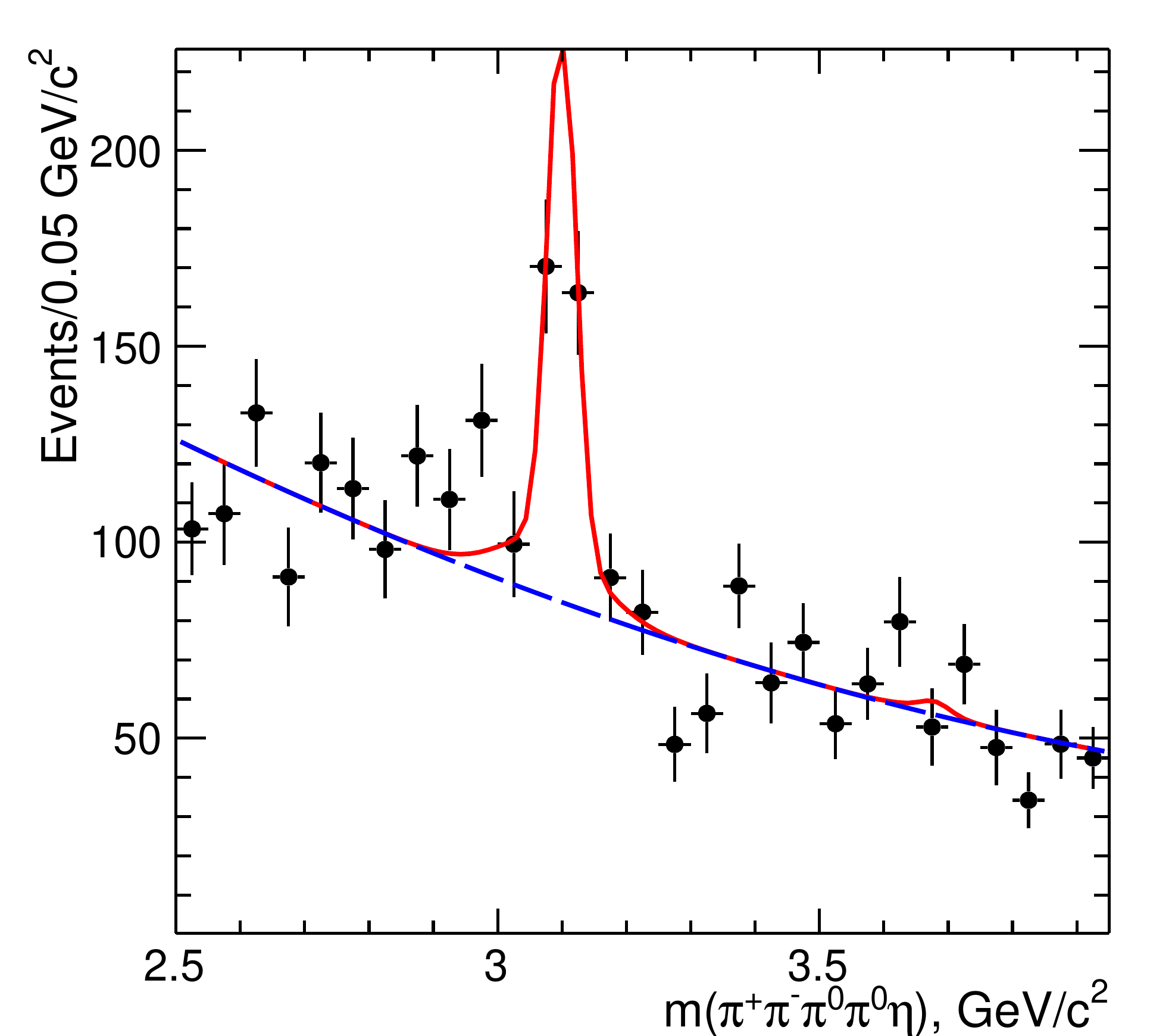}
\put(-50,110){\makebox(0,0)[lb]{\bf(a)}}
\includegraphics[width=0.31\linewidth]{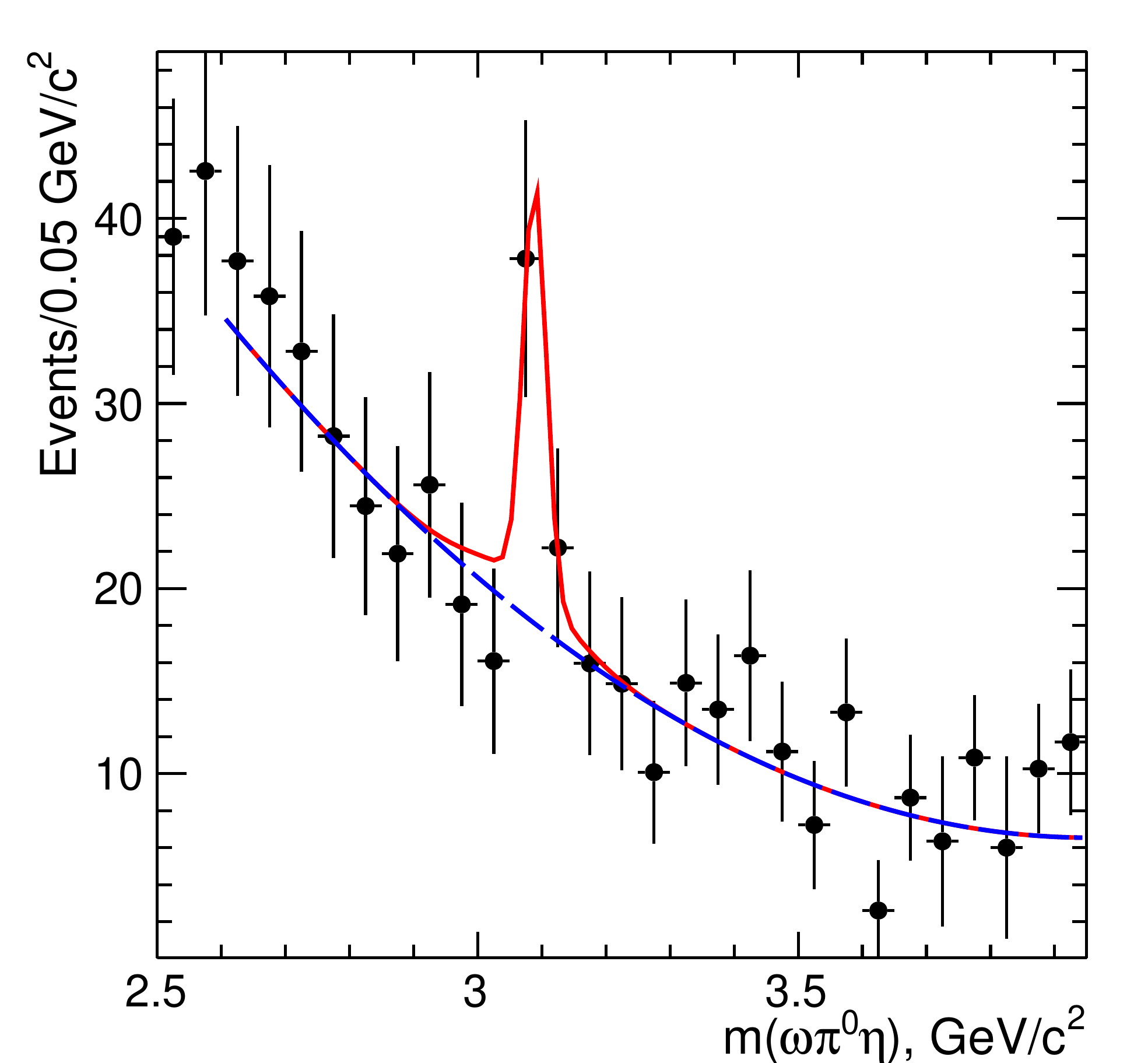}
\put(-50,110){\makebox(0,0)[lb]{\bf(b)}}
\includegraphics[width=0.31\linewidth]{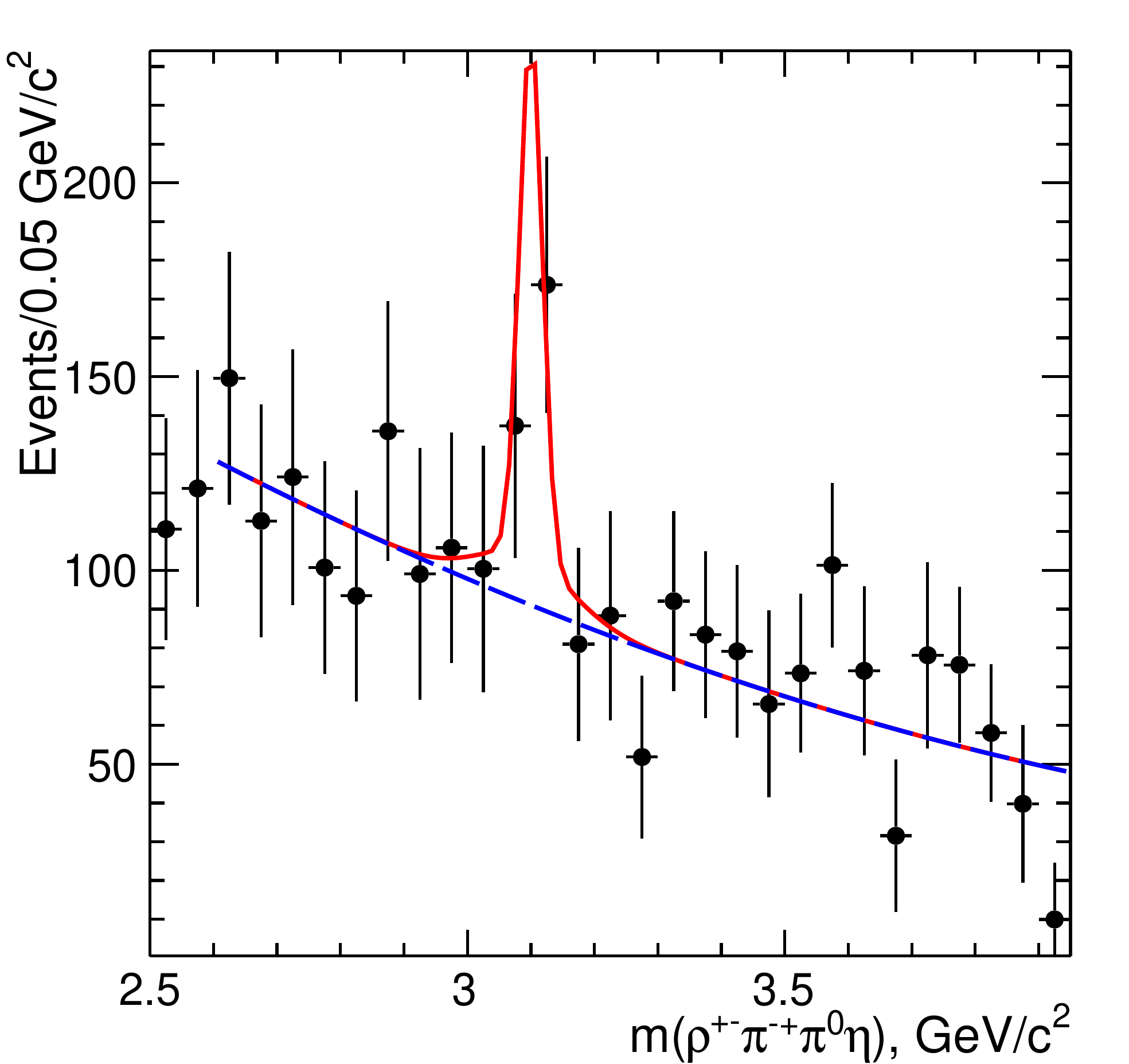}
\put(-50,110){\makebox(0,0)[lb]{\bf(c)}}
\caption{
The $J/\psi$ region for the (a) $\pipi2\piz\eta$,
(b)  $\omega\piz\eta$, and 
(c)  $\rho^{\pm}\pi^{\mp}\piz\eta$ events. The curves show the fit
functions described in the text.
}
\label{2pi2pi0etaatjpsi}
\end{figure*}    
\subsubsection{\bf\boldmath The $\rho^{\pm}\pi^{\mp}\ppz$ intermediate state}
Figure~\ref{etaomega2pi}(b) shows 
an expanded view of Fig.~\ref{sumallev}(a) (squares) for the $\pipi3\piz$ mass,
for events obtained from the fit to the $\rho$ signal in the $\pi^{\pm}\piz$ mass.
The two-Gaussian fit yields $2299\pm201$ and $<88$ events at 90\% C.L.  for
the $J/\psi$ and $\psi(2S)$, respectively. 

The obtained $J/\psi\to\rho^{\pm}\pi^{\mp}\ppz$ result exceeds the total number of
observed $J/\psi$ events. This is  because of  $J/\psi$ decays
to $\rho^+\rho^-\piz$.
Figure~\ref{jpsirhorho}(a) shows a scatter plot
of the $\pi^+\piz$ vs the $\pi^-\piz$ invariant mass for 3051 events in
a $\pm$0.1~\gevcc interval around the $J/\psi$
peak of Fig.~\ref{etaomega2pi}(b). To determine the rate of
correlated $\rho^+\rho^-$ production, we fit the $\pi^+\piz$ invariant mass with a BW and
combinatorial background function in intervals of 0.04~\gevcc in the $\pi^-\piz$
mass distribution. The resulting distribution exibits a clear  $\rho$ peak, shown
in Fig.~\ref{jpsirhorho}(b),  with a correlated $\rho^+\rho^-$ yield of
$703\pm153$ events, corresponding to 46$\pm$8\%  of the
$\rho^{\pm}\pi^{\mp}\ppz$ events.
Using this value we estimate the number of  $J/\psi$ decays to single-
and double-$\rho$ to be  $1241\pm109\pm183$ and $529\pm46\pm92$, respectively.
The second uncertainty is from the uncertainty in
the fraction of $\rho^+\rho^-$ events,  given above.
We obtain:
\begin{eqnarray*}
  B_{J/\psi\to\rho^{\pm}\pi^{\mp}\ppz}\cdot\Gamma^{J/\psi}_{ee}
  &=& (78\pm7\pm8\pm6)\ev\ ,\\
  B_{J/\psi\to\rho^+\rho^-\piz}\cdot\Gamma^{J/\psi}_{ee}
  &=& (33\pm3\pm3\pm3)\ev\ .
\end{eqnarray*} 
Dividing by the value of $\Gamma_{ee}$ from Ref.~\cite{PDG}
then yields:
\begin{eqnarray*}
  B_{J/\psi\to\rho^{\pm}\pi^{\mp}\ppz}
  &=& (1.40\pm0.12\pm0.14\pm0.10)\times10^{-2} ,\\
  B_{J/\psi\to\rho^+\rho^-\piz}
  &=& (0.60\pm0.05\pm0.06\pm0.05)\times10^{-2} ,
\end{eqnarray*} 
where the third uncertainty is associated with the
uncertainty arising from the procedure used to determine
the correlated $\rho^+\rho^-$ rate.
No other measurements for these processes exist.

\subsection{\bf\boldmath The $\pipi2\piz\eta$  final state}

Figure~\ref{2pi2pi0etaatjpsi} shows an expanded view of
Fig.~\ref{sumallev_eta},
with 
a clear $J/\psi$ signal seen in all three distributions:
the inclusive $\pipi2\piz\eta$ mass distribution (Fig.~\ref{2pi2pi0etaatjpsi}(a))
and the mass distributions for the $\omega\piz\eta$ (Fig.~\ref{2pi2pi0etaatjpsi}(b))
and $\rho^{\pm}\pi^{\mp}\piz\eta$ (Fig.~\ref{2pi2pi0etaatjpsi}(c)) intermediate states.
Our fits yield $203\pm29$, $27\pm14$, and $168\pm62$ events for the
$J/\psi$ decays into these final states, respectively. Only an upper
limit with $<12$ events at
90\% C.L. is obtained for  the $\psi(2S)$ decay to $\pipi2\piz\eta$.
We determine:
\begin{eqnarray*}
  B_{J/\psi\to\pipi\ppz\eta}\cdot\Gamma^{J/\psi}_{ee}
  &=& (12.8\pm1.8\pm2.0)\ev\ ,\\
  B_{J/\psi\to\omega\piz\eta}\cdot B_{\omega\to 3\pi}\cdot\Gamma^{J/\psi}_{ee}
  &=& (1.7\pm0.8\pm0.3)\ev\ ,\\
  B_{J/\psi\to\rho^{\pm}\pi^{\mp}\piz\eta}\cdot\Gamma^{J/\psi}_{ee}
  &=& (10.5\pm4.1\pm1.6)\ev\ ,\\
  B_{\psi(2S)\to\pipi\ppz\eta}\cdot\Gamma^{\psi(2S)}_{ee}
  &<& 0.85\ev\ \rm{at ~90\% ~C.L.} .
\end{eqnarray*} 
Dividing by the appropriate $\Gamma_{ee}$ value from Ref.~\cite{PDG},
we find
$B_{J/\psi\to\pipi\ppz\eta} =
(2.30\pm 0.33\pm 0.35)\times 10^{-3}$,  $B_{J/\psi\to\omega\piz\eta} =
(3.4\pm 1.6\pm 0.6)\times 10^{-4}$, $B_{J/\psi\to\rho^{\pm}\pi^{\mp}\piz\eta} =
(1.9\pm 0.7\pm 0.3)\times 10^{-3}$, and $B_{\psi(2S)\to\pipi\ppz\eta} <
3.5\times 10^{-4}$ at 90\% C.L.. 
There are no previous results for these final states.

\begin{table*}[tbh]
\caption{
  Summary of the $J/\psi$ and $\psi(2S)$ branching fractions.
  }
\label{jpsitab}
\begin{tabular}{r@{$\cdot$}l  r@{.}l@{$\pm$}l@{$\pm$}l 
                              r@{.}l@{$\pm$}l@{$\pm$}l
                              r@{.}l@{$\pm$}l } 
\multicolumn{2}{c}{Measured} & \multicolumn{4}{c}{Measured}    &  
\multicolumn{7}{c}{$J/\psi$ or $\psi(2S)$ Branching Fraction  (10$^{-3}$)}\\
\multicolumn{2}{c}{Quantity} & \multicolumn{4}{c}{Value (\ev)} &
\multicolumn{4}{c}{Calculated, this work}    & 
\multicolumn{3}{c}{PDG~\cite{PDG}} \\
\hline
$\Gamma^{J/\psi}_{ee}$  &  $\BR_{J/\psi \to \pipi\ppz\piz}$  &
  150& 0 & 4.0 & 15.0  &   ~~~27&0 & 0.7 & 2.7  &   \multicolumn{3}{c}{no entry}  \\

$\Gamma^{J/\psi}_{ee}$  &  $\BR_{J/\psi  \to \omega\ppz}
                           \cdot \BR_{\omega    \to 3\pi}$  &
  24&8 & 1.8 & 2.5  &   5&04 & 0.37 & 0.50  &   ~~~~3&4 & 0.8  \\

$\Gamma^{J/\psi}_{ee}$  &  $\BR_{J/\psi  \to  \rho^{\pm}\pi^{\mp}\ppz}$&
  78&0 & 9.0 & 8.0  &   14&0 & 1.2 & 1.4  &  \multicolumn{3}{c}{no entry} \\

$\Gamma^{J/\psi}_{ee}$  &  $\BR_{J/\psi  \to \rho^+\rho^-\piz}  $  &
   33&0& 5.0& 3.3 &   6&0 & 0.9 & 0.6  &    \multicolumn{3}{c}{no entry}\\

$\Gamma^{J/\psi}_{ee}$  &  $\BR_{J/\psi  \to \pipi\ppz\eta} $ &
   12&8& 1.8& 2.0 &   2&30 & 0.33 & 0.35  &  \multicolumn{3}{c}{no entry} \\

$\Gamma^{J/\psi}_{ee}$  &  $\BR_{J/\psi  \to \omega\piz\eta}  
                        \cdot \BR_{\omega\to 3\pi} $  &
   1&7& 0.8 & 0.3 &   0&34 & 0.16 & 0.06  &  \multicolumn{3}{c}{no entry}\\

$\Gamma^{J/\psi}_{ee}$  &  $\BR_{J/\psi  \to \rho^{\pm}\pi^{\mp}\piz\eta}$ &  
   10&5& 4.1 & 1.6 &   1&7 & 0.7 & 0.3  &  \multicolumn{3}{c}{no entry}\\

$\Gamma^{\psi(2S)}_{ee}$  &  $\BR_{\psi(2S) \to\pipi\ppz\piz} $  &
   12&4& 1.8& 1.2 &   5&2  & 0.8  & 0.5   &  \multicolumn{3}{c}{no entry} \\

$\Gamma^{\psi(2S)}_{ee}$  &  $\BR_{\psi(2S) \to J/\psi\ppz} 
                             \cdot \BR_{J/\psi \to 3\pi}  $  &
   10&1& 1.5& 1.1 &   22&9  & 2.8  & 2.3   &   ~~~~~~21&1 & 0.7 \\

$\Gamma^{\psi(2S)}_{ee}$  &  $\BR_{\psi(2S) \to \omega\ppz}
                        \cdot \BR_{\omega     \to 3\pi} $  &
   2&3& 0.7& 0.2 &   1&1 & 0.3 & 0.1  &   \multicolumn{3}{c}{no  entry}\\ 

$\Gamma^{\psi(2S)}_{ee}$  &  $\BR_{\psi(2S) \to \rho^{\pm}\pi^{\mp}\ppz} $  &
   ~~~~~~~~$<$6 &\multicolumn{3}{l}{2 at 90\% C.L.} &  $<$2  &
                                                              \multicolumn{3}{l}{6
                                                              at 90\% C.L.}   &    \multicolumn{3}{c}{no entry} \\

$\Gamma^{\psi(2S)}_{ee}$  &  $\BR_{\psi(2S) \to \pipi\ppz\eta} $  &
   ~~~~$<$0 & \multicolumn{3}{l}{85 at 90\% C.L.} & ~~~~ $<$0 &
                                                     \multicolumn{3}{l}{35
                                                     at 90\% C.L.} &      \multicolumn{3}{c}{no entry} \\

\hline
\end{tabular}
\end{table*}

\subsection{Summary of the charmonium region study}
The rates of  $J/\psi$ and $\psi(2S)$ decays to $\pipi3\piz$,
$\pipi2\piz\eta$ and several intermediate  final
states have been measured. A small discrepancy with only one available current PDG
value, measured by the DM2 experiment~\cite{DM2}, is observed for the $J/\psi\to\omega\ppz$ decay rate. The
measured products and calculated 
branching fractions are summarized in Table~\ref{jpsitab}
together with the available PDG values for comparison.

\section{Summary}
\label{sec:Summary}
\noindent
The photon-energy and charged-particle momentum 
resolutions together with the particle
identification capabilities of the \babar\ detector permit the
reconstruction of the 
$\pipi3\piz$ and $\pipi2\piz\eta$
final states produced at low  effective center-of-mass energies 
via initial-state photon radiation in data collected  in $\epem$
annihilation in the $\Upsilon(4S)$  mass region.

The analysis shows that the effective luminosity and efficiency have been understood
with 10--13\% accuracy.
The cross section measurements for the reaction $\epem\to\pipi\ppz\piz$
present  a significant improvement on existing data.
The $\epem\to\pipi\ppz\eta$ cross section
has been measured for the first time.

The selected multi-hadronic final states in the broad range of accessible
energies provide new information on hadron spectroscopy. The  
observed $\ep\en\to \omega\ppz$ and $\epem\to \eta\pipi$  cross
sections provide  evidence of
resonant structures around 1.4 and 1.7~\gevcc, which were previously observed by
DM2 and interpreted as $\omega(1450)$ and $\omega(1650)$
resonances. 

The initial-state radiation events allow a study of $J/\psi$ and
$\psi(2S)$ production and a measurement of the corresponding products of
the decay branching fractions and $\epem$ width for most of
the studied channels, the majority of them for the first time.


\section{Acknowledgments}
\label{sec:Acknowledgments}

We are grateful for the 
extraordinary contributions of our \pep2\ colleagues in
achieving the excellent luminosity and machine conditions
that have made this work possible.
The success of this project also relies critically on the 
expertise and dedication of the computing organizations that 
support \babar.
The collaborating institutions wish to thank 
SLAC for its support and the kind hospitality extended to them. 
This work is supported by the
US Department of Energy
and National Science Foundation, the
Natural Sciences and Engineering Research Council (Canada),
the Commissariat \`a l'Energie Atomique and
Institut National de Physique Nucl\'eaire et de Physique des Particules
(France), the
Bundesministerium f\"ur Bildung und Forschung and
Deutsche Forschungsgemeinschaft
(Germany), the
Istituto Nazionale di Fisica Nucleare (Italy),
the Foundation for Fundamental Research on Matter (The Netherlands),
the Research Council of Norway, the
Ministry of Education and Science of the Russian Federation, 
Ministerio de Econom\'{\i}a y Competitividad (Spain), the
Science and Technology Facilities Council (United Kingdom),
and the Binational Science Foundation (U.S.-Israel).
Individuals have received support from 
the Marie-Curie IEF program (European Union) and the A. P. Sloan Foundation (USA).

\end{document}